\documentclass[11pt,english]{article}
\usepackage[T1]{fontenc}
\usepackage[latin9]{inputenc}
\usepackage{geometry}
\usepackage{color}
\usepackage{babel}
\usepackage{float}
\usepackage{booktabs}
\usepackage{mathrsfs}
\usepackage{mathtools}
\usepackage{url}
\usepackage{amsmath}
\usepackage{amsthm}
\usepackage{amssymb}
\usepackage{graphicx}
\usepackage{setspace}
\usepackage[authoryear]{natbib}
\usepackage[unicode=true,pdfusetitle,
 bookmarks=true,bookmarksnumbered=false,bookmarksopen=false,
 breaklinks=true,pdfborder={0 0 0},pdfborderstyle={},backref=false,colorlinks=true]
 {hyperref}
\hypersetup{
 linkcolor=blue, citecolor=blue}

\makeatletter

\providecommand{\tabularnewline}{\\}

\theoremstyle{plain}
\newtheorem{assumption}{\protect\assumptionname}
\theoremstyle{plain}
\newtheorem{thm}{\protect\theoremname}

\allowdisplaybreaks

\usepackage{bbm}

\date{}

\usepackage{xr}
\newtheoremstyle{remboldstyle}
  {}{}{}{}{\bfseries}{.}{.5em}{{\thmname{#1 }}{\thmnumber{#2}}{\thmnote{ (#3)}}}
\theoremstyle{remboldstyle}
\newtheorem{rembold}{Remark}

\usepackage{appendix}
\appendixtitleon

\usepackage{scalerel,stackengine}
\stackMath
\newcommand\reallywidecheck[1]{%
\savestack{\tmpbox}{\stretchto{%
  \scaleto{%
    \scalerel*[\widthof{\ensuremath{#1}}]{\kern-.6pt\bigwedge\kern-.6pt}%
    {\rule[-\textheight/2]{1ex}{\textheight}}
  }{\textheight}%
}{0.5ex}}%
\stackon[1pt]{#1}{\scalebox{-1}{\tmpbox}}%
}
\date{}
\allowdisplaybreaks



\newtheoremstyle{ntnboldstyle}
  {}{}{}{}{\bfseries}{.}{.5em}{{\thmname{#1}}{\thmnote{(#2)}}}
\theoremstyle{ntnboldstyle}

\makeatletter
\def\@seccntformat#1{\@ifundefined{#1@cntformat}%
   {\csname the#1\endcsname\quad}  
   {\csname #1@cntformat\endcsname}
}
\let\oldappendix\appendix 
\renewcommand\appendix{%
    \oldappendix
    \newcommand{\section@cntformat}{\appendixname~\thesection\quad}
}
\makeatother

\makeatother

\providecommand{\assumptionname}{Assumption}
\providecommand{\theoremname}{Theorem}

\begin{document}
\title{\textbf{Empirical Likelihood Covariate Adjustment for Regression Discontinuity
Designs}\thanks{We thank the coeditor, the associate editor and two anonymous referees,
whose comments have greatly improved the paper. We thank Matias Cattaneo,
Vadim Marmer and Taisuke Otsu for their helpful comments. All errors
are ours. Jun Ma acknowledges the financial support from the National
Natural Science Foundation of China (Grant Numbers 71903190, and 72394392).
Zhengfei Yu acknowledges the financial support from JSPS KAKENHI (Grant
Number 21K01419).}\textbf{ }\let\thefootnote\relax\footnotetext{This version: April 22, 2024}}
\author{Jun Ma\thanks{School of Economics, Renmin University of China} \and
Zhengfei Yu\thanks{Faculty of Humanities and Social Sciences, University of Tsukuba}}
\maketitle
\begin{abstract}
This paper proposes a versatile covariate adjustment method that directly
incorporates covariate balance in regression discontinuity (RD) designs.
The new empirical entropy balancing method reweights the standard
local polynomial RD estimator by using the entropy balancing weights
that minimize the Kullback--Leibler divergence from the uniform weights
while satisfying the covariate balance constraints. Our estimator
can be formulated as an empirical likelihood estimator that efficiently
incorporates the information from the covariate balance condition
as correctly specified over-identifying moment restrictions, and thus
has an asymptotic variance no larger than that of the standard estimator
without covariates. We demystify the asymptotic efficiency gain of
\citet*{calonico2019regression}'s regression-based covariate-adjusted
estimator, as their estimator has the same asymptotic variance as
ours. Further efficiency improvement from balancing over sieve spaces
is possible if our entropy balancing weights are computed using stronger
covariate balance constraints that are imposed on functions of covariates.
We then show that our method enjoys favorable second-order properties
from empirical likelihood estimation and inference: the estimator
has a small (bounded) nonlinearity bias, and the likelihood ratio
based confidence set admits a simple analytical correction that can
be used to improve coverage accuracy. The coverage accuracy of our
confidence set is robust against slight perturbation to the covariate
balance condition, which may happen in cases such as data contamination
and misspecified ``unaffected'' outcomes used as covariates. The
proposed entropy balancing approach for covariate adjustment is applicable
to other RD-related settings. For example, we derive a covariate-adjusted
estimator of the treatment effect derivative of \citet{Dong2015}
and show that it incorporates the covariate information in a more
transparent and flexible way than the regression-based adjustment.
We conduct Monte Carlo simulations to assess our method's finite-sample
performance and also apply it to a real dataset.
\end{abstract}

\section{Introduction}

The RD design resembles a randomized experiment conducted near the
cut-off of the score (forcing variable) and exploits the discontinuous
variation in the probability of treatment to nonparametrically identify
the local average treatment effect (LATE) at the cut-off under mild
continuity assumptions on the latent variables.\footnote{In a recent study, \citet{hyytinen2018does} confirmed that RD produces
estimates that are in line with the results from a comparable experiment
if inference is implemented with the method of \citet{calonico2014robust}.} The transparent close-form identification (\citealp{hahn2001identification})
of the RD LATE calls for nonparametric estimation and inference methods
as they avoid functional form assumptions. See \citet{Cattaneo2019}
for a recent review of RD. In practical implementations, information
from pre-treatment covariates (i.e., variables that have already been
determined before the assignment of the treatment) is incorporated
to enhance efficiency and compensate for the low accuracy of nonparametric
methods. A widely used procedure is augmented local polynomial (LP)
regression, where the covariates enter linearly. \citet[CCFT, hereafter]{calonico2019regression}
formalize this augmented regression approach and derive its (first-order)
asymptotic properties. CCFT shows that the augmented LP regression
estimator consistently estimates the RD LATE under the covariate balance
condition, i.e., the expectations of covariates coincide at both sides
of the cut-off. Apart from CCFT, covariate adjustment for RD has received
much attention in recent literature. See \citet{frolich2019including}
for an alternative approach that requires smoothing over covariates
but allows for the potential failure of covariate balance. \citet{arai2021regression}
and \citet{Kreiss2022} extend CCFT's approach to control for a high-dimensional
covariate vector by regularization. \citet{Noack2021} extend CCFT's
linear regression adjustment to nonparametric adjustment with machine
learning methods. See \citet{Cattaneo2021} for a recent review of
covariate adjustment for RD.

This paper studies a novel and versatile approach based on (generalized)
entropy balancing (EB) to incorporate covariates for RD. The recent
literature on the estimation of the average treatment effect (ATE)
under the unconfoundedness assumption and also broader causal inference
literature (e.g., \citealp{Doudchenko2016}) flourishes with methods
based on balancing. See \citet{BenMichael2021} for a review of this
strand of literature. To the best of our knowledge, the balancing
approach has not been investigated in the RD literature. In this paper,
we follow CCFT to consider a potential outcome and covariate framework.
Here, the covariate balance condition, which is a restriction on the
population feature of the observed covariates, is directly implied
by the predeterminedness (zero RD LATE on covariates) assumption and
standard smoothness assumptions. Our balancing approach adjusts for
covariates by using weights that achieve exact local covariate balance
and have the least Kullback--Leibler (KL) divergence from the uniform
weights. The EB estimator can be constructed in two intuitive steps:
the first step computes the EB weights from a minimum relative entropy
problem subject to the covariate balance constraints, and the second
step replaces the uniform weights in the standard local polynomial
RD estimator (without covariates) with the EB weights. The EB estimator
can also be formulated as an empirical likelihood (EL) estimator,
for which covariate balance translates to a set of over-identifying
LP moment conditions and is used as ``side information.'' Therefore,
our approach explicitly incorporates the covariate balance condition,
which is treated as a maintained assumption in CCFT, into the estimation
and inference procedure. We show in Theorem \ref{thm:normality} that
the EB (EL) estimator is first-order equivalent to the regression
adjustment estimator of CCFT. Although CCFT doubted whether covariate
adjustment can always lead to asymptotic efficiency gain in RD estimation,
it has been pointed out by \citet{Kreiss2022} that this is true.
This paper provides an explanation of the asymptotic efficiency gain
brought by covariate adjustment from the perspective of the generalized
method of moments (GMM): the efficiency gain can be attributed to
the efficient inclusion of covariate balance as side information (Remark
\ref{Rmk: GMM}). We also offer another explanation of the efficiency
gain from the perspective of local randomization (Remark \ref{Rmk: efficiency gain}).
Under CCFT's stronger version of covariate balance (see Page 446 of
CCFT), incorporating functions of baseline covariates can further
improve efficiency. Theorem \ref{thm:sieve} shows that the asymptotic
variance of the EB estimator incorporating basis functions of baseline
covariates attains the lower bound derived in \citet{Noack2021},
if the number of basis functions (i.e., the dimension of the corresponding
linear sieve spaces) grows with the sample size.

Since the EB estimator can be formulated as an EL estimator, we expect
that the favorable second-order properties (\citealp{newey_smith_2004_higher})
may also be shared by the EB estimator. Theorem \ref{thm:nonlin bias}
shows that the EB estimator has a small (bounded) ``nonlinearity
bias''. Such a property is analogous to \citet[Theorem 4.5]{newey_smith_2004_higher}.
Then, we study covariate-adjusted EL inference for RD. A common advantage
of EL inference is that it does not require calculating standard errors
and explicit studentization. Theorem \ref{thm:shape of CS} shows
that the EL confidence set is a finite interval with probability approaching
one. Theorem \ref{thm:Wilks} shows a new uniform-in-bandwidth extension
of the standard Wilks theorem (i.e., the EL ratio is asymptotically
$\chi^{2}$). Our uniform-in-bandwidth version adjusts for specification
search over multiple bandwidths, known as bandwidth snooping (\citealp[AK, hereafter]{armstrong2017simple}),
and takes into account the effects from data-dependent bandwidths
in a robust manner (Remarks \ref{Rmk: specification search} and \ref{Rmk: data-dependent bandwidth}).
It also provides a useful tool for sensitivity analysis in the sense
of AK (Remark \ref{Rmk: sensitivity}). By deriving distributional
expansions, we investigate the second-order properties of our EL inference
method and show that it enjoys a couple of nice properties in this
setting. Theorem \ref{thm:coverage} characterizes the leading coverage
error term (i.e., the discrepancy between the nominal and finite-sample
coverage probabilities; see, e.g., \citealp{calonico2018optimal}
for Wald-type inference). We consider two choices of the LP order:
one less than the assumed smoothness ($p$-th order) and exhausting
the smoothness ($\left(p+1\right)$-th order). In the first case,
the coverage optimal (CO) bandwidth, which is defined as the minimizer
of this leading coverage error, has a simple closed form (Remark \ref{Rmk: coverage}),
which, to the best of our knowledge, cannot be obtained for Wald-type
inference (\citealp{calonico2018optimal}). In both cases, the simple
coverage expansion for the EL confidence sets makes analytical correction
possible. The correction aims to remove the leading term in the coverage
error and does not require resampling. The correction factor has a
very simple form and thus can be estimated with good accuracy in finite
samples. Remark \ref{Rmk: Bartlett correction} proposes Analytically
corrected likelihood ratio statistics and confidence sets for conducting
covariate-adjusted RD inference. Remark \ref{Rmk: doubly corrected}
combines the analytical correction and AK-type correction (Remark
\ref{Rmk: sensitivity}) and provides a more accurate uniform confidence
band that is useful for sensitivity analysis and robust inference.

Theorem \ref{thm:thm local imbalance} considers possible deviations
from covariate balance and shows that the coverage accuracy of our
proposed EL confidence set is highly insensitive to mild deviations
(Remark \ref{Rmk: sensitivity perturbation}), which we refer to as
local imbalance in this paper. Failure of the covariate balance assumption
may happen in a realistic situation when the balance condition holds
for pre-treatment covariates in theory, but our sample observations
on these covariates are contaminated (possibly due to measurement
errors that occur after treatment) so that they are drawn from a perturbed
population (\citealp{kitamura2013robustness}) that slightly violates
the balance condition. When covariate balance does not hold exactly,
the coverage accuracy of the EL confidence set stays relatively unaffected,
while other inference methods may exhibit severe undercoverage (Remark
\ref{Rmk: sensitivity perturbation}). To the best of our knowledge,
such a robustness property is novel in the literature.

Our balancing approach is versatile in dealing with covariate-adjustment
estimation/inference for parameters and/or models beyond the standard
RD, such as the treatment effect derivative (TED) of \citet{Dong2015}
and nonlinear estimators for RD with limited outcome variables (e.g.,
\citealp{xu2017regression,Xu2018}). An algorithmic extension of CCFT's
regression adjustment may not be straightforward in these scenarios.
Indeed, applying our EB approach is about reweighting a sample-analogue-type
estimator (without covariate) in the RD-related context using the
EB weights that are fully determined by the covariate balance condition.
It does not matter if the initial estimator (without covariate) involves
derivative or nonlinear transformation. For this reason, our balancing
approach serves as a useful complement to the regression adjustment.
We consider the following example in this paper. In addition to the
standard RD LATE parameter in the standard RD model, one may be interested
in estimating other parameters that have important causal interpretations,
such as the TED as a measurement of the external validity of RD. Theorem
\ref{thm:normality ted} shows the efficiency gain of the simple TED
estimator using our EB weights in place of uniform weights, for which
the only assumption needed for consistency is covariate balance. Another
class of problems that our approach can tackle is nonlinear estimators
with limited outcome variables (e.g., \citealp{xu2017regression,Xu2018}).
Estimators of \citet{xu2017regression,Xu2018} using the EB weights
achieve desired properties (consistency and potential efficiency gain)
under covariate balance. Lastly, various extensions to the standard
RD model and estimation of the relevant causal parameters have been
considered in the recent literature. Our approach has the potential
to provide easy-to-implement covariate adjustment with clear causal
interpretation. Further investigation is needed in a case-by-case
manner.

\textbf{Related literature.} Our EB estimator resembles the method
of \citet{Hainmueller2012,Chan2016} in the literature on balancing
methods for estimating ATE under unconfoundedness. See \citet{Wong2017,Kallus2020,Hirshberg2021}
for more recent development of this strand of literature. \citet{Graham2012}
show that their balancing-type estimator enjoys a similar small nonlinearity
bias property. EL and generalized EL (\citealp{newey_smith_2004_higher})
are popular alternatives to GMM, and they do not require first-step
estimation of the efficient weighting matrix. See, e.g., \citet{kitamura_2006_el_review}
for a comprehensive review of EL and generalized EL. See, e.g., \citet{chen2000empirical,otsu2013estimation,otsu2015empirical,ma2019minimum}
for EL inference in the context of non-parametric curves. It was shown
that EL has favorable properties relative to GMM. See, e.g., \citet{chen2007second,kitamura_2001_el_optimality,matsushita_otsu_2013_el,newey_smith_2004_higher,otsu_2010_bahadur,ma2017second}
among many others. In relation to the literature, \citet{otsu2015empirical}
proposed EL inference for RD without covariates. Their method was
based on first-order conditions from standard local linear regression.
This paper focuses on covariate adjustment and uses different moment
conditions. In another related paper, \citet{ma2019minimum} studied
EL inference for the parameter of interest in the density discontinuity
design (\citealp{jales_yu_aie}). Our paper uses a similar approach
to covariate adjustment as \citet{Wu2011,Zhang2018} who formulated
covariate balance in randomized experiments as moment conditions and
proposed EL-type methods.  We formulate local imbalance and study
its impact on coverage accuracy by using standard local asymptotic
analysis (e.g., the Pitman approach to local power analysis). Local
imbalance can also be viewed as a special case of local misspecification
in the GMM framework (see, e.g., \citealp{armstrong2021sensitivity}
and references therein). However, the approach we take differs from
those employed by papers in this strand of literature. Our approach
follows \citet{Bravo:2003fj} and is based on the second-order asymptotic
expansion of the coverage probability under drifting data-generating
processes (i.e., local imbalance).

\textbf{Organization.} Section \ref{sec:Preliminaries} quickly reviews
the RD design. Section \ref{sec:Empirical-likelihood-method} introduces
our EB method for RD with covariates. Section \ref{sec:Properties}
provides results on the asymptotic properties of the EB estimator,
including asymptotic normality with a discussion on the efficiency
gain (Section \ref{subsec:Asymptotic-properties}), calculation of
the nonlinearity bias (Section \ref{subsec:Nonlinearity-bias}) and
extension to balancing over sieve spaces (Section \ref{subsec:Balancing-over-functions}).
In Section \ref{sec:Likelihood-ratio-inference}, we consider inference
using the likelihood ratio and show several properties, including
a uniform-in-bandwidth Wilks theorem (Section \ref{subsec:Uniform-in-bandwidth-Wilks-theor}),
derivation of a simple analytical correction (Section \ref{subsec:Coverage-optimal-bandwidth}),
and sensitivity of the coverage probability to the covariate balance
condition (Section \ref{subsec:Local-imbalance}). Section \ref{sec:TED}
proposes a covariate-adjusted estimator of the TED and provides an
asymptotic normality result that shows the efficiency gain. Sections
\ref{sec:Monte-Carlo-Simulations} and \ref{sec:Empirical} present
results from simulation and empirical exercises. Section \ref{sec:Conclusion}
concludes. Proofs are collected in the online appendix (available
at \url{ruc-econ.github.io/supplement_Rev_V12.pdf}).

\textbf{Notation.} $\sum_{i}$ is understood as $\sum_{i=1}^{n}$.
``$a\coloneqq b$'' means that $a$ is defined by $b$ and ``$a\eqqcolon b$''
means that $b$ is defined by $a$. For any $k$-times differentiable
univariate function $f$, let $f^{\left(k\right)}$ denote the $k-$th
order derivative. Let $\mathbbm{1}\left(\cdot\right)$ denote the
indicator function. For a $d$-dimensional vector $x$, let $x^{\left(j\right)}$
denote its $j$-th coordinate, $x^{\top}$ denote its transpose, $x^{\otimes k}$
denote a vector of the distinct entries of $k$-th Kronecker power
for $k=2,3,4$ ($x^{\otimes2}\coloneqq\mathrm{vech}\left(xx^{\top}\right)$,
where $\mathrm{vech}\left(xx^{\top}\right)$ denotes the half vectorization
of $xx^{\top}$, $x^{\otimes3}$ is the vector obtained by stacking
$\left\{ x^{\left(j\right)}\mathrm{vech}\left(x_{j}x_{j}^{\top}\right):j=1,...,d\right\} $,
where $x_{j}\coloneqq\left(x^{\left(j\right)},...,x^{\left(d\right)}\right)^{\top}$,
and $x^{\otimes4}$ can be defined similarly) and $\left\Vert x\right\Vert $
denote its Euclidean norm. Let $\mathrm{I}_{K}$ denote the $K$-dimensional
identity matrix. Let $0_{J\times K}$ denote the $J\times K$ matrix
in which all elements are zeros. Let $0_{J}$ denote the $J$-dimensional
vector in which all elements are zeros. $\mathrm{A}^{\left(jk\right)}$
denotes the $jk$-th element of a matrix $\mathrm{A}$. For a square
matrix $\mathrm{A}$, let $\mathrm{tr}\left(\mathrm{A}\right)$ denote
its trace and $\mathrm{mineig}\left(\mathrm{A}\right)$ and $\mathrm{maxeig}\left(\mathrm{A}\right)$
denote the smallest and the largest eigenvalues, respectively. For
a real-valued function $f:\mathcal{X}\rightarrow\mathbb{R}$, let
$\left\Vert f\right\Vert _{\infty}\coloneqq\mathrm{sup}_{x\in\mathcal{X}}\left|f\left(x\right)\right|$
denote the sup-norm. We write $a_{n}\asymp b_{n}$, if $a_{n}=O\left(b_{n}\right)$
and $b_{n}=O\left(a_{n}\right)$. Let $\mathrm{e}_{k,s}$ denote the
$s$-th unit vector in $\mathbb{R}^{k}$.

\section{Regression discontinuity designs\label{sec:Preliminaries}}

Let $X\in\mathbb{R}$ be a continuous score supported on $\left[\underline{x},\overline{x}\right]$.
Let $f_{X}$ denote its density function. We normalize the cutoff
point to zero (so that $0\in\left[\underline{x},\overline{x}\right]$
without loss of generality) for notational brevity. In this paper,
we assume that $f_{X}$ is continuous at the cutoff. Denote $\varphi\coloneqq f_{X}\left(0\right)$
for simplicity. For a random vector (or matrix) $V$, denote $g_{V}\left(x\right)\coloneqq\mathrm{E}\left[V\mid X=x\right]$,
$m_{V}\left(x\right)\coloneqq g_{V}\left(x\right)f_{X}\left(x\right)$
and $g_{V\mid ZX}\left(z,x\right)\coloneqq\mathrm{E}\left[V\mid Z=z,X=x\right]$.
Denote $\mu_{V,-}^{\left(k\right)}\coloneqq\mathrm{lim}_{x\uparrow0}g_{V}^{\left(k\right)}\left(x\right)$
and $\psi_{V,-}^{\left(k\right)}\coloneqq\mathrm{lim}_{x\uparrow0}m_{V}^{\left(k\right)}\left(x\right)$.
$\left(\mu_{V,+}^{\left(k\right)},\psi_{V,+}^{\left(k\right)}\right)$
are defined similarly with $\mathrm{lim}_{x\uparrow0}$ replaced by
$\mathrm{lim}_{x\downarrow0}$. For simplicity, also denote $\mu_{V,\mathrm{s}}\coloneqq\mu_{V,\mathrm{s}}^{\left(0\right)}$,
$\psi_{V,\mathrm{s}}\coloneqq\psi_{V,\mathrm{s}}^{\left(0\right)}$
($\mathrm{s}\in\left\{ -,+\right\} $), $\mu_{V,\pm}\coloneqq\mu_{V,+}+\mu_{V,-}$,
$\mu_{V,\dagger}\coloneqq\mu_{V,+}-\mu_{V,-}$, $\psi_{V,\pm}\coloneqq\psi_{V,+}+\psi_{V,-}$,
$\psi_{V,\dagger}\coloneqq\psi_{V,+}-\psi_{V,-}$. Let $\mu_{V}$
($\psi_{V}$) denote the common value if $\mu_{V,+}=\mu_{V,-}$ ($\psi_{V,+}=\psi_{V,-}$).
For random vectors $V$ and $U$, $\mathrm{Var}_{\mid0^{+}}\left[U\right]$
is understood as $\mathrm{lim}_{x\downarrow0}\mathrm{Var}\left[U\mid X=x\right]=\mu_{UU^{\top},+}-\mu_{U,+}\mu_{U^{\top},+}$
and $\mathrm{Cov}_{\mid0^{+}}\left[V,U\right]$ is understood as $\mathrm{lim}_{x\downarrow0}\mathrm{Cov}\left[V,U\mid X=x\right]=\mu_{VU^{\top},+}-\mu_{V,+}\mu_{U^{\top},+}$.
Similarly, $\mathrm{Var}_{\mid0^{-}}\left[U\right]\coloneqq\mathrm{lim}_{x\uparrow0}\mathrm{Var}\left[U\mid X=x\right]$
and $\mathrm{Cov}_{\mid0^{-}}\left[V,U\right]\coloneqq\mathrm{lim}_{x\uparrow0}\mathrm{Cov}\left[V,U\mid X=x\right]$.
Also for notational simplicity, let $\mathrm{Var}_{\mid0^{\pm}}\left[U\right]\coloneqq\mathrm{Var}_{\mid0^{+}}\left[U\right]+\mathrm{Var}_{\mid0^{-}}\left[U\right]$
and $\mathrm{Cov}_{\mid0^{\pm}}\left[V,U\right]\coloneqq\mathrm{Cov}_{\mid0^{+}}\left[V,U\right]+\mathrm{Cov}_{\mid0^{-}}\left[V,U\right]$.
$\mathrm{Var}_{\mid0}$ and $\mathrm{Cov}_{\mid0}$ are understood
as $\mathrm{Var}\left[\cdot\mid X=0\right]$ and $\mathrm{Cov}\left[\cdot\mid X=0\right]$.

Let $Y\in\mathbb{R}$ denote the outcome variable, $D\in\left\{ 0,1\right\} $
be the binary treatment and $Z$ be pre-treatment covariates. Variables
in $Z$ can be continuous, discrete or mixed. We observe $\left(Y,D,Z\right)$
and the score $X$. In an RD model, an incentive is assigned if $X\geq0$.
In the sharp RD case $D=I\coloneqq\mathbbm{1}\left(X\geq0\right)$
(i.e., perfect compliance). The more general fuzzy RD model assumes
$D\neq I$ but $g_{D}$ has a jump discontinuity at $x=0$ ($\mu_{D,+}\neq\mu_{D,-}$)
due to the incentive. This is known as limited compliance in the literature.
The RD model can be embedded in the potential outcome and treatment
framework. Let $\left(Y\left(1\right),Y\left(0\right)\right)$ be
the potential outcomes with or without treatment. Let $\left(D\left(1\right),D\left(0\right)\right)$
denote the potential treatments with or without incentives. The observed
outcome $Y$ and treatment $D$ are determined by $Y=D\cdot Y\left(1\right)+\left(1-D\right)Y\left(0\right)$
and $D=I\cdot D\left(1\right)+\left(1-I\right)D\left(0\right)$ respectively.
The complier group is defined to be individuals with $D\left(1\right)>D\left(0\right)$
(i.e., $\left(D\left(1\right),D\left(0\right)\right)=\left(1,0\right)$).
We use ``$\mathsf{co}$'' to denote this event. Following CCFT,
we let $\left(Z\left(1\right),Z\left(0\right)\right)$ denote potential
covariates and then $Z=D\cdot Z\left(1\right)+\left(1-D\right)Z\left(0\right)$.\footnote{\label{fn:RD triangular}The RD design can be represented by a structural
model. See \citet{dong2018alternative}. $\left(Y,D,Z\right)$ are
assumed to be generated by the structural model $Y=g\left(D,X,Z,\epsilon\right)$,
$D=h\left(I,X,\eta\right)$ and $Z=m\left(D,X,\xi\right)$, where
$\left(g,h,m\right)$ are unknown functions and $\left(\epsilon,\eta,\xi\right)$
are (potentially correlated) unobserved disturbances of unrestricted
dimensionality. Then the potential outcomes, covariates and treatments
are given by $Y\left(d\right)=g\left(d,X,Z,\epsilon\right)$, $D\left(d\right)=h\left(d,X,\eta\right)$
and $Z\left(d\right)=m\left(d,X,\xi\right)$.} Let $B\left(d\right)\coloneqq\left(Y\left(d\right),Z\left(d\right)\right)$,
for $d\in\left\{ 0,1\right\} $. Denote $g_{dd'}\left(x\right)\coloneqq\mathrm{Pr}\left[D\left(1\right)=d,D\left(0\right)=d'\mid X=x\right]$
and $g_{B\left(j\right)\mid dd'}\left(x\right)\coloneqq\mathrm{E}\left[B\left(j\right)\mid D\left(1\right)=d,D\left(0\right)=d',X=x\right]$
for $\left(j,d,d'\right)\in\left\{ 0,1\right\} ^{3}$. The RD LATE
(the average treatment effect for individuals with zero score in the
complier group) is given by $\mathrm{E}\left[Y\left(1\right)-Y\left(0\right)\mid X=0,\mathsf{co}\right]$.
The following assumption is sufficient for the identification in RD
and is also imposed in CCFT.
\begin{assumption}
\label{assu:identification}(a) $\left(g_{Y\left(1\right)\mid11},g_{Y\left(0\right)\mid00},g_{Y\left(1\right)\mid10},g_{Y\left(0\right)\mid10}\right)$
are all continuous at the threshold $0$; (b) $g_{dd'}$ is continuous
at the threshold $0$ for all $\left(d,d'\right)\in\left\{ 0,1\right\} ^{2}$;
(c) $\mathrm{Pr}\left[D\left(1\right)\geq D\left(0\right)\mid X=0\right]=1$;
(d) $\mathrm{Pr}\left[\mathsf{co}\mid X=0\right]>0$; (e) $\left(g_{Z\left(1\right)\mid11},g_{Z\left(0\right)\mid00},g_{Z\left(1\right)\mid10},g_{Z\left(0\right)\mid10}\right)$
are all continuous at the threshold $0$; (f) $g_{Z\left(1\right)\mid10}\left(0\right)=g_{Z\left(0\right)\mid10}\left(0\right)$.
\end{assumption}
It can be shown that under (a,b,c,d), the RD LATE is identified by
the standard RD estimand $\vartheta\coloneqq\mu_{Y,\dagger}/\mu_{D,\dagger}$
(i.e., $\mathrm{E}\left[Y\left(1\right)-Y\left(0\right)\mid X=0,\mathsf{co}\right]=\vartheta$,
see \citealp{hahn2001identification,dong2018alternative} and \citealp{Arai2021}
for testable implications of these assumptions), where $\vartheta$
is a population feature of the observed variables.\footnote{In the sharp RD model ($\mu_{D,+}=1$ and $\mu_{D,-}=0$ in this case)
or under a stronger conditional independence assumption (\citealp{hahn2001identification}),
a causal parameter that corresponds to a broader subpopulation (conditional
average treatment effect) is identified by the same ratio: $\mathrm{E}\left[Y\left(1\right)-Y\left(0\right)\mid X=0\right]=\vartheta$
.} As in \citet{frolich2019including}, the continuity assumption (a)
can be viewed as an exclusion restriction. Intuitively, continuity
of $g_{Y\left(j\right)\mid dd'}$ essentially requires that $Y\left(j\right)$
cannot depend on $I$ or (observed or unobserved) variables related
to $I$ (so that their distributions change discontinuously at the
cutoff). Since $Y\left(j\right)$ often depends on $Z$, continuity
of $g_{Y\left(j\right)\mid dd'}$ also implicitly requires that the
conditional distributions of $Z$ given $\left(D\left(1\right),D\left(0\right),X\right)=\left(d,d',x\right)$
change smoothly at $x=0$. Since the distribution of $Z$ coincides
with that of $Z\left(d\right)$ ($Z\left(d'\right)$), given $\left(D\left(1\right),D\left(0\right),X\right)=\left(d,d',x\right)$
with $x\geq0$ ($x<0$), continuity of the conditional distribution
of $Z$ given $\left(D\left(1\right),D\left(0\right),X\right)=\left(d,d',x\right)$,
for $\left(d,d'\right)\in\left\{ \left(1,1\right),\left(0,0\right),\left(1,0\right)\right\} $,
holds if the conditional distributions of the potential covariates
change smoothly at $x=0$ and the distribution of $Z\left(1\right)$
given $\mathsf{co}$ and $X=0$ is the same as that of $Z\left(0\right)$
given $\mathsf{co}$ and $X=0$. Following CCFT, we consider using
weaker versions of these assumptions in (e,f). We consider using the
strong versions in Section \ref{subsec:Balancing-over-functions}.
(e) essentially requires that the covariates satisfy the same exclusion
restriction (not affected by $I$). It is clear from $g_{Z}\left(x\right)=\sum_{d,d'}g_{Z\mid dd'}\left(x\right)g_{dd'}\left(x\right)$,
where ``$\sum_{d,d'}$'' is understood as ``$\sum_{\left(d,d'\right)\in\left\{ 0,1\right\} ^{2}}$'',
that covariate balance $\mu_{Z,+}=\mu_{Z,-}$ holds as a testable
implication for the population of the observed variables.\footnote{Indeed, $\mu_{Z,+}=\mu_{Z,-}$ is the null hypothesis of a popular
falsification or placebo test for the RD model. See, e.g., \citet{lee2008randomized,Canay2017}.
Evidence against $\mu_{Z,+}=\mu_{Z,-}$ in the data (so that a hypothesis
test of $\mu_{Z,+}=\mu_{Z,-}$ is rejected) casts doubts on the validity
of the key identifying assumption of the RD design (i.e., Assumption
\ref{assu:identification}(a)). While most empirical works conduct
the balance test separately for each covariate, some researchers have
noted that the problem of multiple testing may generate statistical
imbalance of some covariates by chance. See, e.g., \citet{hyytinen2018does}.}

\section{Empirical entropy balancing\label{sec:Empirical-likelihood-method}}

This section introduces the EB method. We quickly review the idea
of entropy balancing and reweighting in the literature on ATE estimation
under unconfoundedness (i.e., conditional independence of the potential
outcomes and the treatment given the covariates). Then, we utilize
the idea of EB to propose a new balancing-based method for covariate
adjustment for RD.

In observational studies, because of the selection bias, the difference
in the sample means corresponding to the treatment and control groups
does not consistently estimate the ATE. The balancing weights satisfy
the requirement that the weighted control (treatment) group sample
moments of the covariates match the unweighted sample moments of the
covariates of all units. Within all balancing weights, \citet{Hainmueller2012}
defines the EB weights as those being as close as possible to the
uniform weights in the sense of minimal relative entropy (KL divergence).
\citet{Hainmueller2012} replaces the uniform weights used by the
simple sample means with the EB weights. \citet{Chan2016} construct
EB weights that equalize weighted and unweighted sample means of transformations
of the covariates via basis functions. \citet{Chan2016} show that
the estimator using these EB weights overcomes the selection bias
under the unconfoundedness assumption if the number of basis functions
of the covariates increases with the sample size.

From a GMM/EL perspective, in \citet{Hainmueller2012,Chan2016}, the
entropy balancing and reweighting approach uses weights under which
some intentionally misspecified (biased) moment restrictions are satisfied
to correct for the selection bias.\footnote{In observational data, the population moments of covariates in the
control or treatment group may not be the same as the unconditional
population moments, since the treatment status is not independent
from the covariates.} In our RD case, the moment restrictions (balancing constraints) are
correctly specified and entropy balancing and reweighting aim at enhancing
efficiency (Section \ref{subsec:Entropy-balancing-for}). In our case,
the EB estimator can be formulated as a standard EL estimator (Section
\ref{subsec:Connection-to-empirical}).

\subsection{Entropy balancing for covariate adjustment in RD\label{subsec:Entropy-balancing-for}}

Now we elaborate on the entropy balancing and reweighting approach
to covariate adjustment in the RD context. Firstly, we introduce some
notations. Let $K$ denote the kernel function and let $h$ denote
the bandwidth. We assume that $h=h_{n}$ decreases with the sample
size $n$. For notational simplicity, we suppress the dependence of
$h$ on $n$. Let the data $\left\{ \left(Y_{i},D_{i},X_{i},Z_{i}\right)\right\} _{i=1}^{n}$
be i.i.d. copies of $\left(Y,D,X,Z\right)$. We drop the subscript
$i$ when we refer to population-level estimands. Let $p\geq1$ be
the integer-valued LP order. Let $r_{p}\left(t\right)\coloneqq\left(1,t,\ldots,t^{p}\right)^{\top}$
and let $\mathrm{H}$ be the $\left(p+1\right)\times\left(p+1\right)$
diagonal matrix with $\left(1,h,...,h^{p}\right)$ being on the diagonal.
Denote
\begin{equation}
\widehat{\Pi}_{p,-}\coloneqq\frac{1}{nh}\sum_{i}r_{p}\left(\frac{X_{i}}{h}\right)r_{p}^{\top}\left(\frac{X_{i}}{h}\right)K\left(\frac{X_{i}}{h}\right)\mathbbm{1}\left(X_{i}<0\right).\label{eq:PI -}
\end{equation}
Let $\widehat{\Pi}_{p,+}$ be defined similarly by the right-hand
side of (\ref{eq:PI -}) with $\mathbbm{1}\left(X_{i}<0\right)$ replaced
by $\mathbbm{1}\left(X_{i}>0\right)$. Let 
\begin{equation}
\widehat{W}_{p;-,i}\coloneqq\mathrm{e}_{p+1,1}^{\top}\widehat{\Pi}_{p,-}^{-1}r_{p}\left(\frac{X_{i}}{h}\right)K\left(\frac{X_{i}}{h}\right)\mathbbm{1}\left(X_{i}<0\right).\label{eq:regression weight}
\end{equation}
Let $\widehat{W}_{p;+,i}$ be defined similarly by the right-hand
side of (\ref{eq:PI -}) with $\mathbbm{1}\left(X_{i}<0\right)$ and
$\widehat{\Pi}_{p,-}$ replaced by $\mathbbm{1}\left(X_{i}>0\right)$
and $\widehat{\Pi}_{p,+}$.\footnote{We restrict the bandwidths on the left and the right of the cut-off
to be the same. It is possible to extend all of the theorems in this
paper to accommodate different bandwidths on different sides.} 

Let $\widehat{W}_{p,i}\coloneqq\widehat{W}_{p;+,i}-\widehat{W}_{p;-,i}$.
The standard LP regression estimator of $\vartheta$ is
\begin{equation}
\widehat{\vartheta}_{p}^{\mathsf{lp}}\coloneqq\frac{\frac{1}{nh}\sum_{i}\widehat{W}_{p,i}Y_{i}}{\frac{1}{nh}\sum_{i}\widehat{W}_{p,i}D_{i}},\label{eq:standard}
\end{equation}
where the numerator $\left(nh\right)^{-1}\text{\ensuremath{\sum_{i}\widehat{W}_{p,i}Y_{i}}}$
is the LP regression estimator of $\mu_{Y,\dagger}$ and the denominator
is the LP regression estimator of $\mu_{D,\dagger}$. 

Now we incorporate the covariate information to the standard LP estimator
$\widehat{\vartheta}_{p}^{\mathsf{lp}}$ by reweighting its numerator
and denominator using the EB weights computed from the covariate balance
constraints . Denote $\bar{Z}_{i}\coloneqq\left(1,Z_{i}^{\top}\right)^{\top}$.
We define EB weights $\left(w_{1}^{\mathsf{eb}},...,w_{n}^{\mathsf{eb}}\right)$
as the solution to the following minimum relative entropy problem:
\begin{flalign}
 & \underset{w_{1},...,w_{n}}{\mathrm{min}}\mathit{KL}\left(w_{1},...,w_{n}\parallel\frac{1}{n},...,\frac{1}{n}\right)\nonumber \\
 & \textrm{subject to }\sum_{i}w_{i}\begin{array}{c}
\widehat{W}_{p,i}\bar{Z}_{i}\end{array}=0_{d_{z}+1},\,\ensuremath{\sum_{i}w_{i}=1},\label{eq:entropy balancing}
\end{flalign}
where $\mathit{KL}\left(w_{1},...,w_{n}\parallel1/n,...,1/n\right)\coloneqq-\sum_{i}\mathrm{log}\left(n\cdot w_{i}\right)/n$
is the KL divergence from $\left(w_{1},...,w_{n}\right)$ to the uniform
weights $\left(1/n,...,1/n\right)$. The construction of these balancing
weights is similar to those in \citet{Hainmueller2012}. By solving
the minimization problem (\ref{eq:entropy balancing}), we find the
set of weights with the least KL divergence from the uniform weights
among these balancing weights. The uniform weights satisfy the important
finite-sample property of $\sum_{i}\widehat{W}_{p,i}=0$. By requiring
$\sum_{i}w_{i}\begin{array}{c}
\widehat{W}_{p,i}\end{array}=0$ in the constraint of (\ref{eq:entropy balancing}), we require that
the balancing weights satisfy the same property. The balancing weights
should also satisfy $\sum_{i}w_{i}\begin{array}{c}
\widehat{W}_{p;+,i}Z_{i}\end{array}=\sum_{i}w_{i}\begin{array}{c}
\widehat{W}_{p;-,i}Z_{i}\end{array}$. This requires that the (kernel-weighted) local averages of the covariates
on both sides of the thresholds coincide in finite samples under the
new weights for the data points. To solve for the optimal weights,
we use strong duality and concentrate out $\left(w_{1},...,w_{n}\right)$
to obtain the following dual characterization 
\begin{equation}
w_{i}^{\mathsf{eb}}=\frac{1}{n}\cdot\frac{1}{1+\left(\lambda_{p}^{\mathsf{eb}}\right)^{\top}\left(\widehat{W}_{p,i}\bar{Z}_{i}\right)},\label{eq:entropy balancing weight definition}
\end{equation}
where
\begin{equation}
\lambda_{p}^{\mathsf{eb}}\coloneqq\underset{\lambda}{\mathrm{argmax}}\sum_{i}\mathrm{log}\left(1+\lambda^{\top}\left(\widehat{W}_{p,i}\bar{Z}_{i}\right)\right)\label{eq:lambda_hat}
\end{equation}
is the Lagrangian multiplier. Computing the EB weights requires dealing
with a well-understood convex optimization problem (\ref{eq:lambda_hat})
that can be solved by the Newton algorithm. The domain of its objective
function is the convex set $\left\{ \lambda:1+\lambda^{\top}\left(\widehat{W}_{p,i}\bar{Z}_{i}\right)>0\textrm{ for all }i\right\} $.
The algorithm should either take these constraints into account or
use a modified objective function defined for all $\lambda\in\mathbb{R}^{d_{z}+1}$.
(\ref{eq:lambda_hat}) has no solution if the origin is not an interior
point of the convex hull of $\left\{ \widehat{W}_{p,1}\bar{Z}_{1},...,\widehat{W}_{p,n}\bar{Z}_{n}\right\} $.\footnote{By arguments similar to those in \citet[Chapter 11.2]{owen_2001_el_book},
we can show that if covariate balance holds, the origin lies in the
convex hull with probability approaching one.} See \citet{kitamura_2006_el_review} and \citet{owen_2001_el_book}
for more algorithmic details.\footnote{In the ``no solution'' scenario, the Newton algorithm trying to
solve (\ref{eq:lambda_hat}) returns a sequence of vectors with diverging
lengths. In this scenario, after the algorithm terminates (either
the gradient is sufficiently small or the maximal number of iterations
is reached), we would get weights not summing up to one in the former
case (\citealp[Chapter 3.14]{owen_2001_el_book}) or a large gradient
in the latter case. In our simulation studies and computation for
the empirical application, we use the Matlab code written by Kirill
Evdokimov and Yuichi Kitamura (\url{https://kitamura.sites.yale.edu/matlabstata-codes-el})
and hardly see any ``no solution'' case.}

We propose the following empirical entropy balancing estimator by
reweighting the numerator and denominator of $\widehat{\vartheta}_{p}^{\mathsf{lp}}$
in (\ref{eq:standard}) using the EB weights $w_{i}^{\mathsf{eb}}$
defined by (\ref{eq:entropy balancing weight definition}):
\begin{equation}
\widehat{\vartheta}_{p}^{\mathsf{eb}}\coloneqq\frac{\sum_{i}w_{i}^{\mathsf{eb}}\widehat{W}_{p,i}Y_{i}}{\sum_{i}w_{i}^{\mathsf{eb}}\widehat{W}_{p,i}D_{i}}.\label{eq:entropy balancing estimator definition}
\end{equation}
The reweighting form of our EB estimator $\widehat{\vartheta}_{p}^{\mathsf{eb}}$
has a clear causal interpretation. The continuity and predeterminedness
assumptions in Assumption \ref{assu:identification} imply a restriction
$\mu_{Z,+}=\mu_{Z,-}$ on the population distribution of the observed
covariates. $\widehat{\vartheta}_{p}^{\mathsf{eb}}$ directly uses
weights that explicitly exploit such information from the covariates.

\subsection{Generalized entropy balancing\label{subsec:Generalized-entropy-balancing}}

The entropy balancing approach looks for balancing weights closest
to the uniform weights, where ``closeness'' is measured by the KL
divergence. It is useful to consider the following extension. Let
$\left(p_{1},...,p_{n}\right)$ and $\left(p_{1}',...,p_{n}'\right)$
be two sets of probability masses. For any $\varrho\in\mathbb{R}$,
let 
\begin{equation}
D_{\varrho}\left(p_{1},...,p_{n}\parallel p_{1}',...,p_{n}'\right)\coloneqq\frac{1}{\varrho\left(1+\varrho\right)}\sum_{i}\left\{ \left(\frac{p_{i}}{p_{i}'}\right)^{-\varrho}-1\right\} p_{i}'\label{eq:Cressie-Read}
\end{equation}
be the Cressie-Read divergence from $\left(p_{1},...,p_{n}\right)$
to $\left(p_{1}',...,p_{n}'\right)$. Taking $\varrho=0$ gives the
KL divergence from $\left(p_{1},...,p_{n}\right)$ to $\left(p_{1}',...,p_{n}'\right)$.
Taking $\varrho=-1$ gives the KL divergence from $\left(p_{1}',...,p_{n}'\right)$
to $\left(p_{1},...,p_{n}\right)$.\footnote{$D_{0}\left(p_{1},...,p_{n}\parallel p_{1}',...,p_{n}'\right)$ (or
$D_{-1}\left(p_{1},...,p_{n}\parallel p_{1}',...,p_{n}'\right)$)
is defined as the limit of the right hand side of (\ref{eq:Cressie-Read})
as $\varrho\rightarrow0$ (or $\varrho\rightarrow-1$).} The generalized balancing estimator is based on the weights $\left(w_{\varrho,1}^{\mathsf{gb}},...,w_{\varrho,n}^{\mathsf{gb}}\right)$
that solve
\begin{align}
 & \underset{w_{1},...,w_{n}}{\mathrm{min}}D_{\varrho}\left(w_{1},...,w_{n}\parallel\frac{1}{n},...,\frac{1}{n}\right)\nonumber \\
 & \textrm{subject to }\sum_{i}w_{i}\begin{array}{c}
\widehat{W}_{p,i}\bar{Z}_{i}\end{array}=0_{d_{z}+1},\,\ensuremath{\sum_{i}w_{i}=1}.\label{eq:generalized balancing}
\end{align}
We define the generalized balancing estimator $\widehat{\vartheta}_{\varrho,p}^{\mathsf{gb}}$
by the right hand side of (\ref{eq:entropy balancing estimator definition})
with $w_{i}^{\mathsf{eb}}$ replaced by $w_{\varrho,i}^{\mathsf{gb}}$.

Interestingly, although CCFT takes an augmented regression approach
to incorporate pre-treatment covariates, a slight modification of
CCFT's estimator can be written as a generalized balancing estimator
with $\varrho=-2$. It is easy to see that $D_{-2}\left(w_{1},...,w_{n}\parallel1/n,...,1/n\right)$
is proportional to the square of the Euclidean distance between $\left(w_{1},...,w_{n}\right)$
and the uniform weights. CCFT's covariate adjusted estimator $\widehat{\vartheta}_{Y,p}^{\mathsf{CCFT}}$
for $\mu_{Y,\dagger}$ is given by the regression coefficient of $I_{i}\coloneqq\mathbbm{1}\left(X_{i}\geq0\right)$
in
\begin{equation}
\widehat{\vartheta}_{Y,p}^{\mathsf{CCFT}}\coloneqq\mathrm{e}_{2\left(p+1\right)+d_{z},p+2}^{\top}\underset{b_{0},b_{1},b_{2}}{\mathrm{argmin}}\,\sum_{i}K\left(\frac{X_{i}}{h}\right)\left\{ Y_{i}-r_{p}^{\top}\left(X_{i}\right)b_{0}-I_{i}\cdot r_{p}^{\top}\left(X_{i}\right)b_{1}-Z_{i}^{\top}b_{2}\right\} ^{2}.\label{eq:CCFT estimator}
\end{equation}
Similarly, CCFT's estimator $\widehat{\vartheta}_{D,p}^{\mathsf{CCFT}}$
for $\mu_{D,\dagger}$ is defined by the right hand side of the above
equation with $Y_{i}$ replaced by $D_{i}$. Then CCFT's estimator
of $\vartheta$ is $\widehat{\vartheta}_{p}^{\mathsf{CCFT}}\coloneqq\widehat{\vartheta}_{Y,p}^{\mathsf{CCFT}}/\widehat{\vartheta}_{D,p}^{\mathsf{CCFT}}$.
CCFT shows that by the partitioned regression argument, the numerator
$\widehat{\vartheta}_{Y,p}^{\mathsf{CCFT}}$ can be written as
\begin{equation}
\widehat{\vartheta}_{Y,p}^{\mathsf{CCFT}}=\frac{1}{nh}\sum_{i}\widehat{W}_{p,i}\left(Y_{i}-Z_{i}^{\top}\widehat{\gamma}_{Y}^{\mathsf{CCFT}}\right),\label{eq:CCFT estimator 2}
\end{equation}
where $\widehat{\gamma}_{Y}^{\mathsf{CCFT}}$ is a consistent estimator
of $\gamma_{Y}\coloneqq\left(\mathrm{Var}_{\mid0^{\pm}}\left[Z\right]\right)^{-1}\mathrm{Cov}_{\mid0^{\pm}}\left[Z,Y\right]$.
A similar result holds for $\widehat{\vartheta}_{D,p}^{\mathsf{CCFT}}$.
To find the optimal weights that solve (\ref{eq:generalized balancing})
with $\varrho=-2$, we again apply the Lagrangian multiplier method.
What differs from EB is that in this case the Lagrangian multiplier
has an explicit form. Then we can see that $\sum_{i}w_{-2,i}^{\mathsf{gb}}\widehat{W}_{p,i}Y_{i}/h$
(or $\sum_{i}w_{-2,i}^{\mathsf{gb}}\widehat{W}_{p,i}D_{i}/h$) can
be written in the form of (\ref{eq:CCFT estimator 2}) with a slightly
different estimator of $\gamma_{Y}$ (or $\gamma_{D}$). More details
can be found in Section S9 of our online supplement. Therefore, our
formulation provides information-theoretic and balancing interpretation
of CCFT's estimator. The generalized balancing estimators are all
first-order equivalent to the EB estimator in the sense that the conclusion
of Theorem \ref{thm:normality} also holds for them.

\subsection{Connection to empirical likelihood\label{subsec:Connection-to-empirical}}

We now show that the EB estimator can be formulated as an EL estimator
which incorporates covariate balance as side information. The RD estimand
$\vartheta$, which has causal interpretation under the identifying
assumptions of the RD model, can be approximately identified by a
moment condition. Note that
\[
\underset{x\downarrow0}{\mathrm{lim}}\,\mathrm{E}\left[Y-\theta D\mid X=x\right]=\underset{x\uparrow0}{\mathrm{lim}}\,\mathrm{E}\left[Y-\theta D\mid X=x\right]\textrm{ if and only if }\theta=\vartheta.
\]
By the standard LP regression theory, we have
\[
\frac{1}{nh}\sum_{i}\widehat{W}_{p,i}\left(Y_{i}-\theta D_{i}\right)\rightarrow_{p}\underset{x\downarrow0}{\mathrm{lim}}\,\mathrm{E}\left[Y-\theta D\mid X=x\right]-\underset{x\uparrow0}{\mathrm{lim}}\,\mathrm{E}\left[Y-\theta D\mid X=x\right],
\]
under standard assumptions. Under covariate balance, $\left(nh\right)^{-1}\sum_{i}\widehat{W}_{p,i}\bar{Z}_{i}\rightarrow_{p}0_{d_{z}+1}$.
We consider the following EL-type criterion function:
\begin{eqnarray}
\ell_{p}^{\mathsf{\mathsf{el}}}\left(\theta\mid h\right) & \coloneqq & \underset{w_{1},...,w_{n}}{\mathrm{min}}\mathit{KL}\left(w_{1},...,w_{n}\parallel\frac{1}{n},...,\frac{1}{n}\right)\nonumber \\
 &  & \textrm{subject to }\sum_{i}w_{i}\begin{array}{c}
\widehat{W}_{p,i}\left(\begin{array}{c}
Y_{i}-\theta D_{i}\\
\bar{Z}_{i}
\end{array}\right)\end{array}=0_{d_{z}+2},\,\ensuremath{\sum_{i}w_{i}=1}.\label{eq:EL criterion regression}
\end{eqnarray}
Note that we have $2+d_{z}$ LP moment conditions that approximately
identify one parameter of interest $\vartheta$. Note that the covariate
balance condition provides a set of over-identifying moment restrictions.
We can easily see that the EB estimator is also an EL estimator, defined
as a minimizer of the EL criterion function. It is clear that $\ell_{p}^{\mathsf{\mathsf{el}}}\left(\theta\mid h\right)\geq-n^{-1}\sum_{i}\mathrm{log}\left(n\cdot w_{i}^{\mathsf{eb}}\right)$
for all $\theta$, since $-n^{-1}\sum_{i}\mathrm{log}\left(n\cdot w_{i}^{\mathsf{eb}}\right)$
is the minimum corresponding to a larger constraint set. Since the
constraint set of (\ref{eq:EL criterion regression}) with $\theta=\widehat{\vartheta}_{p}^{\mathsf{\mathsf{eb}}}$
contains the EB weights, we have $\ell_{p}^{\mathsf{\mathsf{el}}}\left(\widehat{\vartheta}_{p}^{\mathsf{\mathsf{eb}}}\mid h\right)\leq-n^{-1}\sum_{i}\mathrm{log}\left(n\cdot w_{i}^{\mathsf{eb}}\right)$.
Therefore, $\widehat{\vartheta}_{p}^{\mathsf{\mathsf{eb}}}$ is a
minimizer of $\ell_{p}^{\mathsf{\mathsf{el}}}\left(\cdot\mid h\right)$.

We consider replacing the kernel-dependent weight $\widehat{W}_{p,i}$
in (\ref{eq:EL criterion regression}) by the weight from population-level
LP fitting (or solving a minimum contrast (MC) problem) and define
the MC-EL estimator as the minimizer. See \citet[Chapter 11.3]{bickel2015mathematical}
for more details about the construction of the population-level LP
fitting. Denote $\mathrm{V}_{p;-}\coloneqq\int_{-1}^{0}r_{p}\left(t\right)r_{p}^{\top}\left(t\right)K\left(t\right)\mathrm{d}t$
and $\mathcal{K}_{p;-}\left(t\right)\coloneqq\mathrm{e}_{p+1,1}^{\top}\mathrm{V}_{p;-}^{-1}r_{p}\left(t\right)K\left(t\right)$.
Let $\left(\mathrm{V}_{p;+},\mathcal{K}_{p;+}\right)$ be defined
by the same equations with the integral range $\left[-1,0\right]$
replaced by $\left[0,1\right]$. $\left(\mathcal{K}_{p;+},\mathcal{K}_{p;-}\right)$
coincide with the ``equivalent kernel'' associated with the LP regression.
See, e.g., Section S2.1 of AK. Let $W_{p;-,i}\coloneqq\mathbbm{1}\left(X_{i}<0\right)\mathcal{K}_{p;-}\left(X_{i}/h\right)$,
$W_{p;+,i}\coloneqq\mathbbm{1}\left(X_{i}>0\right)\mathcal{K}_{p;+}\left(X_{i}/h\right)$
and $W_{p,i}\coloneqq W_{p;+,i}-W_{p;-,i}$. By Taylor expansion (see
\citealp{Jiang_2003}), $\mathrm{E}\left[W_{p}\left(Y-\theta D\right)\right]=O\left(h^{p+2}\right)$
if and only if $\theta=\vartheta$ and $\mathrm{E}\left[W_{p}\cdot\bar{Z}\right]=O\left(h^{p+2}\right)$
under suitable smoothness assumptions. Let the MC-EL criterion function
$\ell_{p}^{\mathsf{mc}}$ be defined by the right hand side of (\ref{eq:EL criterion regression})
with $\widehat{W}_{p,i}$ replaced by $W_{p,i}$. The $p$-th order
MC-EL estimator is given by $\widehat{\vartheta}_{p}^{\mathsf{mc}}\coloneqq\mathrm{argmin}_{\theta}\ell_{p}^{\mathsf{mc}}\left(\theta\mid h\right)$.
Similar derivations show that the estimator can be written as a balancing-type
estimator. $\widehat{\vartheta}_{p}^{\mathsf{mc}}$ is equal to the
right hand side of (\ref{eq:entropy balancing estimator definition})
with $\left(\widehat{W}_{p,i},w_{i}^{\mathsf{eb}}\right)$ replaced
by $\left(W_{p,i},w_{i}^{\mathsf{mc}}\right)$, where $\left(w_{i}^{\mathsf{mc}},\lambda_{p}^{\mathsf{mc}}\right)$
are defined by the right hand sides of (\ref{eq:entropy balancing weight definition})
and (\ref{eq:lambda_hat}) with $\widehat{W}_{p,i}$ replaced by $W_{p,i}$.
The MC-EL estimator has similar asymptotic properties as the EB estimator
(see Theorems \ref{thm:normality} and \ref{thm:nonlin bias} ahead).
The MC-EL criterion function is useful for constructing confidence
sets for $\vartheta$ with favorable second-order properties (see
Section \ref{sec:Likelihood-ratio-inference} ahead).

\section{Properties of the empirical likelihood (balancing) estimator\label{sec:Properties}}

In this section, we show several large-sample properties of the proposed
empirical balancing (likelihood) estimator. Section \ref{subsec:Asymptotic-properties}
gives the asymptotic normality result of the estimators proposed in
the preceding section. We compare our result with that of CCFT's estimator
and discuss the efficiency gain brought by the covariates. Section
\ref{subsec:Nonlinearity-bias} gives a result on the ``nonlinearity
bias'' of the proposed estimators. We argue that our estimators have
small nonlinearity biases, especially in the situation when a relatively
large number of valid covariates satisfying the balance condition
are available. In Section \ref{subsec:Balancing-over-functions},
we consider the situation when strong continuity and predeterminedness
assumptions hold. The main result in Section \ref{subsec:Balancing-over-functions}
shows that an extension of our EB estimator whose weights balance
functions of covariates in a sequence of linear sieve spaces achieves
the variance lower bound derived in \citet{Noack2021}.

\subsection{Efficiency gain\label{subsec:Asymptotic-properties}}

This section shows asymptotic normality of the EB and MC-EL estimators,
and gives the expression for the asymptotic mean square error (AMSE).
We then compare our results with the asymptotic result from CCFT.
Let $B\coloneqq\left(Y,Z^{\top}\right)^{\top}$, $\bar{B}\coloneqq\left(Y,D,Z^{\top}\right)^{\top}$
and $M\coloneqq Y-\vartheta D$. The following assumptions are imposed
on the population distribution of the observed variables. Let $\mathbb{B}\subseteq\left[\underline{x},\overline{x}\right]$
denote a neighborhood around 0.
\begin{assumption}
\label{assu:data generating process}(a) $g_{\bar{B}}$ is $\left(p+1\right)$-times
continuously differentiable on $\mathbb{B}\setminus\left\{ 0\right\} $
and $g_{\bar{B}}^{\left(p+1\right)}$ is Hölder continuous with unknown
exponent $\mathfrak{h}\in\left(0,1\right]$; (b) $g_{\bar{B}{}^{\otimes2}}$
is uniformly continuous on $\mathbb{B}\setminus\left\{ 0\right\} $;
(c) $f_{X}$ is $\left(p+1\right)$-times continuously differentiable
on $\mathbb{B}$ and $f_{X}^{\left(p+1\right)}$ is is Hölder continuous
with unknown exponent $\mathfrak{h}\in\left(0,1\right]$; (d) $\mathrm{Var}_{\mid0^{+}}\left[\left(M,Z^{\top}\right)^{\top}\right]$
and $\mathrm{Var}_{\mid0^{-}}\left[\left(M,Z^{\top}\right)^{\top}\right]$
are positive definite.
\end{assumption}
Assumption \ref{assu:data generating process} parallels Assumption
SA-5 of CCFT. We will invoke it directly in the proofs. These assumptions
are satisfied under suitable conditions imposed on the population
distribution of the latent variables as in Assumption \ref{assu:identification}.
Since $B=D\left(1\right)B\left(1\right)+\left(1-D\left(1\right)\right)B\left(0\right)$
if $X\geq0$ and $B=D\left(0\right)B\left(1\right)+\left(1-D\left(0\right)\right)B\left(0\right)$
if $X<0$, by the law of iterated expectations (LIE), for any function
$\varphi\left(\cdot,\cdot\right)$, we have
\begin{equation}
g_{\varphi\left(B,D\right)}\left(x\right)=\begin{cases}
\sum_{d,d'}g_{dd'}\left(x\right)g_{\varphi\left(B\left(d\right),d\right)\mid dd'}\left(x\right) & \textrm{if }x\geq0\\
\sum_{d,d'}g_{dd'}\left(x\right)g_{\varphi\left(B\left(d'\right),d'\right)\mid dd'}\left(x\right) & \textrm{if }x<0.
\end{cases}\label{eq:g_phi_B}
\end{equation}
(a) is satisfied if $\left(g_{B\left(d\right)\mid dd'},g_{B\left(d'\right)\mid dd'},g_{dd'}\right)$
are $\left(p+1\right)$-times continuously differentiable on $\mathbb{B}$
with uniformly continuous derivatives for all $\left(d,d'\right)\in\left\{ 0,1\right\} ^{2}$.
The smoothness level in (a) is similar to that commonly assumed in
the literature, i.e., the minimal smoothness level ($p+1$) such that
that the leading smoothing bias term of the estimator (using $p$-th
order LP) can be explicitly characterized. (b) is satisfied if we
impose the additional condition that for all $\left(d,d'\right)\in\left\{ 0,1\right\} ^{2}$,
$\left(g_{B\left(d\right){}^{\otimes2}\mid dd'},g_{B\left(d'\right){}^{\otimes2}\mid dd'}\right)$
are uniformly continuous on $\mathbb{B}$. (a,c) also guarantee that
$m_{B}=g_{B}f_{X}$ have uniformly continuous derivatives up to $\left(p+1\right)$-th
order on the left and right neighborhoods of $0$. Existence of $\mathrm{Var}_{0^{+}}\left[\left(M,Z^{\top}\right)^{\top}\right]$
and $\mathrm{Var}_{\mid0^{-}}\left[\left(M,Z^{\top}\right)^{\top}\right]$
is guaranteed by (b). By the law of total variance (writing $\mathrm{Var}\left[\left(M,Z^{\top}\right)^{\top}\mid X\right]$
as the sum of $\mathrm{E}\left[\mathrm{Var}\left[\left(M,Z^{\top}\right)^{\top}\mid D\left(1\right),D\left(0\right),X\right]\mid X\right]$
and $\mathrm{Var}\left[\mathrm{E}\left[\left(M,Z^{\top}\right)^{\top}\mid D\left(1\right),D\left(0\right),X\right]\mid X\right]$)
and (b), $\mathrm{Var}_{\mid0^{+}}\left[\left(M,Z^{\top}\right)^{\top}\right]$
(or $\mathrm{Var}_{\mid0^{-}}\left[\left(M,Z^{\top}\right)^{\top}\right]$)
is guaranteed to be positive definite if $\mathrm{Var}\left[B\left(1\right)\mid X=0,\mathsf{co}\right]$
(or $\mathrm{Var}\left[B\left(0\right)\mid X=0,\mathsf{co}\right]$)
is positive definite.
\begin{assumption}
\label{assu:kernel}(a) $K$ is a symmetric continuous probability
density function (PDF) supported on $\left[-1,1\right]$; (b) $\mathcal{K}_{p;+}$
is differentiable with bounded first-order derivatives on $\left(-1,0\right)$
and $\left(0,1\right)$.
\end{assumption}
(a) is standard and also imposed in CCFT. (b) is also found in AK.
(a) implies that $\mathcal{K}_{p;+}\left(t\right)=\mathcal{K}_{p;-}\left(-t\right)$
and therefore (b) also holds for $\mathcal{K}_{p;-}$. Denote $\omega_{p;+}^{j,k}\coloneqq\int_{0}^{1}t^{j}\mathcal{K}_{p;+}^{k}\left(t\right)\mathrm{d}t$
and $\omega_{p;-}^{j,k}\coloneqq\int_{-1}^{0}t^{j}\mathcal{K}_{p;-}^{k}\left(t\right)\mathrm{d}t$.
It is easy to see that $\omega_{p;-}^{0,k}=\omega_{p;+}^{0,k}\eqqcolon\omega_{p}^{0,k}$.
Let $\gamma_{M}\coloneqq\left(\mathrm{Var}_{\mid0^{\pm}}\left[Z\right]\right)^{-1}\mathrm{Cov}_{\mid0^{\pm}}\left[Z,M\right]$,
$\epsilon\coloneqq M-Z^{\top}\gamma_{M}$ and $\sigma^{2}\coloneqq\mathrm{Var}_{\mid0^{\pm}}\left[M\right]-\mathrm{Cov}_{\mid0^{\pm}}\left[M,Z\right]\cdot\gamma_{M}=\mathrm{Var}_{\mid0^{\pm}}\left[\epsilon\right]$.
Existence of these quantities is guaranteed by Assumption \ref{assu:data generating process}(b).
Under Assumption \ref{assu:data generating process}(d), $\sigma^{2}$
is strictly positive. Under Assumption \ref{assu:identification},
$\mu_{\epsilon,+}=\mu_{\epsilon,-}\eqqcolon\mu_{\epsilon}$. Assumption
\ref{assu:data generating process}(a) guarantees that $g_{\epsilon}$
and $m_{\epsilon}$ admit continuous derivatives up to $\left(p+1\right)$-th
order on the left and right neighborhoods of $0$ so that the leading
bias terms can be characterized. For any $j\in\mathbb{N}$, $g_{\left\Vert B\right\Vert ^{j}}$
is bounded on $\mathbb{B}\setminus\left\{ 0\right\} $ if $\left(g_{\left\Vert B\left(d\right)\right\Vert ^{j}\mid dd'},g_{\left\Vert B\left(d'\right)\right\Vert ^{j}\mid dd'},g_{dd'}\right)$
are bounded on $\mathbb{B}$, for all $\left(d,d'\right)\in\left\{ 0,1\right\} ^{2}$.
The following result shows the asymptotic normality of the EB and
MC-EL estimators.
\begin{thm}
\label{thm:normality}Suppose that Assumptions \ref{assu:identification},
\ref{assu:data generating process} and \ref{assu:kernel} hold. Assume
that $g_{\left\Vert B\right\Vert ^{4}}$ is bounded on $\mathbb{B}\setminus\left\{ 0\right\} $.
Assume that the bandwidth satisfies $nh^{2p+3}=O\left(1\right)$ and
$nh\rightarrow\infty$. Then,
\[
\sqrt{nh}\left(\widehat{\vartheta}_{p}^{\mathsf{eb}}-\vartheta-\mathscr{B}_{p}^{\mathsf{eb}}h^{p+1}\right)\rightarrow_{d}\mathrm{N}\left(0,\mathscr{V}_{p}\right),
\]
where
\[
\mathscr{B}_{p}^{\mathsf{\mathsf{eb}}}\coloneqq\frac{\mu_{\epsilon,+}^{\left(p+1\right)}\omega_{p;+}^{p+1,1}-\mu_{\epsilon,-}^{\left(p+1\right)}\omega_{p;-}^{p+1,1}}{\mu_{D,\dagger}\left(p+1\right)!}\textrm{ \textit{and} }\mathscr{V}_{p}\coloneqq\frac{\omega_{p}^{0,2}\sigma^{2}}{\varphi\mu_{D,\dagger}^{2}}.
\]
And,
\[
\sqrt{nh}\left(\widehat{\vartheta}_{p}^{\mathsf{mc}}-\vartheta-\mathscr{B}_{p}^{\mathsf{mc}}h^{p+1}\right)\rightarrow_{d}\mathrm{N}\left(0,\mathscr{V}_{p}\right),
\]
where
\[
\mathscr{B}_{p}^{\mathsf{\mathsf{mc}}}\coloneqq\frac{\left(\psi_{\epsilon,+}^{\left(p+1\right)}-\mu_{\epsilon}\varphi^{\left(p+1\right)}\right)\omega_{p;+}^{p+1,1}-\left(\psi_{\epsilon,-}^{\left(p+1\right)}-\mu_{\epsilon}\varphi^{\left(p+1\right)}\right)\omega_{p;-}^{p+1,1}}{\psi_{D,\dagger}\left(p+1\right)!}.
\]
\end{thm}
\begin{rembold}\label{Rmk: bias}The asymptotic smoothing bias $\mathscr{B}_{p}^{\mathsf{eb}}h^{p+1}$
of the EB estimator is exactly the same as that of CCFT's estimator.
The standard LP regression theory (see, e.g., \citealp{imbens2011optimal})
shows that for the standard estimator $\widehat{\vartheta}_{p}^{\mathsf{lp}}$
defined by (\ref{eq:standard}) without using covariates, we have
\[
\sqrt{nh}\left(\widehat{\vartheta}_{p}^{\mathsf{lp}}-\vartheta-\mathscr{B}_{p}^{\mathsf{lp}}h^{p+1}\right)\rightarrow_{d}\mathrm{N}\left(0,\mathscr{V}_{p}^{\mathsf{lp}}\right),
\]
where
\[
\mathscr{B}_{p}^{\mathsf{lp}}\coloneqq\frac{\mu_{M,+}^{\left(p+1\right)}\omega_{p;+}^{p+1,1}-\mu_{M,-}^{\left(p+1\right)}\omega_{p;-}^{p+1,1}}{\mu_{D,\dagger}\left(p+1\right)!}\textrm{ and }\mathscr{V}_{p}^{\mathsf{lp}}\coloneqq\frac{\omega_{p}^{0,2}\mathrm{Var}_{\mid0^{\pm}}\left[M\right]}{\varphi\mu_{D,\dagger}^{2}}.
\]
Without further assumptions, the ranking of $\left|\mathscr{B}_{p}^{\mathsf{eb}}\right|$
versus $\left|\mathscr{B}_{p}^{\mathsf{lp}}\right|$ is undetermined,
in general. Consider the case of $p=1$, which is the usual choice
of LP order for point estimation. It is easy to see that $\omega_{1;+}^{2,1}=\omega_{1;-}^{2,1}\eqqcolon\omega_{1}^{2,1}$
in this case. By linearity of the conditional expectation, we have
$\mu_{\epsilon,\mathrm{s}}^{\left(p+1\right)}=\mu_{M,\mathrm{s}}^{\left(p+1\right)}-\left(\mu_{Z,\mathrm{s}}^{\left(p+1\right)}\right)^{\top}\gamma_{M}$
for $\mathrm{s}\in\left\{ -,+\right\} $. If $g_{Z}$ is twice continuously
differentiable on $\mathbb{B}$ so that $\mu_{Z,+}^{\left(2\right)}=\mu_{Z,-}^{\left(2\right)}$,
the constant part $\mathscr{B}_{1}^{\mathsf{eb}}$ of the leading
smoothing bias term coincides with $\mathscr{B}_{1}^{\mathsf{lp}}$.\footnote{\label{fn:bias term}If $\left(g_{Z\left(d\right)\mid dd'},g_{Z\left(d'\right)\mid dd'},g_{dd'}\right)$
are smooth, it is clear from (\ref{eq:g_phi_B}) that $g_{Z}$ is
twice continuously differentiable on $\mathbb{B}$ if and only if
$\left.\left(\mathrm{d}/\mathrm{d}x\right)^{j}g_{Z\left(1\right)\mid10}\left(x\right)\right|_{x=0}=\left.\left(\mathrm{d}/\mathrm{d}x\right)^{j}g_{Z\left(0\right)\mid10}\left(x\right)\right|_{x=0}$
for $j=0,1,2$. A causal interpretation of this condition is that
the TED's up to the second order of the treatment on covariates are
zero (i.e., $\left.\left(\mathrm{d}/\mathrm{d}x\right)^{j}\mathrm{E}\left[Z\left(1\right)-Z\left(0\right)\mid X=x,\mathsf{co}\right]\right|_{x=0}=0$
for $j=0,1,2$, see Section \ref{sec:TED} ahead).} The MC-EL estimator has a different asymptotic bias term. It can
be seen that $\mathscr{B}_{p}^{\mathsf{\mathsf{mc}}}$ can be written
as the sum of $\mathscr{B}_{p}^{\mathsf{\mathsf{eb}}}$ and additional
terms. However, the ranking of $\left|\mathscr{B}_{p}^{\mathsf{eb}}\right|$
versus $\left|\mathscr{B}_{p}^{\mathsf{\mathsf{mc}}}\right|$ is undetermined,
since the additional terms may have signs opposite to that of $\mathscr{B}_{p}^{\mathsf{eb}}$
and cancellation may happen. When $p=1$, under the additional assumption
$\mu_{Z,+}^{\left(2\right)}=\mu_{Z,-}^{\left(2\right)}$, we have
$\mathscr{B}_{1}^{\mathsf{\mathsf{mc}}}=\left(\psi_{M,+}^{\left(2\right)}-\psi_{M,-}^{\left(2\right)}\right)\omega_{1}^{2,1}/\left(2\psi_{D,\dagger}\right)$.\end{rembold}

\begin{rembold}\label{Rmk: variance comp}The asymptotic variance
$\mathscr{V}_{p}$ of the EB and MC-EL estimators is also the same
as that of CCFT's estimator.\footnote{Indeed, it can be shown that the EB and CCFT's estimators are first-order
equivalent in a stronger sense: $\widehat{\vartheta}_{p}^{\mathsf{CCFT}}-\widehat{\vartheta}_{p}^{\mathsf{eb}}=o_{p}\left(\left(nh\right)^{-1/2}\right)$.} \citet{Kreiss2022} show that CCFT's estimator achieves efficiency
gain $\mathscr{V}_{p}\leq\mathscr{V}_{p}^{\mathsf{lp}}$ by using
$\mathrm{Var}_{\mid0^{\pm}}\left[M-Z^{\top}\gamma_{M}\right]=\mathrm{min}_{\gamma}\,\mathrm{Var}_{\mid0^{\pm}}\left[M-Z^{\top}\gamma\right]\leq\mathrm{Var}_{\mid0^{\pm}}\left[M\right]$.
Consider the case of $p=1$ and assume that $\mu_{Z,+}^{\left(2\right)}=\mu_{Z,-}^{\left(2\right)}$
holds. Since we have $\mathscr{B}_{1}^{\mathsf{lp}}=\mathscr{B}_{1}^{\mathsf{eb}}$
in this case, the AMSE of $\widehat{\vartheta}_{p}^{\mathsf{eb}}$,
which equals $\left(\mathscr{B}_{1}^{\mathsf{eb}}\right)^{2}h^{4}+\mathscr{V}_{p}/\left(nh\right)$
, is always less than or equal to the AMSE of $\widehat{\vartheta}_{p}^{\mathsf{lp}}$,
which equals $\left(\mathscr{B}_{1}^{\mathsf{lp}}\right)^{2}h^{4}+\mathscr{V}_{p}^{\mathsf{lp}}/\left(nh\right)$.
It is noted in \citet[Section 4]{Noack2021} that the AMSE-minimizing
bandwidth for $\widehat{\vartheta}_{p}^{\mathsf{eb}}$ is also always
less than or equal to that of $\widehat{\vartheta}_{p}^{\mathsf{lp}}$.
As a result, the smoothing bias of $\widehat{\vartheta}_{p}^{\mathsf{eb}}$
is also smaller than that of $\widehat{\vartheta}_{p}^{\mathsf{lp}}$,
when the AMSE-minimizing bandwidths are used for both estimators.\end{rembold}

\begin{rembold}\label{Rmk: GMM}Theorem \ref{thm:normality} and
the first-order equivalence between the EL and CCFT estimators explain
the asymptotic efficiency ranking from a GMM perspective: CCFT's estimator
can be interpreted as efficiently incorporating the side information
from the covariate balance condition, which will typically reduce
the asymptotic variance, and in the worst scenario, will yield the
same asymptotic variance if the side information is irrelevant. Such
an argument is analogous to that of \citet{Hirano:2003cz}, which
explains the puzzling phenomenon that the inverse probability weighting
estimator using the nonparametrically estimated propensity score has
a smaller asymptotic variance relative to that uses the true propensity
score. \citet{Hirano:2003cz} show that the former is equivalent to
an EL estimator that incorporates the side information from knowing
the true propensity score efficiently.\end{rembold}

\begin{rembold}\label{Rmk: efficiency gain}When CCFT claim no definite
ranking between their estimator and the standard LP estimator without
covariates, they interpret such an indeterminacy as ``in perfect
agreement with those in the literature on analysis of experiments,...,
where it is also found that incorporating covariates in randomized
controlled trials using linear regression leads to efficiency gains
only under particular assumptions''. As the RD design is often viewed
as local randomization, let us reconcile the asymptotic efficiency
gain and CCFT's comment from the perspective of randomized experiments.
In RD designs, the continuity of the density of the score $X$ implies
that the shares of units with $X$ being in small neighborhoods to
the left and right of the cutoff are equal (\citealp[Section 5.2]{Noack2021}).
Therefore, the RD design is analogous to a randomized experiment with
equal probabilities of being in treatment and control groups. In the
literature of randomized experiments, \citet[Theorem 5.2(iv)]{Negi2014}
show that when the assignment probability is equal to $1/2$, the
pooled regression adjustment (see \citealp{Negi2014} for its definition),
whose algorithm is analogous to that of the CCFT estimator, always
leads to a smaller or equal asymptotic variance. The assignment probability
assumption is automatically fulfilled in RD designs.\end{rembold}

\begin{rembold}\label{Rmk: irrelevant}Theorem \ref{thm:normality}
also implies that including a covariate will not change the asymptotic
variance if and only if the corresponding element in $\gamma_{M}$
is zero. Note that the (true) projection coefficients $\gamma_{M}$
are the probabilistic limits of the regression coefficients of $Z_{i}$
in the ``long'' regression (\ref{eq:CCFT estimator}) including
all covariates. Consider the partition $Z=\left(Z_{1}^{\top},Z_{2}^{\top}\right)^{\top}$
of $Z$ and let $\gamma_{M}^{\top}=\left(\gamma_{1}^{\top},\gamma_{2}^{\top}\right)^{\top}$
be the conformable partition of $\gamma_{M}$ such that the dimension
of $\gamma_{j}^{\top}$ coincides with that of $Z_{j}$, $j=1,2$.
Using Theorem \ref{thm:normality} and $\sigma^{2}=\mathrm{Var}_{\mid0^{\pm}}\left[M\right]-\mathrm{Cov}_{\mid0^{\pm}}\left[M,Z\right]\left(\mathrm{Var}_{\mid0^{\pm}}\left[Z\right]\right)^{-1}\mathrm{Cov}_{\mid0^{\pm}}\left[Z,M\right]$,
then writing $\mathrm{Var}_{\mid0^{\pm}}\left[Z\right]$ as a block
matrix and inverting it, we can easily show that $\mathscr{V}_{p}$
is equal to the asymptotic variance of the covariate-adjusted estimator
using only $Z_{1}$ if and only if $\gamma_{2}=0$. In this case,
$Z_{2}$ is irrelevant in the sense that dropping $Z_{2}$ has no
first-order impact: it neither leads to efficiency loss nor changes
the asymptotic smoothing bias. In conclusion, if we say that an estimator
achieves efficiency gain when its asymptotic variance is smaller than
that of the standard estimator without covariates, then EB, MC-EL,
and CCFT estimators achieve efficiency gain as long as the coefficients
of some covariates are nonzero.\end{rembold}

\subsection{Nonlinearity bias\label{subsec:Nonlinearity-bias}}

This section carries out a higher-order analysis of the MC-EL and
EB estimators $\widehat{\vartheta}_{p}^{\mathsf{mc}}$ and $\widehat{\vartheta}_{p}^{\mathsf{eb}}$.
We apply the quadratic stochastic expansion (\citealp[Section 3]{newey_smith_2004_higher})
to the estimator and write it as the sum of a quadratic function of
centered sample averages and a remainder term of a smaller order of
magnitude. E.g., for CCFT's estimator, using the expression (\ref{eq:CCFT estimator 2}),
we simply write
\[
\widehat{\vartheta}_{Y,p}^{\mathsf{CCFT}}=\frac{1}{nh}\sum_{i}\widehat{W}_{p,i}\left(Y_{i}-Z_{i}^{\top}\gamma_{Y}\right)-\left(\frac{1}{nh}\sum_{i}\widehat{W}_{p,i}Z_{i}\right)^{\top}\left(\widehat{\gamma}_{Y}^{\mathsf{CCFT}}-\gamma_{Y}\right).
\]
The first-order asymptotic analysis is based on the linear term $\left(nh\right)^{-1}\sum_{i}\widehat{W}_{p,i}\left(Y_{i}-Z_{i}^{\top}\gamma_{Y}\right)$.
Using the second term on the right hand side of above equation and
replacing $\widehat{\gamma}_{Y}^{\mathsf{CCFT}}-\gamma_{Y}$ by its
linearization, we extract the quadratic terms. For $\widehat{\vartheta}_{p}^{\mathsf{mc}}$
and $\widehat{\vartheta}_{p}^{\mathsf{eb}}$, more complicated derivations
are needed.

In our nonparametric context, we write the leading (linear and quadratic)
terms as the sum of the first-order stochastic variability term, the
first-order smoothing bias term, the second-order smoothing bias term,
the second-order stochastic variability term, and a (smoothing) bias-variability
interaction term. The first-order (second-order) stochastic variability
term is a linear (quadratic) function of centralized sample averages.
The first-order stochastic variability term is approximately distributed
as $\mathrm{N}\left(0,\mathscr{V}_{p}/\left(nh\right)\right)$. The
first-order smoothing bias has a leading term given by $\mathscr{B}_{p}^{\mathsf{eb}}h^{p+1}$
(or $\mathscr{B}_{p}^{\mathsf{mc}}h^{p+1}$). The expectation of the
second-order stochastic variability term is referred to as the nonlinearity
bias.\footnote{Such a bias is referred to as ``higher-order bias'' by \citet{newey_smith_2004_higher}
and \citet{Graham2012}. We use terminology similar to \citet{Cattaneo2013}
to distinguish such a bias incurred by (second-order) stochastic variability
from smoothing bias in our nonparametric context.} The following theorem provides an asymptotic representation for the
nonlinearity bias.
\begin{thm}
\label{thm:nonlin bias}Suppose that Assumptions \ref{assu:identification},
\ref{assu:data generating process} and \ref{assu:kernel} hold. Assume
that $g_{\left\Vert B\right\Vert ^{6}}$ is bounded on $\mathbb{B}\setminus\left\{ 0\right\} $.
The nonlinearity bias of $\widehat{\vartheta}_{p}^{\mathsf{eb}}$
is given by
\[
\frac{1}{nh}\cdot\left\{ \omega_{p}^{0,2}\cdot\frac{\mathrm{Cov}_{\mid0^{\pm}}\left[\epsilon,D\right]}{\varphi\mu_{D,\dagger}^{2}}+o\left(1\right)\right\} .
\]
The nonlinearity bias of $\widehat{\vartheta}_{p}^{\mathsf{mc}}$
has the same asymptotic representation.
\end{thm}
\begin{rembold}We consider the situation when a relatively large
number of valid covariates that satisfy the covariate balance condition
are available. The first-order asymptotic theory (Theorem \ref{thm:normality}
and Remark \ref{Rmk: variance comp}) shows that the covariate-adjusted
estimator using more covariates should have a smaller asymptotic variance.
Since the covariate adjustment methods can be viewed as effectively
incorporating covariate balance as overidentifying moment restrictions,
second-order asymptotic analysis (\citealp{newey_smith_2004_higher})
reveals that using more covariates could be costly in terms of increased
nonlinearity bias. In our case, we can see that the leading term in
the nonlinearity bias admits an upper bound independent of the number
of covariates since it follows easily from Cauchy-Schwarz inequality
that $\left|\mathrm{Cov}_{\mid0^{\pm}}\left[\epsilon,D\right]\right|\leq\sqrt{2}\cdot\sqrt{\mathrm{Var}_{\mid0^{\pm}}\left[\epsilon\right]}\leq\sqrt{2}\cdot\sqrt{\mathrm{Var}_{\mid0^{\pm}}\left[M\right]}$.
Such a property is analogous to the small bias properties given by
\citet[Theorem 4.5]{newey_smith_2004_higher} and \citet[Theorem 4.1]{Graham2012}.\end{rembold}

\begin{rembold}\label{Rmk: nonlin bias}Let $T\coloneqq\left(\epsilon-\mu_{\epsilon}\right)\left(Z-\mu_{Z}\right)^{\top}\left(\mathrm{Var}_{\mid0^{\pm}}\left[Z\right]\right)^{-1}\left(Z-\mu_{Z}\right)$.
By adapting the proof arguments, we can show that for the generalized
balancing estimator defined in Section \ref{subsec:Generalized-entropy-balancing},
the nonlinearity bias is 
\begin{equation}
\frac{1}{nh}\cdot\left\{ \omega_{p}^{0,2}\cdot\frac{\mathrm{Cov}_{\mid0^{\pm}}\left[\epsilon,D\right]}{\varphi\mu_{D,\dagger}^{2}}-\frac{\varrho}{2}\cdot\omega_{p}^{0,3}\cdot\frac{\mu_{T,\dagger}}{\varphi\mu_{D,\dagger}}+o\left(1\right)\right\} ,\label{eq:nonlin bias generalized}
\end{equation}
where $\varrho\in\mathbb{R}$ is the parameter in the definition of
Cressie-Read divergence in (\ref{eq:Cressie-Read}). It is possible
to construct examples where the absolute value of the extra term increases
linearly with the number of covariates.\footnote{E.g., we consider a modification of the simulation design in Section
\ref{sec:Monte-Carlo-Simulations}. The outcome and the $l$ covariates
are generated by $Y=\mathbbm{1}\left(X\geq0\right)\left(\mu_{y1}\left(X\right)+0.28\cdot\sum_{j=1}^{l}Z^{\left(j\right)}\right)+\mathbbm{1}\left(X<0\right)\left(\mu_{y0}\left(X\right)+0.22\cdot\sum_{j=1}^{l}Z^{\left(j\right)}\right)+\varepsilon_{y}$
and $Z^{\left(j\right)}=\mathbbm{1}\left(X\geq0\right)\mu_{z1}\left(X\right)+\mathbbm{1}\left(X<0\right)\mu_{z0}\left(X\right)+\varepsilon_{z}^{\left(j\right)}$
for all $j=1,...,l$, where $\left(\varepsilon_{z}^{\left(1\right)},...,\varepsilon_{z}^{\left(l\right)}\right)$
are i.i.d and $\varepsilon_{z}^{\left(j\right)}\sim\chi_{1}^{2}-1$
for all $j=1,...,l$. Then we get $\mu_{T,\dagger}=1.44\cdot l$ by
straightforward calculation.} With a relatively large number of valid covariates, a generalized
balancing estimator could have a large nonlinearity bias, while the
leading term in the nonlinearity bias of the EB and MC-EL estimators
are guaranteed to be bounded. Since CCFT's estimator is a slight modification
of the generalized balancing estimator with $\varrho=-2$, we expect
that its nonlinearity bias should admit an asymptotic expansion in
a form similar to (\ref{eq:nonlin bias generalized}) with an extra
unbounded term.\end{rembold}

\subsection{Balancing over functions in linear sieve spaces\label{subsec:Balancing-over-functions}}

The EB approach can incorporate information from not only the covariate
balance conditions imposed on $Z$ but also on those imposed on functions
of $Z.$ This improves efficiency relative to CCFT and can achieve
the best attainable asymptotic variance derived in \citet[Theorem 3]{Noack2021}.
Let $\mathcal{Z}\subseteq\mathbb{R}^{d_{z}}$ denote the support of
$Z$. We assume that $\mathcal{Z}$ is compact. Let $\left(b_{1},...,b_{k},...\right)$
be approximating basis functions defined on $\mathcal{Z}$ (typically,
$b_{1}=1$). Denote $\rho\coloneqq\left(b_{1},...,b_{k}\right)^{\top}$.
We assume that $k=k_{n}$ increases with the sample size $n$. For
notational simplicity, we suppress the dependence of $k$ on $n$
and also the dependence of $\rho$ on $k$. Examples of such basis
functions commonly used in econometrics include algebraic polynomials
(and their transformations), trigonometric polynomials, and B-spline
functions, among others. See, e.g., \citet{chen2007large} and \citet{Belloni:2015fo}
for more details. Consider the following problem of balancing functions
in the linear sieve space $\mathcal{M}_{k}\coloneqq\left\{ \rho^{\top}\gamma:\gamma\in\mathbb{R}^{k}\right\} $
and define the EB weights $\left(w_{1}^{\mathsf{sieve}},...,w_{n}^{\mathsf{sieve}}\right)$
as the solution to
\begin{flalign}
 & \underset{w_{1},...,w_{n}}{\mathrm{min}}\mathit{KL}\left(w_{1},...,w_{n}\parallel\frac{1}{n},...,\frac{1}{n}\right)\nonumber \\
 & \textrm{subject to }\sum_{i}w_{i}\begin{array}{c}
\widehat{W}_{p,i}\rho\left(Z_{i}\right)\end{array}=0_{k},\,\ensuremath{\sum_{i}w_{i}=1}.\label{eq:sieve balancing}
\end{flalign}
The constraint in (\ref{eq:sieve balancing}) imposes the balancing
constraint that $\sum_{i}w_{i}\begin{array}{c}
\widehat{W}_{p,i}f\left(Z_{i}\right)\end{array}=0$ for all $f\in\mathcal{M}_{k}$ for the balancing weights. Then, by
the Lagrangian multiplier method, we get the optimal weights and the
associated Lagrangian multiplier $\left(w_{i}^{\mathsf{sieve}},\lambda_{p}^{\mathsf{sieve}}\right)$
defined by the right-hand sides of (\ref{eq:entropy balancing weight definition})
and (\ref{eq:lambda_hat}) with $\bar{Z}_{i}$ replaced by $\rho\left(Z_{i}\right)$.
Define the sieve EB estimator by 
\begin{equation}
\widehat{\vartheta}_{p}^{\mathsf{sieve}}\coloneqq\frac{\sum_{i}w_{i}^{\mathsf{sieve}}\widehat{W}_{p,i}Y_{i}}{\sum_{i}w_{i}^{\mathsf{sieve}}\widehat{W}_{p,i}D_{i}}.\label{eq:sieve EB}
\end{equation}

The weak continuity and predeterminedness assumptions (Assumption
\ref{assu:identification}(e,f)) allow us to use only the margins
in constructing the balancing weights. Now, in order to use functions
in broader classes for further efficiency gain, we essentially need
CCFT's strong predeterminedness assumption $F_{Z\left(1\right)\mid10}\left(\cdot\mid0\right)=F_{Z\left(0\right)\mid10}\left(\cdot\mid0\right)$
(see Section III of CCFT for discussion), where $F_{Z\left(j\right)\mid dd'}\left(\cdot\mid x\right)$
denotes the conditional cumulative distribution function (CDF) of
$Z\left(j\right)$ given $\left(D\left(1\right),D\left(0\right),X\right)=\left(d,d',x\right)$
and $F_{Z\mid X}$ denotes the conditional CDF of $Z$ given $X$.
In addition, we need to replace Assumption \ref{assu:identification}(e)
with the stronger assumption that the conditional distributions of
$\left(Z\left(d\right),Z\left(d'\right)\right)$ given $\left(D\left(1\right)=d,D\left(0\right)=d',X=x\right)$
change smoothly around the threshold $x=0$. Under the strong continuity
and predeterminedness assumptions, $F_{Z\mid X}$ changes smoothly
($\mathrm{lim}_{x\downarrow0}F_{Z\mid X}\left(z\mid x\right)=\mathrm{lim}_{x\uparrow0}F_{Z\mid X}\left(z\mid x\right)$
for all $z\in\mathcal{Z}$). Then $\left(g_{f\left(Z\left(d\right)\right)\mid dd'},g_{f\left(Z\left(d'\right)\right)\mid dd'}\right)$
are continuous at 0 for all $\left(d,d'\right)\in\left\{ 0,1\right\} ^{2}$
if $f\in\mathcal{M}_{k}$ satisfies some mild conditions.\footnote{\label{fn:sieve}If for all $\left(j,d,d',x\right)\in\left\{ 0,1\right\} ^{3}\times\left[\underline{x},\overline{x}\right]$,
the conditional distribution of $Z\left(j\right)$ given $\left(D\left(1\right),D\left(0\right),X\right)=\left(d,d',x\right)$
admits a density $f_{Z\left(j\right)\mid dd'}\left(\cdot\mid x\right)$
with respect to the $\sigma$-finite dominating measure $\nu$ with
$\nu\left(\mathcal{Z}\right)<\infty$ such that $f_{Z\left(j\right)\mid dd'}\left(z\mid\cdot\right)$
is continuous at 0 for all $z\in\mathcal{Z}$, we can write $g_{f\left(Z\left(j\right)\right)\mid dd'}\left(x\right)=\int f\left(z\right)f_{Z\left(j\right)\mid dd'}\left(z\mid x\right)\nu\left(\mathrm{d}z\right)$.
If for all $x$ in an open neighborhood of 0, $f_{Z\left(j\right)\mid dd'}\left(\cdot\mid x\right)$
is uniformly bounded, this condition is satisfied if $\int_{\mathcal{Z}}\left|f\right|\mathrm{d}\nu<\infty$.} And the strong predeterminedness assumption implies that $g_{f\left(Z\left(1\right)\right)\mid10}\left(0\right)=g_{f\left(Z\left(0\right)\right)\mid10}\left(0\right)$.
It is clear from $g_{f\left(Z\right)}\left(x\right)=\sum_{d,d'}g_{f\left(Z\right)\mid dd'}\left(x\right)g_{dd'}\left(x\right)$
that the covariate balance condition for $f\left(Z\right)$ (i.e.,
$\mu_{f\left(Z\right),+}=\mu_{f\left(Z\right),-}$) is fulfilled.

Theorem \ref{thm:sieve} below shows that the sieve EB estimator
given in (\ref{eq:sieve EB}) achieves further efficiency gain relative
to CCFT's estimator and our EB estimator in (\ref{eq:entropy balancing estimator definition}).
Interestingly, the asymptotic variance of the sieve EB estimator coincides
with \citet{Noack2021}'s best attainable asymptotic variance of their
LP estimator in which a flexible function is subtracted from the dependent
variable. It is clear from (\ref{eq:CCFT estimator 2}) that one can
write CCFT's estimator for $\mu_{Y,\dagger}$ ($\mu_{D,\dagger}$)
as a standard LP regression estimator using $Y_{i}-Z_{i}^{\top}\widehat{\gamma}_{Y}^{\mathsf{CCFT}}$
($Y_{i}-Z_{i}^{\top}\widehat{\gamma}_{D}^{\mathsf{CCFT}}$) as the
dependent variable. \citet{Noack2021} consider replacing the linear
adjustment $Z_{i}^{\top}\widehat{\gamma}_{Y}^{\mathsf{CCFT}}$ in
(\ref{eq:CCFT estimator 2}) with a nonlinear transformation of the
baseline covariates $Z_{i}$. Let $\left(\eta_{Y},\eta_{D}\right)$
be a real-valued adjustment functions defined on $\mathcal{Z}$ and
we consider the standard LP regression estimator $\widehat{\vartheta}_{p}\left(\eta_{Y},\eta_{D}\right)$
using $Y_{i}-\eta_{Y}\left(Z_{i}\right)$ (or $D_{i}-\eta_{D}\left(Z_{i}\right)$)
as the dependent variable in (\ref{eq:standard}). Such an estimator
is consistent if $\mu_{\eta_{Y}\left(Z\right),+}=\mu_{\eta_{Y}\left(Z\right),-}$
and $\mu_{\eta_{D}\left(Z\right),+}=\mu_{\eta_{D}\left(Z\right),-}$.
See \citet[Footnote 6]{Noack2021} for more discussion. Denote $\mu_{+}^{*}\left(z\right)\coloneqq\mathrm{lim}_{x\downarrow0}g_{M\mid ZX}\left(z,x\right)$,
$\mu_{-}^{*}\left(z\right)\coloneqq\mathrm{lim}_{x\uparrow0}g_{M\mid ZX}\left(z,x\right)$,
$\eta^{*}\left(z\right)\coloneqq\left(\mu_{+}^{*}\left(z\right)+\mu_{-}^{*}\left(z\right)\right)/2$,
and $\epsilon^{*}\coloneqq M-\eta^{*}\left(Z\right)$. \citet[Theorem 3]{Noack2021}
show that under some mild conditions on $\left(\eta_{Y},\eta_{D}\right)$,
the asymptotic variance of $\widehat{\vartheta}_{p}\left(\eta_{Y},\eta_{D}\right)$
cannot be smaller than the best attainable asymptotic variance $\mathscr{V}_{p}^{\mathsf{opt}}\coloneqq\omega_{p}^{0,2}\sigma_{\mathsf{opt}}^{2}/\left(\varphi\mu_{D,\dagger}^{2}\right)$,
where $\sigma_{\mathsf{opt}}^{2}\coloneqq\mathrm{Var}_{\mid0^{\pm}}\left[\epsilon^{*}\right]$.\footnote{\label{fn:sieve optimality} Let $\eta^{\dagger}\coloneqq\eta_{Y}-\vartheta\eta_{D}$.
$\mathscr{V}_{p}^{\mathsf{opt}}$ is an asymptotic variance lower
bound for all $\widehat{\vartheta}_{p}\left(\eta_{Y},\eta_{D}\right)$
with adjustment functions $\left(\eta_{Y},\eta_{D}\right)$ fulfilling
the condition that $\mathrm{Cov}_{\mid0^{+}}\left[\eta^{\dagger}\left(Z\right),\mu_{\mathrm{s}}^{*}\left(Z\right)\right]=\mathrm{Cov}_{\mid0^{-}}\left[\eta^{\dagger}\left(Z\right),\mu_{\mathrm{s}}^{*}\left(Z\right)\right]$
for $\mathrm{s}\in\left\{ -,+\right\} $ and $\mathrm{Var}_{\mid0^{+}}\left[\eta^{\dagger}\left(Z\right)-\eta^{*}\left(Z\right)\right]$
and $\mathrm{Var}_{\mid0^{-}}\left[\eta^{\dagger}\left(Z\right)-\eta^{*}\left(Z\right)\right]$
exist. Under the conditions imposed on the densities in Footnote \ref{fn:sieve}
, this assumption is satisfied, if $\int_{\mathcal{Z}}\left(\eta^{\dagger}\right)^{2}\mathrm{d}\nu<\infty$
and $\int_{\mathcal{Z}}\left(\mu_{\mathrm{s}}^{*}\right)^{2}\mathrm{d}\nu<\infty$
for $\mathrm{s}\in\left\{ -,+\right\} $. \citet{Noack2021} show
how to construct estimators that attain the optimal asymptotic variance.}

To show that (\ref{eq:sieve EB}) is asymptotically normally distributed
with the asymptotic variance $\mathscr{V}_{p}^{\mathsf{opt}}$, we
impose the following assumption on the distribution of the observed
variables, which we invoke directly in the proof of Theorem \ref{thm:sieve}.
$\mathbb{B}$ is defined in Assumption \ref{assu:data generating process}.
Let $\bar{B}_{\eta}\coloneqq\left(Y,D,\eta^{*}\left(Z\right)\right)^{\top}$
and let $f_{X\mid Z}$ be the conditional PDF of $X$ given $Z$.
\begin{assumption}
\label{assu:sieve}(a) $\mu_{\eta^{*}\left(Z\right),+}=\mu_{\eta^{*}\left(Z\right),-}$;
(b) Let $\left\{ \varepsilon_{n}\right\} _{n=1}^{\infty}$ denote
a sequence of real-valued functions defined on $\mathcal{Z}$ such
that $\left\Vert \varepsilon_{n}\right\Vert _{\infty}\downarrow0$
as $n\uparrow\infty$ and $g_{\varepsilon_{n}\left(Z\right)}$ is
$\left(p+1\right)$-times continuously differentiable on $\mathbb{B}\setminus\left\{ 0\right\} $,
then, $\mathrm{sup}_{x\in\left(-h,0\right)\cup\left(0,h\right)}\left|g_{\varepsilon_{n}\left(Z\right)}^{\left(p+1\right)}\left(x\right)\right|\downarrow0$
as $n\uparrow\infty$; (c) $g_{\bar{B}_{\eta}}$ has uniformly continuous
derivatives up to the $\left(p+1\right)$-th order on $\mathbb{B}\setminus\left\{ 0\right\} $;
(d) $g_{\bar{B}_{\eta}^{\otimes2}}$ is uniformly continuous on $\mathbb{B}\setminus\left\{ 0\right\} $;
(e) For all $z\in\mathcal{Z}$, $f_{X\mid Z}\left(\cdot\mid z\right)$
and $g_{M\mid ZX}\left(z,\cdot\right)$ are Lipschitz continuous on
$\mathbb{B}\setminus\left\{ 0\right\} $ with Lipschitz constants
$L_{f},L_{g}>0$ respectively; (f) $\mathrm{Var}_{\mid0^{+}}\left[\epsilon^{*}\right]>0$
and $\mathrm{Var}_{\mid0^{-}}\left[\epsilon^{*}\right]>0$.
\end{assumption}
Sufficient and easy-to-interpret conditions can be imposed on the
population distribution of the latent variables to guarantee that
Assumption \ref{assu:sieve} holds. Under the strong predeterminedness
assumption $F_{Z\left(1\right)\mid10}\left(\cdot\mid0\right)=F_{Z\left(0\right)\mid10}\left(\cdot\mid0\right)$,
(a) is satisfied if the assumptions on the densities in Footnote \ref{fn:sieve}
hold and $\eta^{*}$ satisfies the integrability condition in Footnote
\ref{fn:sieve}. (b) is a mild regularity condition similar to \citet[Assumption 2]{Noack2021}.\footnote{\label{fn:sieve 2}Under the existence of the densities of the latent
variables defined in Footnote \ref{fn:sieve}, the conditional distribution
of $Z$ given $X=x$ admits a density $f_{Z\mid X}\left(\cdot\mid x\right)$
with respect to $\nu$ as a mixture:
\begin{equation}
f_{Z\mid X}\left(z\mid x\right)=\begin{cases}
\sum_{d,d'}g_{dd'}\left(x\right)f_{Z\left(d\right)\mid dd'}\left(z\mid x\right) & \textrm{if }x\geq0\\
\sum_{d,d'}g_{dd'}\left(x\right)f_{Z\left(d'\right)\mid dd'}\left(z\mid x\right) & \textrm{if }x<0.
\end{cases}\label{eq:density Z given X mixture}
\end{equation}
The assumption that $f_{Z\mid X}\left(z\mid\cdot\right)$ is $\left(p+1\right)$-times
continuously differentiable on $\mathbb{B}\setminus\left\{ 0\right\} $
for all $z\in\mathcal{Z}$ and $\left(\partial/\partial x\right)^{j}f_{Z\mid X}\left(\cdot\mid x\right)$
is uniformly bounded for all $\left(x,j\right)\in\left(\mathbb{B}\setminus\left\{ 0\right\} \right)\times\left\{ 0,1,...,p+1\right\} $
is satisfied if for all $\left(d,d'\right)\in\left\{ 0,1\right\} ^{2}$,
(1) $\left(f_{Z\left(d\right)\mid dd'}\left(z\mid\cdot\right),f_{Z\left(d'\right)\mid dd'}\left(z\mid\cdot\right)\right)$
are $\left(p+1\right)$-times continuously differentiable on $\mathbb{B}$
with uniformly (in $z\in\mathcal{Z}$) bounded derivatives; (2) $g_{dd'}$
is $\left(p+1\right)$-times continuously differentiable on $\mathbb{B}$
with bounded derivatives. Then under these assumptions, we have $g_{\varepsilon_{n}\left(Z\right)}^{\left(p+1\right)}\left(x\right)=\int\varepsilon_{n}\left(z\right)\left(\left(\partial/\partial x\right)^{p+1}f_{Z\mid X}\left(z\mid x\right)\right)\nu\left(\mathrm{d}z\right)$
and Part (b) holds. } (c,d) are similar to Assumption \ref{assu:data generating process}(a,b).
By (\ref{eq:g_phi_B}), these are satisfied under suitable smoothness
assumptions on $\left(g_{B_{\eta}\left(d\right)\mid dd'},g_{B_{\eta}\left(d'\right)\mid dd'},g_{dd'}\right)$
and $\left(g_{B_{\eta}\left(d\right)^{\otimes2}\mid dd'},g_{B_{\eta}\left(d'\right)^{\otimes2}\mid dd'}\right)$
for $\left(d,d'\right)\in\left\{ 0,1\right\} ^{2}$, where $B_{\eta}\left(d\right)\coloneqq\left(Y\left(d\right),\eta^{*}\left(Z\left(d\right)\right)\right)^{\top}$.
The first part of Assumption \ref{assu:sieve}(e) is satisfied if
$f_{X\mid Z}\left(\cdot\mid z\right)$ are differentiable on $\mathbb{B}\setminus\left\{ 0\right\} $
with uniformly (in $z\in\mathcal{Z}$) bounded derivatives.\footnote{By the Bayes theorem, we can show that the first part is satisfied
if the assumptions discussed in Footnote \ref{fn:sieve 2} hold and
$f_{X}$ is continuously differentiable on $\mathbb{B}$ with uniformly
continuous derivatives.} The second part of Assumption \ref{assu:sieve}(e) is satisfied if
the conditional PDF $f_{M\mid ZX}\left(y\mid z,\cdot\right)$ of $M$
given $\left(Z,X\right)$ is differentiable on $\mathbb{B}\setminus\left\{ 0\right\} $
with derivatives that satisfy some dominance and integrability condition.\footnote{Assume for simplicity that the support $\mathcal{Y}$ of $Y$ is bounded.
Let $f_{Y\left(j\right)Z\left(j\right)\mid dd'}\left(\cdot\mid x\right)$
denote the conditional joint density of $\left(Y\left(j\right),Z\left(j\right)\right)$
given $\left(D\left(1\right),D\left(0\right),X\right)=\left(d,d',x\right)$,
for $\left(j,d,d',x\right)\in\left\{ 0,1\right\} ^{3}\times\left[\underline{x},\overline{x}\right]$.
Then we can write the conditional joint density $f_{MZ\mid X}$ of
$\left(M,Z\right)$ given $X$ as a mixture similar to (\ref{eq:density Z given X mixture})
and write $f_{M\mid ZX}=f_{MZ\mid X}/f_{Z\mid X}$. It is clear that
the second part is satisfied, if for all $\left(d,d'\right)\in\left\{ 0,1\right\} ^{2}$
(1) $\left(f_{Y\left(d\right)Z\left(d\right)\mid dd'}\left(y,z\mid\cdot\right),f_{Y\left(d\right)Z\left(d\right)\mid dd'}\left(y,z\mid\cdot\right)\right)$
are differentiable on $\mathbb{B}$ with uniformly (in $\left(y,z\right)\in\mathcal{Y}\times\mathcal{Z}$)
bounded derivatives and similar assumptions hold for $\left(f_{Z\left(d\right)\mid dd'}\left(z\mid\cdot\right),f_{Z\left(d'\right)\mid dd'}\left(z\mid\cdot\right)\right)$
and $g_{dd'}$; (2) $\left(f_{Z\left(d\right)\mid dd'}\left(\cdot\mid\cdot\right),f_{Z\left(d'\right)\mid dd'}\left(\cdot\mid\cdot\right)\right)$
are bounded away from zero on $\mathcal{Z}\times\mathbb{B}$ and a
similar assumption holds for $g_{dd'}$.} Assumption \ref{assu:sieve}(f) is similar to Assumption \ref{assu:data generating process}(d).
Under Assumption \ref{assu:identification}(c), it is satisfied as
long as $\mathrm{Var}\left[Y\left(d\right)-\eta^{*}\left(Z\left(d\right)\right)\mid X=0,\mathsf{co}\right]>0$
for $d\in\left\{ 0,1\right\} $.

We also impose the following assumption on the basis functions. For
notational simplicity, let $P\coloneqq\rho\left(Z\right)$. 
\begin{assumption}
\label{assu:sieve basis}(a) $\mu_{b_{j}\left(Z\right),+}=\mu_{b_{j}\left(Z\right),-}$
and $\mu_{b_{j}\left(Z\right)\mu_{\mathrm{s}}^{*}\left(Z\right),+}=\mu_{b_{j}\left(Z\right)\mu_{\mathrm{s}}^{*}\left(Z\right),-}$
for all $\left(j,\mathrm{s}\right)\in\mathbb{N}\times\left\{ -,+\right\} $;
(b) There exists constants $0<\underline{\sigma}<\overline{\sigma}<\infty$
independent of $k$ such that for all $x\in\mathbb{B}\setminus\left\{ 0\right\} $
and uniformly over all $k$, $\mathrm{mineig}\left(\mathrm{E}\left[PP^{\top}\mid X=x\right]\right)>\underline{\sigma}$,
$\mathrm{maxeig}\left(\mathrm{E}\left[PP^{\top}\mid X=x\right]\right)<\overline{\sigma}$,
$\mathrm{mineig}\left(\mathrm{E}\left[PP^{\top}\right]\right)>\underline{\sigma}$
and $\mathrm{maxeig}\left(\mathrm{E}\left[PP^{\top}\right]\right)<\overline{\sigma}$;
(c) $g_{b_{j}\left(Z\right)}$ has uniformly continuous derivatives
up to the $\left(p+1\right)$-th order on $\mathbb{B}\setminus\left\{ 0\right\} $,
for all $j\in\mathbb{N}$; (d) There exists a constant $c_{P}>0$
such that $\mathrm{sup}_{x\in\mathbb{B}\setminus\left\{ 0\right\} }\left\Vert g_{P}^{\left(p+1\right)}\left(x\right)\right\Vert \leq c_{P}\sqrt{k}$.
\end{assumption}
(a) is satisfied by all commonly used basis functions, as long as
the assumption in Footnote \ref{fn:sieve} is fulfilled and $\left(\mu_{+}^{*},\mu_{-}^{*}\right)$
satisfy the integrability condition in Footnote \ref{fn:sieve optimality}.
(b) is a standard regularity condition imposing restrictions on the
collinearity of the basis functions for which mild sufficient conditions
are available (see \citealp[Proposition 2.1]{Belloni:2015fo}).\footnote{\label{fn:sieve 3}Part (b) is satisfied if (1) the conditional distribution
of $Z$ given $X=x$ admits a Lebesgue density that is uniformly (for
all $x\in\mathbb{B}\setminus\left\{ 0\right\} $) bounded above and
away from zero; (2) the marginal distribution of $Z$ admits a Lebesgue
density that is bounded above and away from zero; (3) the basis functions
are orthonormal with respect to the Lebesgue measure. By (\ref{eq:density Z given X mixture}),
Condition (1) is satisfied if for all $\left(d,d'\right)\in\left\{ 0,1\right\} ^{2}$,
$\left(f_{Z\left(d\right)\mid dd'}\left(\cdot\mid\cdot\right),f_{Z\left(d'\right)\mid dd'}\left(\cdot\mid\cdot\right)\right)$
are bounded above and away from zero on $\mathcal{Z}\times\mathbb{B}$
and a similar assumption holds for $g_{dd'}$.} (c) is analogous to Assumption \ref{assu:sieve}(c). (d) is satisfied
under (b) and some other mild regularity conditions.\footnote{Under conditions in Footnote \ref{fn:sieve 2}, $g_{P}^{\left(p+1\right)}\left(x\right)=\int\rho\left(z\right)\left(\left(\partial/\partial x\right)^{p+1}f_{Z\mid X}\left(z\mid x\right)\right)\nu\left(\mathrm{d}z\right)$.
Then, by Jensen's and Cauchy-Schwarz inequalities, 
\[
\left\Vert g_{P}^{\left(p+1\right)}\left(x\right)\right\Vert ^{2}\leq\mathrm{E}\left[\left(\frac{\left(\partial/\partial x\right)^{p+1}f_{Z\mid X}\left(Z\mid x\right)}{f_{Z\mid X}\left(Z\mid x\right)}\right)^{2}\right]\cdot\mathrm{E}\left[\left\Vert P\right\Vert ^{2}\mid X=x\right].
\]
Under Part (b), $\mathrm{E}\left[\left\Vert P\right\Vert ^{2}\mid X=x\right]=\mathrm{tr}\left(\mathrm{E}\left[PP^{\top}\mid X=x\right]\right)\leq k\cdot\overline{\sigma}$.
Part (d) holds if the first term is bounded. This holds if (1) $f_{Z\mid X}$
is bounded away from zero on $\mathcal{Z}\times\left(\mathbb{B}\setminus\left\{ 0\right\} \right)$;
(2) $\left(\partial/\partial x\right)^{p+1}f_{Z\mid X}\left(\cdot\mid x\right)$
is uniformly bounded for all $x\in\mathbb{B}\setminus\left\{ 0\right\} $.
Sufficient conditions for these assumptions are discussed in Footnotes
\ref{fn:sieve 2} and \ref{fn:sieve 3}.}

Let $\alpha_{k}\coloneqq\mathrm{inf}_{\gamma\in\mathbb{R}^{k}}\left\Vert \eta^{*}-\rho^{\top}\gamma\right\Vert _{\infty}$
and $\beta_{k}\coloneqq\mathrm{sup}_{x\in\mathcal{Z}}\left\Vert \rho\left(z\right)\right\Vert $.
Bounds for $\alpha_{k}$ under commonly used basis functions are available
from the approximation theory. E.g., if we take $\left(b_{1},...,b_{k},...\right)$
to be the algebraic polynomials and $\eta^{*}$ is $s$-smooth (see,
e.g., \citealt[Section 2.3.1]{chen2007large} for its definition),
then $\alpha_{k}$ is bounded by $k^{-s/d_{z}}$, up to a constant.
Bounds for $\beta_{k}$ are also available in the literature for commonly
used basis functions. For the algebraic polynomials, $\beta_{k}$
is bounded by $k$ up to a constant. If $\left(b_{1},...,b_{k},...\right)$
are B-splines, then an upper bound is $\sqrt{k}$. See \citet{chen2007large}
and \citet{Belloni:2015fo} for results for other basis functions.
In the statement of the following theorem, we impose Assumptions \ref{assu:sieve}
and \ref{assu:sieve basis} in place of Assumption \ref{assu:data generating process}.
\begin{thm}
\label{thm:sieve}Suppose that Assumptions \ref{assu:identification},
\ref{assu:kernel}, \ref{assu:sieve} and \ref{assu:sieve basis}
hold. Assume $\mathrm{E}\left[\left(\mu_{\mathrm{s}}^{*}\right)^{2}\left(Z\right)\right]<\infty$
for $\mathrm{s}\in\left\{ -,+\right\} $, for some $r\geq4$ and $\varsigma\in\left(0,1\right)$,
$g_{\left\Vert Y\right\Vert ^{r}}$ and $g_{\left|\mu_{\mathrm{s}}^{*}\left(Z\right)\right|^{2+\varsigma}}$
($\mathrm{s}\in\left\{ -,+\right\} $) are bounded on $\mathbb{B}\setminus\left\{ 0\right\} $.
Assume that the tuning parameters $\left(h,k\right)$ satisfy $nh^{2p+3}=O\left(1\right)$,
$nh\rightarrow\infty$, $\left(\alpha_{k}+h\right)\beta_{k}\downarrow0$
and $\left(\beta_{k}+\left(nh\right)^{1/r}\right)k/\sqrt{nh}\downarrow0$.
Then, 
\[
\sqrt{nh}\left(\widehat{\vartheta}_{p}^{\mathsf{sieve}}-\vartheta-\mathscr{B}_{p}^{\mathsf{opt}}h^{p+1}\right)\rightarrow_{d}\mathrm{N}\left(0,\mathscr{V}_{p}^{\mathsf{opt}}\right),
\]
where
\[
\mathscr{B}_{p}^{\mathsf{\mathsf{opt}}}\coloneqq\frac{\mu_{\epsilon^{*},+}^{\left(p+1\right)}\omega_{p;+}^{p+1,1}-\mu_{\epsilon^{*},-}^{\left(p+1\right)}\omega_{p;-}^{p+1,1}}{\mu_{D,\dagger}\left(p+1\right)!}.
\]
\end{thm}
\begin{rembold}Theorem \ref{thm:sieve} is analogous to \citet[Theorem 2]{Noack2021}.
Consider the case of $p=1$ as in Remark \ref{Rmk: bias}. If we assume
that $g_{\eta^{*}\left(Z\right)}$ is twice continuously differentiable
on the neighborhood $\mathbb{B}$ of 0 so that $\mu_{\eta^{*}\left(Z\right),+}^{\left(2\right)}=\mu_{\eta^{*}\left(Z\right),-}^{\left(2\right)}$
as in \citet[Assumption 1]{Noack2021}, $\mathscr{B}_{1}^{\mathsf{\mathsf{opt}}}$
coincides with $\mathscr{B}_{1}^{\mathsf{\mathsf{lp}}}$ (i.e., the
constant part of the asymptotic smoothing bias of the standard LP
regression estimator without covariates).\footnote{By arguments similar to those in Footnote \ref{fn:bias term}, $g_{\eta^{*}\left(Z\right)}$
is twice continuously differentiable on $\mathbb{B}$ if and only
if $\left.\left(\mathrm{d}/\mathrm{d}x\right)^{j}g_{\eta^{*}\left(Z\left(1\right)\right)\mid10}\left(x\right)\right|_{x=0}=\left.\left(\mathrm{d}/\mathrm{d}x\right)^{j}g_{\eta^{*}\left(Z\left(0\right)\right)\mid10}\left(x\right)\right|_{x=0}$
for $j=0,1,2$. A causal interpretation is that the TED's up to the
second order of the treatment on $\eta^{*}\left(Z\right)$ are zero.
This condition holds under $\left.\left(\partial/\partial x\right)^{j}f_{Z\left(1\right)\mid10}\left(\cdot\mid x\right)\right|_{x=0}=\left.\left(\partial/\partial x\right)^{j}f_{Z\left(0\right)\mid10}\left(\cdot\mid x\right)\right|_{x=0}$
for $j=0,1,2$, where $f_{Z\left(j\right)\mid dd'}\left(\cdot\mid x\right)$
is the density defined in Footnote \ref{fn:sieve}.}\end{rembold}

\begin{rembold}Theorem \ref{thm:sieve} is analogous to \citet[Theorem 5.6]{Donald:2003ci}.
Viewed as an EL estimator based on a set of over-identified moment
restrictions whose dimension grows with the sample size, $\widehat{\vartheta}_{p}^{\mathsf{sieve}}$
attains the variance lower bound derived by \citet{Noack2021} asymptotically.
Theorem \ref{thm:sieve} also parallels the main result of \citet{Chan2016},
which shows that the sieve-based generalized EB estimator for the
ATE under unconfoundedness attains the semiparametric efficiency bound.
As discussed in the remark following \citet[Theorem 4.5]{newey_smith_2004_higher}
(also see \citealp{Donald:2009cu}), the calculation and conclusion
in Theorem \ref{thm:nonlin bias} and Remark \ref{Rmk: nonlin bias}
are still valid if the number of effective covariates is allowed to
grow with the sample size. The calculation implies that the nonlinearity
bias of the sieve EB estimator is of order $O\left(\left(nh\right)^{-1}\right)$,
while other sieve-based estimators can have nonlinearity bias of order
$O\left(k/\left(nh\right)\right)$.\end{rembold}

\begin{rembold}As in \citet{Donald:2003ci}, we can consider a generalization
using the Cressie-Read divergence defined by (\ref{eq:Cressie-Read}).
The conclusion of Theorem \ref{thm:sieve} holds for any sieve-based
generalized balancing estimator. If $\varrho=-2$, the condition $\left(\beta_{k}+\left(nh\right)^{1/r}\right)k/\sqrt{nh}\downarrow0$
can be weakened to $\beta_{k}\sqrt{\mathrm{log}\left(k\right)k}/\sqrt{nh}\downarrow0$.
Since the generalized balancing estimator with $\varrho=-2$ is a
slight modification of CCFT's estimator. We expect that a ``LP-series''
regression extension (i.e., replacing $Z_{i}$ by $\rho\left(Z_{i}\right)$
in (\ref{eq:CCFT estimator})) of CCFT's estimator has the same asymptotic
distribution under the weaker conditions imposed on the pair of tuning
parameters.\end{rembold}

\section{Likelihood ratio based inference\label{sec:Likelihood-ratio-inference}}

In this section, we consider inference using the likelihood ratio
statistics. Denote $M_{i}\left(\theta\right)\coloneqq Y_{i}-\theta D_{i}$,
$U_{i}\left(\theta\right)\coloneqq\left(M_{i}\left(\theta\right),\bar{Z}_{i}^{\top}\right)^{\top}$
and $U_{i}\coloneqq U_{i}\left(\vartheta\right)$ for notational simplicity.
Let $\tau\in\left(0,1\right)$ be the significance level. Let $F_{\chi_{1}^{2}}$
and $f_{\chi_{1}^{2}}$ denote the CDF and the PDF of a $\chi_{1}^{2}$
($\chi^{2}$ with one degree of freedom) random variable respectively.
Let $c_{\tau}\coloneqq F_{\chi_{1}^{2}}^{-1}\left(1-\tau\right)$
be the $\left(1-\tau\right)$ quantile of the $\chi_{1}^{2}$ distribution.
The standard EL ratio statistic is given by $\mathit{LR}_{p}\left(\theta\mid h\right)\coloneqq2n\left(\ell_{p}^{\mathsf{mc}}\left(\theta\mid h\right)-\ell_{p}^{\mathsf{mc}}\left(\widehat{\vartheta}_{p}^{\mathsf{mc}}\mid h\right)\right)$,
which is a function of $\theta$. An EL confidence set for $\vartheta$
with nominal coverage probability $1-\tau$ is $\mathit{CS}_{p,\tau}\left(h\right)\coloneqq\left\{ \theta:\mathit{LR}_{p}\left(\theta\mid h\right)\leq c_{\tau}\right\} $.\footnote{For fuzzy RD, as \citet{noack2019bias}'s method, the EL confidence
set avoids a ``delta method'' argument used by the Wald-type inference
of CCFT.} When $p=2$ is taken, our smoothness assumption and construction
of $\mathit{CS}_{p,\tau}\left(h\right)$ parallel CCFT in that $\mathit{CS}_{p,\tau}\left(h\right)$
uses the same LP order as CCFT's inference method and Assumption \ref{assu:data generating process}(a)
assumes the same (three-times differentiability) smoothness as CCFT's
Assumption SRD.\footnote{CCFT proposes Wald-type inference using their local linear estimator
with bias correction and standard errors that take into account estimation
of the bias. \citet[Remark 7]{calonico2014robust} show that subtracting
the $p$-th order LP estimator by the nonparametric estimator for
the leading bias term with the same bandwidth is the same as a $\left(p+1\right)$-th
order LP estimator. CCFT's bias-corrected local linear estimator (with
common bandwidths) is the same as a local quadratic regression estimator.} By the Lagrangian multiplier method and strong duality, for fixed
$\theta$,
\begin{equation}
\ell_{p}^{\mathsf{mc}}\left(\theta\mid h\right)=\underset{\lambda}{\mathrm{sup}}\,\frac{1}{n}\sum_{i}\mathrm{log}\left(1+\lambda^{\top}\left(W_{p,i}U_{i}\left(\theta\right)\right)\right).\label{eq:l_constrained}
\end{equation}
By similar derivations as those in Section \ref{sec:Empirical-likelihood-method},
\begin{equation}
\ell_{p}^{\mathsf{mc}}\left(\widehat{\vartheta}_{p}^{\mathsf{mc}}\mid h\right)=-\frac{1}{n}\sum_{i}\mathrm{log}\left(n\cdot w_{i}^{\mathsf{mc}}\right)=\underset{\lambda}{\mathrm{sup}}\,\frac{1}{n}\sum_{i}\mathrm{log}\left(1+\lambda^{\top}\left(W_{p,i}\bar{Z}_{i}\right)\right).\label{eq:l_mc unconstrained}
\end{equation}

Computation of $\mathit{LR}_{p}\left(\theta\mid h\right)$ only requires
solving convex optimization problems. The right hand side of the second
equality in (\ref{eq:l_mc unconstrained}) can be $\infty$ in the
``no solution'' scenario discussed in Section \ref{sec:Empirical-likelihood-method}.
If our algorithm finds a solution for the maximization problem in
(\ref{eq:l_mc unconstrained}), then we proceed to compute (\ref{eq:l_constrained})
for fixed $\theta$ using a similar algorithm. The right hand side
of (\ref{eq:l_constrained}) is $\infty$ if the origin is not in
the interior of the convex hull of $\left\{ W_{p,1}U_{1}\left(\theta\right),...,W_{p,n}U_{n}\left(\theta\right)\right\} $.
In this case, the Newton algorithm would return a very large value
for $\ell_{p}^{\mathsf{mc}}\left(\theta\mid h\right)$ and $\theta$
is excluded from the confidence set. We have the following result
on the shape of $\mathit{CS}_{p,\tau}\left(h\right)$.
\begin{thm}
\label{thm:shape of CS}Suppose that Assumptions \ref{assu:identification},
\ref{assu:data generating process} and \ref{assu:kernel} hold. Assume
that $g_{\left\Vert B\right\Vert ^{4}}$ is bounded on $\mathbb{B}\setminus\left\{ 0\right\} $.
Assume that the bandwidth satisfies $nh^{2p+3}=o\left(1\right)$ and
$nh\rightarrow\infty$. Then, $\mathit{CS}_{p,\tau}\left(h\right)$
is a finite interval with probability approaching one.
\end{thm}
\begin{rembold}Theorem \ref{thm:shape of CS} is an extension of
\citet[Theorem 2.2]{hall1990methodology}. It shows that when the
sample size is large, with high probability, $\mathit{CS}_{p,\tau}\left(h\right)$
must be a finite interval. In general, EL confidence sets may not
satisfy such a property in finites samples. See \citet[Section 3]{otsu2015empirical}
for more discussion.\footnote{In the proof of Theorem \ref{thm:shape of CS}, we show that in finite
samples, $\mathit{CS}_{p,\tau}\left(h\right)$ is unbounded if and
only if some covariate-adjusted EL confidence set for $\psi_{D,\dagger}$
contains zero. In our case, we have the same observation as \citet[Section 3]{otsu2015empirical}.
Unboundedness of $\mathit{CS}_{p,\tau}\left(h\right)$ is indicative
of weak identification in the sense of \citet{feir2016weak}.}\end{rembold}

In the rest of this section, we give several large-sample properties
of the EL inference method. Section \ref{subsec:Uniform-in-bandwidth-Wilks-theor}
establishes uniform-in-bandwidth (first-order) validity of the EL
confidence set. Sections \ref{subsec:Coverage-optimal-bandwidth}
and \ref{subsec:Local-imbalance} are devoted to second-order properties.
Section \ref{subsec:Coverage-optimal-bandwidth} shows the distributional
expansion for the likelihood ratio and proposes a simple analytical
correction to improve coverage accuracy. Section \ref{subsec:Local-imbalance}
considers a scenario in which covariate balance fails to hold and
analyze the sensitivity of the coverage accuracy to this assumption.
We derive the distributional expansion under local perturbation to
the covariate balance condition.

\subsection{Uniform-in-bandwidth Wilks theorem\label{subsec:Uniform-in-bandwidth-Wilks-theor}}

The following theorem parallels the main result of AK and is a substantial
extension of the standard Wilks theorem, which states that $\mathit{LR}_{p}\left(\vartheta\mid h\right)\rightarrow_{d}\chi_{1}^{2}$.
Our result incorporates covariates and accommodates unbounded outcomes.
The proof techniques we use differ from those employed by AK. Let
$\ell^{\infty}\left(\left[1,\overline{h}/\underline{h}\right]\right)$
denote the space of all bounded real-valued functions defined on $\left[1,\overline{h}/\underline{h}\right]$
endowed with the sup-norm. Let $\mathbb{H}\coloneqq\left[\underline{h},\overline{h}\right]$
be a compact bandwidth set where $\underline{h}>0$ and $\overline{h}>0$
($\underline{h}<\overline{h}$) are bandwidths that depend on the
sample size.\footnote{As the main result of AK, Theorem \ref{thm:Wilks} assumes deterministic
upper and lower bounds. Let $\left(\underline{h}^{*},\overline{h}^{*}\right)$
denote some deterministic bounds that some data-dependent bounds $\left(\underline{h},\overline{h}\right)$
capture. As argued by AK, the conclusion of Theorem \ref{thm:Wilks}
still holds under data-dependent bounds if the orders of $\overline{h}/\overline{h}^{*}-1$
and $\underline{h}/\underline{h}^{*}-1$ are sufficiently small and
$\left(\underline{h}^{*},\overline{h}^{*}\right)$ satisfy the assumptions
of Theorem \ref{thm:Wilks}.}
\begin{thm}
\label{thm:Wilks}Suppose that Assumptions \ref{assu:identification},
\ref{assu:data generating process} and \ref{assu:kernel} hold. Suppose
that $\left(\underline{h},\overline{h}\right)$ satisfy $\mathrm{log}\left(n\right)\cdot\overline{h}=o\left(1\right)$,
$n\overline{h}^{2p+3}=o\left(1\right)$ and $n^{1/r}/\left(n\underline{h}\right)^{1/2}+\left(n\underline{h}\right)^{-1/6}=o\left(\mathrm{log}\left(n\right)^{-3}\right)$.
Assume that $g_{\bar{B}^{\otimes2}}$ is Lipschitz continuous and
$g_{\left\Vert B\right\Vert ^{r}}$ is bounded for some $r\geq4$.
There exists a zero-mean Gaussian process $\left\{ \varGamma_{G}\left(s\right):s\in\left[1,\overline{h}/\underline{h}\right]\right\} $
which is a tight random element in $\ell^{\infty}\left(\left[1,\overline{h}/\underline{h}\right]\right)$
with the covariance structure given by
\begin{equation}
\mathrm{E}\left[\varGamma_{G}\left(s\right)\varGamma_{G}\left(t\right)\right]=\sqrt{\frac{s}{t}}\cdot\frac{\int_{0}^{\infty}\mathcal{K}_{p;+}\left(z\right)\mathcal{K}_{p;+}\left(\left(s/t\right)z\right)\mathrm{d}z}{\int_{0}^{\infty}\mathcal{K}_{p;+}\left(z\right)^{2}\mathrm{d}z}.\label{eq:covariance structure}
\end{equation}
Then, $\mathrm{Pr}\left[\mathit{LR}_{p}\left(\vartheta\mid h\right)\leq z_{\tau}\left(\overline{h}/\underline{h}\right)^{2},\ensuremath{\forall}\ensuremath{h\in\mathbb{H}}\right]\rightarrow1-\tau$,
as $n\uparrow\infty$, where $z_{\tau}\left(\overline{h}/\underline{h}\right)$
denotes the $1-\tau$ quantile of $\left\Vert \varGamma_{G}\right\Vert _{\left[1,\overline{h}/\underline{h}\right]}$. 
\end{thm}
\begin{rembold}\label{Rmk: specification search}Theorem \ref{thm:Wilks}
generalizes the standard Wilks theorem with a single bandwidth. It
implies that when $h=\underline{h}=\overline{h}$, $\mathrm{Pr}\left[\vartheta\in\mathit{CS}_{p,\tau}\left(h\right)\right]\rightarrow1-\tau$.
The standard EL confidence set $\mathit{CS}_{p,\tau}\left(h\right)$
may undercover if the bandwidth is selected after specification search
over $\mathbb{H}$. As an example, suppose that $\widehat{h}\coloneqq\mathrm{argmax}_{h\in\mathbb{H}}\mathit{LR}_{p}\left(0\mid h\right)$
is selected to maximize the $p$-value for the two-sided hypothesis
test of $\vartheta=0$. AK shows that $z_{\tau}\left(\overline{h}/\underline{h}\right)^{2}>c_{\tau}$
when $\overline{h}/\underline{h}>1$ but $z_{\tau}\left(\overline{h}/\underline{h}\right)$
grows at a logarithmic speed as $\overline{h}/\underline{h}\uparrow\infty$.
It is clear from Theorem \ref{thm:Wilks} that under $\vartheta=0$,
$\mathrm{Pr}\left[\vartheta\in\mathit{CS}_{p,\tau}\left(\widehat{h}\right)\right]\rightarrow1-\tilde{\tau}$,
where $\tilde{\tau}>\tau$ solves $z_{\tilde{\tau}}\left(\overline{h}/\underline{h}\right)^{2}=c_{\tau}$
and the test does not have asymptotically correct size. Theorem \ref{thm:Wilks}
justifies a simple correction for bandwidth snooping as AK by replacing
the critical value $c_{\tau}$ used by $\mathit{CS}_{p,\tau}\left(h\right)$
with $z_{\tau}\left(\overline{h}/\underline{h}\right)^{2}$. Let $\mathit{CS}_{p,\tau}^{\mathsf{sc}}\left(h\mid\overline{h}/\underline{h}\right)\coloneqq\left\{ \theta:\mathit{LR}_{p}\left(\theta\mid h\right)\leq z_{\tau}\left(\overline{h}/\underline{h}\right)^{2}\right\} $
be the snooping corrected confidence set. Then, $\mathit{CS}_{p,\tau}^{\mathsf{sc}}\left(h\mid\overline{h}/\underline{h}\right)$
has asymptotically correct coverage no matter how $h$ is selected
from $\mathbb{H}$, i.e., $\mathrm{liminf}_{n\uparrow\infty}\mathrm{Pr}\left[\vartheta\in\mathit{CS}_{p,\tau}^{\mathsf{sc}}\left(h\mid\overline{h}/\underline{h}\right)\right]\geq1-\tau$,
for all $h\in\mathbb{H}$. The critical value $z_{\tau}\left(\overline{h}/\underline{h}\right)$
can be easily simulated.\footnote{See the $\mathtt{R}$ package $\mathtt{BWSnooping}$ from \href{http://github.com/kolesarm/BWSnooping}{github.com/kolesarm/BWSnooping}.
If $\overline{h}/\underline{h}\uparrow\infty$ as $n\uparrow\infty$,
then $z_{\tau}\left(\overline{h}/\underline{h}\right)$ can be replaced
by its asymptotic counterpart. See AK for more detailed discussion
on the critical values.}\end{rembold}

\begin{rembold}Note that Theorem \ref{thm:Wilks} uses undersmoothing
to guarantee that the bias term is asymptotically negligible, so it
requires the rate of $\overline{h}$ to be smaller than that optimally
trades off bias and variance. The bias-aware inference approaches
(\citealp{Armstrong2018,Armstrong2020,Imbens2019}) that explicitly
characterize the worst-case bias can give shorter confidence intervals.
This paper considers a different criterion in bandwidth selection
and proposes in Remark \ref{Rmk: coverage} a bandwidth that minimizes
the coverage error of $\mathit{CS}_{p,\tau}\left(h\right)$ and satisfies
the rate requirement for $\overline{h}$.\end{rembold}

\begin{rembold}\label{Rmk: sensitivity}Theorem \ref{thm:thm local imbalance}
shows that $\left\{ \mathit{CS}_{p,\tau}^{\mathsf{sc}}\left(h\mid\overline{h}/\underline{h}\right):h\in\mathbb{H}\right\} $
is an asymptotically valid confidence band for the constant $\vartheta$,
which uses multiple bandwidth choices. Therefore, such an inference
procedure is more robust and less sensitive to bandwidth choice. The
uniform confidence band can also be used for sensitivity analysis
of the result from the confidence set to bandwidth choice. Let $h_{\mathsf{rf}}$
denote a reference bandwidth, and one computes $\mathit{CS}_{p,\tau}\left(h_{\mathsf{rf}}\right)$.
In case of a statistically insignificant result (i.e., $0\in\mathit{CS}_{p,\tau}\left(h_{\mathsf{rf}}\right)$),
it can be argued that using a smaller (larger) bandwidth is necessary
due to high bias (variance) incurred by $h_{\mathsf{rf}}$. However,
the specification search or multiple testing issue undermines the
validity of a significant result ($\mathit{CS}_{p,\tau}\left(h\right)\subseteq\left(0,\infty\right)$
or $\mathit{CS}_{p,\tau}\left(h\right)\subseteq\left(-\infty,0\right)$)
corresponding to some $h\neq h_{\mathsf{rf}}$. In such a case, with
suitable lower and upper bounds $\left(\underline{h},\overline{h}\right)$
such that $\underline{h}<h_{\mathsf{rf}}<\overline{h}$, one may follow
AK's approach and use the band $\left\{ \mathit{CS}_{p,\tau}^{\mathsf{sc}}\left(h\mid\overline{h}/\underline{h}\right):h\in\mathbb{H}\right\} $.
If there exists $h\in\mathbb{H}$ such that $\mathit{CS}_{p,\tau}^{\mathsf{sc}}\left(h\mid\overline{h}/\underline{h}\right)\subseteq\left(0,\infty\right)$
or $\mathit{CS}_{p,\tau}^{\mathsf{sc}}\left(h\mid\overline{h}/\underline{h}\right)\subseteq\left(-\infty,0\right)$,
one may conclude that the RD LATE is different from zero, and the
validity of such a result is guaranteed by Theorem \ref{thm:Wilks}.
On the other hand, if $0\in\mathit{CS}_{p,\tau}^{\mathsf{sc}}\left(h\mid\overline{h}/\underline{h}\right)$
for all $h\in\mathbb{H}$, we conclude that the insignificant result
is insensitive to bandwidth choice. In the case of $0\notin\mathit{CS}_{p,\tau}\left(h_{\mathsf{rf}}\right)$,
it is still necessary to examine the sensitivity of such a significant
result to bandwidth choice (\citealp{imbens2008regression}). With
suitable $\left(\underline{h},\overline{h}\right)$, one may conclude
that $\vartheta>0$ in a robust sense if there exists $h\in\mathbb{H}$
such that $\mathit{CS}_{p,\tau}^{\mathsf{sc}}\left(h\mid\overline{h}/\underline{h}\right)\subseteq\left(0,\infty\right)$
and for all $h\in\mathbb{H}$, $\mathit{CS}_{p,\tau}^{\mathsf{sc}}\left(h\mid\overline{h}/\underline{h}\right)\cap\left(0,\infty\right)\neq\emptyset$.
Compared with AK, our confidence band incorporates information from
covariates, so the robust inference based on it is more powerful.
\end{rembold}

\begin{rembold}\label{Rmk: data-dependent bandwidth}Let $\widehat{h}$
be the minimizer of some data-dependent criterion function defined
on $\left[\underline{h},\overline{h}\right]$. By Theorem \ref{thm:Wilks},
the asymptotic validity of the confidence set $\mathit{CS}_{p,\tau}^{\mathsf{sc}}\left(\widehat{h}\mid\overline{h}/\underline{h}\right)$
is guaranteed without assuming that $\widehat{h}$ fulfills any property,
such as the stochastic order of $\widehat{h}/h-1$ is sufficiently
small so that the noise in $\widehat{h}$ is negligible, where $h$
is some deterministic bandwidth that $\widehat{h}$ tries to capture.\end{rembold}

\subsection{Analytical correction\label{subsec:Coverage-optimal-bandwidth}}

This section provides coverage expansions of the EL confidence sets.
Similar to \citet[Theorem 3.1(a)]{calonico2018optimal}, Theorem \ref{thm:coverage}
below considers two scenarios under the given smoothness assumption
(Assumption \ref{assu:data generating process}(a)). The first scenario
uses the LP order $p$ so that the leading bias term in the coverage
error of the confidence set $\mathit{CS}_{p,\tau}\left(h\right)$
can be characterized. The second scenario exhausts the smoothness
by setting LP order to $p+1$. The smoothing bias, in this case, is
of a smaller order $O\left(h^{p+1+\mathfrak{h}}\right)$ but its leading
term can not be explicitly characterized. The following mild assumption
on the kernel function is used when establishing the validity of the
Edgeworth expansions in the proofs of Theorems \ref{thm:coverage}
and \ref{thm:thm local imbalance}.
\begin{assumption}
\label{assu:kernel 2} $\left(1,\mathcal{K}_{p;+},\mathcal{K}_{p;+}^{2},\mathcal{K}_{p;+}^{3}\right)$
are linearly independent as elements in the vector space of continuous
functions on $\left(0,1\right)$.
\end{assumption}
Since $K\left(\cdot\right)$ is assumed to be symmetric, an analogous
property holds for $\left(1,\mathcal{K}_{p;-},\mathcal{K}_{p;-}^{2},\mathcal{K}_{p;-}^{3}\right)$
as functions on $\left(-1,0\right)$ under this assumption. It is
clear that the assumption is satisfied if $\mathcal{K}_{p;+}$ is
a non-constant polynomial on $\left[-1,1\right]$. If $p\geq1$, this
condition is satisfied if $K\left(\cdot\right)$ is any of the commonly
used kernel functions (triangular, biweight, triweight, etc.) including
the uniform kernel.\footnote{Suppose that $p=1$ and $K$ is the uniform kernel, i.e., $K\left(t\right)=\mathbbm{1}\left(\left|t\right|\leq1\right)/2$.
Then, by simple calculation, $\mathcal{K}_{p;+}\left(t\right)=\left(4-6t\right)\mathbbm{1}\left(\left|t\right|\leq1\right)$
and $\mathcal{K}_{p;-}\left(t\right)=\left(4+6t\right)\mathbbm{1}\left(\left|t\right|\leq1\right)$.} Denote $\Xi\coloneqq\mu_{UU^{\top},\pm}^{-1}$, $\Psi_{1}^{\mathsf{kl}}\coloneqq\mathrm{tr}\left(\Xi\cdot\mu_{U^{\left(\mathsf{k}\right)}U^{\left(\mathsf{l}\right)}UU^{\top},\pm}\right)$
and $\Psi_{2}^{\mathsf{kl}}\coloneqq\mathrm{tr}\left(\Xi\cdot\mu_{U^{\left(\mathsf{k}\right)}UU^{\top},\dagger}\Xi\cdot\mu_{U^{\left(\mathsf{l}\right)}UU^{\top},\dagger}\right)$.
Let
\begin{equation}
\mathscr{V}_{p}^{\dagger}\coloneqq\sum_{\mathsf{k},\mathsf{l}=1,...,d_{z}+2}\left(\omega_{p}^{0,2}\varphi\right)^{-1}\left\{ \frac{1}{2}\cdot\frac{\omega_{p}^{0,4}}{\omega_{p}^{0,2}}\Xi^{\left(\mathsf{kl}\right)}\Psi_{1}^{\mathsf{kl}}-\frac{1}{3}\cdot\frac{\left(\omega_{p}^{0,3}\right)^{2}}{\left(\omega_{p}^{0,2}\right)^{2}}\Xi^{\left(\mathsf{kl}\right)}\Psi_{2}^{\mathsf{kl}}\right\} .\label{eq:V_0 definition}
\end{equation}
Let $\mathscr{V}_{p}^{\ddagger}$ be defined by the same formula with
$U$ replaced by $\bar{Z}$ and the range changed to $1,...,d_{z}+1$
accordingly. Let $\mathscr{V}_{p}^{\mathsf{LR}}\coloneqq\mathscr{V}_{p}^{\dagger}-\mathscr{V}_{p}^{\ddagger}$
and $\mathscr{B}_{p}^{\mathsf{LR}}\coloneqq\left(\mathscr{B}_{p}^{\mathsf{mc}}\right)^{2}/\mathscr{V}_{p}$.

Now we provide distributional expansions for both $\mathit{LR}_{p}\left(\vartheta\mid h\right)$
and $\mathit{LR}_{p+1}\left(\vartheta\mid h\right)$. Asymptotic expansions
of the coverage probabilities follow from these results (e.g, $\mathrm{Pr}\left[\vartheta\in\mathit{CS}_{p,\tau}\left(h\right)\right]=\mathrm{Pr}\left[\mathit{LR}_{p}\left(\vartheta\mid h\right)\leq c_{\tau}\right]$).
The proof uses the method of \citet{calonico2018coverage} and calculations
in \citet{chen2007second}.

\begin{thm}
\label{thm:coverage}Suppose that Assumptions \ref{assu:identification},
\ref{assu:data generating process}, \ref{assu:kernel} and \ref{assu:kernel 2}
hold. Assume that $g_{\bar{B}^{\otimes j}}$ is Lipschitz continuous
on $\mathbb{B}\setminus\left\{ 0\right\} $ for $j=2,3,4$ and $g_{\left\Vert B\right\Vert ^{20}}$
is bounded on $\mathbb{B}\setminus\left\{ 0\right\} $. Suppose that
$h$ satisfies $nh^{2p+3}=o\left(1\right)$ and $nh\rightarrow\infty$.
Then,
\begin{equation}
\mathrm{Pr}\left[\mathit{LR}_{p}\left(\vartheta\mid h\right)\leq x\right]=F_{\chi_{1}^{2}}\left(x\right)-\left(nh^{2p+3}\mathscr{B}_{p}^{\mathsf{LR}}+\frac{\mathscr{V}_{p}^{\mathsf{LR}}}{nh}\right)xf_{\chi_{1}^{2}}\left(x\right)+O\left(\upsilon_{p,n}\right)\label{eq:p-th order distributional expansion}
\end{equation}
and
\[
\mathrm{Pr}\left[\mathit{LR}_{p+1}\left(\vartheta\mid h\right)\leq x\right]=F_{\chi_{1}^{2}}\left(x\right)-\frac{\mathscr{V}_{p+1}^{\mathsf{LR}}}{nh}\cdot xf_{\chi_{1}^{2}}\left(x\right)+O\left(\upsilon_{p+1,n}\right),
\]
where $\upsilon_{p,n}\coloneqq h^{p+1}/\sqrt{nh}+\left(\mathrm{log}\left(n\right)\right)^{5/2}/\left(nh\right)^{3/2}+h^{p+2}+n^{-1}+\left(nh\right)^{2}\left(h^{p+1}\right)^{4}+nh^{2p+3+\mathfrak{h}}$
and $\upsilon_{p+1,n}\coloneqq nh^{2p+3+2\mathfrak{h}}+h^{p+1+\mathfrak{h}}/\sqrt{nh}+\left(\mathrm{log}\left(n\right)\right)^{5/2}/\left(nh\right)^{3/2}+h^{p+2+\mathfrak{h}}+n^{-1}$.
\end{thm}
\begin{rembold}\label{Rmk: coverage}In (\ref{eq:p-th order distributional expansion})
, $nh^{2p+3}\mathscr{B}_{p}^{\mathsf{LR}}$ is the ``bias'' term
that is brought by the smoothing bias and $\left(nh\right)^{-1}\mathscr{V}_{p}^{\mathsf{LR}}$
is the ``variability term'' that stems from the stochastic variability.
Since $h\asymp n^{-1/\left(p+2\right)}$ gives the best coverage error
decay rate, following CCFT we restrict our attention to bandwidths
that satisfy $h=H\cdot n^{-1/\left(p+2\right)}$ for some constant
$H>0$. The leading coverage error is proportional to $-n^{-\left(p+1\right)/\left(p+2\right)}\left(\mathscr{B}_{p}^{\mathsf{LR}}H^{2p+3}+\mathscr{V}_{p}^{\mathsf{LR}}H^{-1}\right)$.\footnote{Note that typically the distributional expansion corresponding to
a nonparametric kernel-based Wald-type statistic (e.g., \citealp[Theorem 3.1]{calonico2018optimal})
is more complicated and involves another ``bias-variability'' interaction
term of order $h^{p+1}$.} Parallel to \citet{calonico2018effect}, we define $H_{\mathsf{co}}\coloneqq\mathrm{argmin}_{H>0}\left|\mathscr{B}_{p}^{\mathsf{LR}}H^{2p+3}+\mathscr{V}_{p}^{\mathsf{LR}}H^{-1}\right|$
to be the optimal constant. Note that $H_{\mathsf{co}}$ is independent
of the nominal coverage probability $1-\tau$ and has a simple closed
form.\footnote{If $\mathscr{V}_{p}^{\mathsf{LR}}>0$, the unique minimizer $H_{\mathsf{co}}$
satisfies the first-order condition. An explicit solution is available
from solving it: $H_{\mathsf{co}}=\left(\mathscr{V}_{p}^{\mathsf{LR}}/\left(\left(2p+3\right)\mathscr{B}_{p}^{\mathsf{LR}}\right)\right)^{1/\left(2p+4\right)}$.
If $\mathscr{V}_{p}^{\mathsf{LR}}<0$, it is easy to see that $H_{\mathsf{co}}=\left(-\mathscr{V}_{p}^{\mathsf{LR}}/\mathscr{B}_{p}^{\mathsf{LR}}\right)^{1/\left(2p+4\right)}$
and $\mathscr{B}_{p}^{\mathsf{LR}}H_{\mathsf{co}}^{2p+3}+\mathscr{V}_{p}^{\mathsf{LR}}H_{\mathsf{co}}^{-1}=0$.
In this case, the first-order coverage error vanishes at the optimal
bandwidth.} These properties are not shared by the CO bandwidths for the Wald-type
inference methods. In practical implementation, $H_{\mathsf{co}}$
has to be estimated. A simple plug-in estimator $\widehat{\mathscr{V}}_{p}^{\mathsf{LR}}$
of $\mathscr{V}_{p}^{\mathsf{LR}}$ that is based on local linear
regression with standard rule-of-thumb (ROT) bandwidths (\citealp[Chapter 21.6]{Hansen2021})
has a relatively fast $O_{p}\left(n^{-2/5}\right)$ rate of convergence.
On the other hand, since $\mathscr{B}_{p}^{\mathsf{LR}}$ involves
higher-order derivatives up to the $\left(p+1\right)$-th order, estimation
of derivatives using a working parametric model is recommended for
bandwidth selection (see, e.g., \citealp[Chapter 21.6]{Hansen2021}).\footnote{\label{fn:B_LR nonparametric}If $\mathfrak{h}$ is known and the
bandwidth is chosen to guarantee the fastest rate of convergence,
the rate of a fully nonparametric estimator of $\mathscr{B}_{p}^{\mathsf{LR}}$
based on the $\left(p+1\right)$-th order LP regression is $O_{p}\left(n^{-\mathfrak{h}/\left(2p+3+2\mathfrak{h}\right)}\right)$
under our smoothness assumption.}\end{rembold}

\begin{rembold}The confidence set $\mathit{CS}_{p,\tau}\left(h\right)$
uses the same bandwidth $h$ for all values of $\theta$ and can be
considered as being obtained by inversion of a test of $\mathrm{H}_{0}:\vartheta=\theta$
using the test statistic $\mathit{LR}_{p}\left(\theta\mid h\right)$.
We can consider a bandwidth dependent on the hypothesized value $\theta$
under $\mathrm{H}_{0}$. Let $\left(\mathscr{B}_{p}^{\mathsf{LR}}\left(\theta\right),\mathscr{V}_{p}^{\mathsf{LR}}\left(\theta\right)\right)$
be defined by the formulae of $\left(\mathscr{B}_{p}^{\mathsf{LR}},\mathscr{V}_{p}^{\mathsf{LR}}\right)$
with $M$ replaced by $M\left(\theta\right)$. By Theorem \ref{thm:coverage}
and similar arguments as those in Remark \ref{Rmk: coverage}, the
size-distortion-minimizing bandwidth is given by $H_{\mathsf{co}}\left(\theta\right)\cdot n^{-1/\left(p+2\right)}$,
where $H_{\mathsf{co}}\left(\theta\right)\coloneqq\mathrm{argmin}_{H>0}\left|\mathscr{B}_{p}^{\mathsf{LR}}\left(\theta\right)H^{2p+3}+\mathscr{V}_{p}^{\mathsf{LR}}\left(\theta\right)H^{-1}\right|$.\footnote{In the inference part of this paper, we mainly focus on improving
the coverage accuracy. If the size of the confidence set is concerned,
one may consider local alternatives for a given hypothesized value
$\theta$ under $\mathrm{H}_{0}$ and choose the $\theta$-dependent
power-optimal constant under a criterion from the distributional expansion
of the test statistic $\mathit{LR}_{p}\left(\theta\mid h\right)$
under the local alternatives.} Note that the constant $H_{\mathsf{co}}$ defined in Remark \ref{Rmk: coverage}
is just $H_{\mathsf{co}}\left(\vartheta\right)$. Clearly, the coverage
expansion of the confidence set $\text{\ensuremath{\widetilde{\mathit{CS}}_{p,\tau}}}\coloneqq\left\{ \theta:\mathit{LR}_{p}\left(\theta\mid H_{\mathsf{co}}\left(\theta\right)\cdot n^{-1/\left(p+2\right)}\right)\leq c_{\tau}\right\} $
has the same second-order term as $\mathit{CS}_{p,\tau}\left(H_{\mathsf{co}}\cdot n^{-1/\left(p+2\right)}\right)$.
A preliminary estimator of $\vartheta$ is required for estimation
of $H_{\mathsf{co}}$ but not for estimation of $H_{\mathsf{co}}\left(\theta\right)$.\footnote{For this reason, $\text{\ensuremath{\widetilde{\mathit{CS}}_{p,\tau}}}$
with estimated $H_{\mathsf{co}}\left(\theta\right)$ is likely to
have better coverage accuracy in finite samples since the estimator
of $H_{\mathsf{co}}\left(\theta\right)$ is less variable than that
of $H_{\mathsf{co}}$. Noise in the selection of bandwidth will translate
into coverage error of the confidence set (see \citealp[Theorem 4 and Remark 5]{Ma2023}).
Also see \citet[Chapter 21.6]{Hansen2021}.} However, in light of Theorem \ref{thm:shape of CS}, $\mathit{CS}_{p,\tau}\left(H_{\mathsf{co}}\cdot n^{-1/\left(p+2\right)}\right)$
has a more interpretable form, while $\text{\ensuremath{\widetilde{\mathit{CS}}_{p,\tau}}}$
can be disconnected.\end{rembold}

\begin{rembold}\label{Rmk: Bartlett correction}The simple expression
on the right hand side of (\ref{eq:p-th order distributional expansion})
suggests that analytical correction can be implemented to improve
coverage accuracy. E.g., it follows from (\ref{eq:p-th order distributional expansion})
and Taylor expansion that the distribution of $\mathit{LR}_{p}\left(\vartheta\mid h\right)/\left(1+nh^{2p+3}\mathscr{B}_{p}^{\mathsf{LR}}+\left(nh\right)^{-1}\mathscr{V}_{p}^{\mathsf{LR}}\right)$
is $F_{\chi_{1}^{2}}\left(x\right)+O\left(\upsilon_{p,n}\right)$
(i.e., rescaling completely removes the leading terms). Feasible correction
uses nonparametric estimators $\left(\widehat{\mathscr{B}}_{p}^{\mathsf{LR}},\widehat{\mathscr{V}}_{p}^{\mathsf{LR}}\right)$
of $\left(\mathscr{B}_{p}^{\mathsf{LR}},\mathscr{V}_{p}^{\mathsf{LR}}\right)$.
Let $\mathit{LR}_{p}^{\mathsf{bc}}\left(\theta\mid h\right)\coloneqq\mathit{LR}_{p}\left(\theta\mid h\right)/\left(1+nh^{2p+3}\widehat{\mathscr{B}}_{p}^{\mathsf{LR}}+\left(nh\right)^{-1}\widehat{\mathscr{V}}_{p}^{\mathsf{LR}}\right)$
be the likelihood ratio with analytical (Bartlett) correction and
let $\mathit{CS}_{p,\tau}^{\mathsf{bc}}\left(h\right)\coloneqq\left\{ \theta:\mathit{LR}_{p}^{\mathsf{bc}}\left(\theta\mid h\right)\leq c_{\tau}\right\} $
be the corrected confidence set. This correction approach removes
the leading bias term in (\ref{eq:p-th order distributional expansion})
by using an estimator $\widehat{\mathscr{B}}_{p}^{\mathsf{LR}}$.
We  also consider implicit bias removal based on the idea of \citet{calonico2014robust}
by increasing the LP order by one. Let $\mathit{LR}_{p+1}^{\mathsf{bc}}\left(\theta\mid h\right)\coloneqq\mathit{LR}_{p+1}\left(\theta\mid h\right)/\left(1+\left(nh\right)^{-1}\widehat{\mathscr{V}}_{p+1}^{\mathsf{LR}}\right)$
be the likelihood ratio with analytical (partial Bartlett) correction
(\citealp{chen1996empirical}) and let $\mathit{CS}_{p+1,\tau}^{\mathsf{bc}}\left(h\right)\coloneqq\left\{ \theta:\mathit{LR}_{p+1}^{\mathsf{bc}}\left(\theta\mid h\right)\leq c_{\tau}\right\} $
be the corrected confidence set. This approach essentially trades
bias for variability, as the latter can be estimated with good accuracy.
By the second part of Theorem \ref{thm:coverage}, under the assumption
that $\widehat{\mathscr{V}}_{p+1}^{\mathsf{LR}}-\mathscr{V}_{p+1}^{\mathsf{LR}}=O_{p}\left(n^{-2/5}\right)$
and $\left(nh^{3}\right)^{-1}=O\left(1\right)$, $\mathrm{Pr}\left[\mathit{LR}_{p+1}^{\mathsf{bc}}\left(\vartheta\mid h\right)\leq x\right]=F_{\chi_{1}^{2}}\left(x\right)+O\left(\upsilon_{p+1,n}\right)$.
The confidence set $\mathit{CS}_{p+1,\tau}^{\mathsf{bc}}\left(h\right)$
has a faster coverage error decay rate than $\mathit{CS}_{p,\tau}^{\mathsf{bc}}\left(h\right)$
for all $h$.\footnote{If $\widehat{\mathscr{B}}_{p}^{\mathsf{LR}}$ is the fully nonparametric
estimator in Footnote \ref{fn:B_LR nonparametric}, the coverage error
of $\left\{ \theta:\mathit{LR}_{p}^{\mathsf{bc}}\left(\theta\mid h\right)\leq c_{\tau}\right\} $
is of order $n^{1-\mathfrak{h}/\left(2p+3+2\mathfrak{h}\right)}h^{2p+3}+\upsilon_{p,n}$,
which converges to zero at a rate slower than $\upsilon_{p+1,n}$.
It is easy to check that $\upsilon_{p+1,n}=O\left(n^{-1}\right)$
under $h\asymp n^{-1/\left(p+2\right)}$ if $p\geq1$ and $\mathfrak{h}\geq1/2$.
However, we note that this does not imply that the finite-sample coverage
accuracy of $\mathit{CS}_{p+1,\tau}^{\mathsf{bc}}\left(h\right)$
is always better than that of $\mathit{CS}_{p,\tau}^{\mathsf{bc}}\left(h\right)$,
since the constant terms in the coverage errors are different.} Viewed differently, $\mathit{CS}_{p+1,\tau}^{\mathsf{bc}}\left(h\right)$
with $h\asymp n^{-1/\left(p+2\right)}$ follows the idea of partial
Bartlett correction of \citet{chen1996empirical} in that upon removal
of the leading variability term, undersmoothing relative to its CO
rate ($n^{-1/\left(p+2+\mathfrak{h}\right)}$) reduces the effects
from the smoothing bias on the coverage accuracy and gives a faster
coverage error decay rate.\end{rembold}

\begin{rembold}\label{Rmk: doubly corrected}By using the AK-type
correction proposed in Theorem \ref{thm:Wilks}, we can also construct
a confidence band that uses a continuous range of bandwidths to analyze
the sensitivity of the result from $\mathit{CS}_{p,\tau}^{\mathsf{bc}}\left(h\right)$
or $\mathit{CS}_{p+1,\tau}^{\mathsf{bc}}\left(h\right)$ to bandwidth
choice. The conclusion of Theorem \ref{thm:Wilks} still holds for
$\mathit{LR}_{p+1}\left(\vartheta\mid h\right)$ and also for $\mathit{LR}_{p}^{\mathsf{bc}}\left(\vartheta\mid h\right)$
and $\mathit{LR}_{p+1}^{\mathsf{bc}}\left(\vartheta\mid h\right)$
since they are first-order equivalent to $\mathit{LR}_{p}\left(\vartheta\mid h\right)$
and $\mathit{LR}_{p+1}\left(\vartheta\mid h\right)$, uniformly in
$h\in\mathbb{H}$. We can take the lower and upper bounds in $\mathbb{H}$
to be proportional to some commonly used reference bandwidths. The
``doubly corrected'' confidence sets can be constructed by following
the procedure in Remark \ref{Rmk: specification search}. We also
expect a small coverage error for the corrected EL confidence band.\footnote{In the proof of the asymptotic validity of the confidence band, we
show that the distribution of $\mathrm{sup}_{h\in\mathbb{H}}\mathit{LR}_{p}^{\mathsf{bc}}\left(\vartheta\mid h\right)$
is approximated by the distribution of $\left\Vert \varGamma_{G}\right\Vert _{\left[1,\overline{h}/\underline{h}\right]}^{2}=\mathrm{sup}_{h\in\mathbb{H}}\varGamma_{G}\left(h/\underline{h}\right)^{2}$
with a vanishing error, where $\varGamma_{G}\left(h/\underline{h}\right)^{2}$
follows the $\chi_{1}^{2}$ distribution for all $h\in\mathbb{H}$.
We expect that the distributional approximation of $\mathrm{sup}_{h\in\mathbb{H}}\varGamma_{G}\left(h/\underline{h}\right)^{2}$
to $\mathrm{sup}_{h\in\mathbb{H}}\mathit{LR}_{p}^{\mathsf{bc}}\left(\vartheta\mid h\right)$
inherits the good accuracy of the pointwise-in-bandwidth distributional
approximation of $\varGamma_{G}\left(h/\underline{h}\right)^{2}$
to $\mathit{LR}_{p}^{\mathsf{bc}}\left(\vartheta\mid h\right)$.}\end{rembold}

\subsection{Local imbalance\label{subsec:Local-imbalance}}

This section shows that the coverage performance of the EL confidence
set is maintained even if the covariate balance assumption is slightly
violated, a scenario we call ``local imbalance''. Specially, we
assume that the observed covariates $Z$ are subject to data contamination
(measurement errors) that occurs after treatment. The contaminated
covariates may not satisfy the predeterminedness assumption and can
be drawn from some perturbed probability law that generates a slight
imbalance (\citealp{kitamura2013robustness}). On the other hand,
the genuine but unobserved predetermined covariates $Z^{\star}\in\mathbb{R}^{d_{z}}$,
which typically affects $Y\left(d\right)$, still satisfy the balance
condition. The continuity of $\left(g_{Y\left(d\right)\mid dd'},g_{Y\left(d'\right)\mid dd'}\right)$
remains to hold. In other words, the imbalance is caused by measurement
errors that are known to be excluded from the data-generating process
of $Y\left(d\right)$. In this case, the standard RD estimand $\vartheta$,
which confidence sets try to cover, remains to identify a causal parameter
of interest.\footnote{\citet{frolich2019including} consider a different scenario where
conditionally on $Z$, the continuity (exclusion) assumption is satisfied.
This essentially assumes that there are no unobserved variables that
both affect the potential outcomes and are affected by $I$. \citet{frolich2019including}
show that the RD LATE is still identifiable (under additional assumptions),
but smoothing over $Z$ is required for estimation.}

Formally, let $\zeta\in\mathbb{R}^{d_{z}}$ denote the measurement
errors realized after treatment. The measurement error $\zeta$ is
nonclassical in the sense that it relates to $\left(D,X,Z^{\star}\right)$.
Let $\left(Z^{\star}\left(1\right),Z^{\star}\left(0\right),\zeta\left(1\right),\zeta\left(0\right)\right)$
be potential covariates and measurement errors.\footnote{The structural model representation in Footnote \ref{fn:RD triangular}
can be extended to $Y=g\left(D,X,Z^{\star},\epsilon\right)$, $D=h\left(I,X,\eta\right)$,
$Z^{\star}=m\left(D,X,\xi\right)$ and $\zeta=q\left(D,Z^{\star},X,\nu\right)$
for some unknown functions $\left(g,h,m,q\right)$ and unobserved
disturbances $\left(\epsilon,\eta,\xi,\nu\right)$.} The contaminated potential covariates $\left(Z\left(1\right),Z\left(0\right)\right)$
are generated by $Z\left(d\right)=Z^{\star}\left(d\right)+\zeta\left(d\right)$
for $d=0,1$. And the observed contaminated covariates are $Z=D\cdot Z\left(1\right)+\left(1-D\right)Z\left(0\right)$.
We assume that the true covariates satisfy the ``predeterminedness''
assumption $g_{Z^{\star}\left(1\right)\mid10}\left(0\right)=g_{Z^{\star}\left(0\right)\mid10}\left(0\right)$,
but the measurement errors fail to satisfy it. As a result, local
imbalance in essence assumes that $g_{Z\left(1\right)\mid10}\left(0\right)-g_{Z\left(0\right)\mid10}\left(0\right)$
approaches 0 at the rate of $\left(nh\right)^{-1/2}$. We are interested
in the coverage probability $\mathrm{Pr}\left[\vartheta\in\mathit{CS}_{p+1,\tau}^{\mathsf{bc}}\left(h\right)\right]$,
which is expected to have a limit in $\left(0,1-\tau\right)$ and
thus captures the phenomenon that covariate imbalance results in undercoverage.We
set the bandwidth to $h\asymp n^{-1/\left(p+2\right)}$ as discussed
in Remark \ref{Rmk: Bartlett correction}. Let $l_{n}\coloneqq n^{-\left(p+1\right)/\left(2p+4\right)}\asymp\left(nh\right)^{-1/2}$.
The following assumption formalizes local imbalance.
\begin{assumption}
\label{assu:imbalance}(a) $\left(g_{Z^{\star}\left(d\right)\mid dd'},g_{\zeta\left(d\right)\mid dd'}\right)$
and $\left(g_{Z^{\star}\left(d'\right)\mid dd'},g_{\zeta\left(d'\right)\mid dd'}\right)$
are all continuous at the threshold $0$ for all $\left(d,d'\right)\in\left\{ 0,1\right\} ^{2}$;
(b) $g_{Z^{\star}\left(1\right)\mid10}\left(0\right)=g_{Z^{\star}\left(0\right)\mid10}\left(0\right)$;
(c) $g_{\zeta\left(1\right)\mid10}\left(0\right)-g_{\zeta\left(0\right)\mid10}\left(0\right)=\delta\cdot l_{n}$
for some localizing parameter $\delta\in\mathbb{R}^{d_{z}}$.
\end{assumption}
Part (a) essentially assumes that no other variables depending on
$I$ affect $\left(Z^{\star}\left(d\right),\zeta\left(d\right)\right)$.
Under Assumption \ref{assu:identification}(a,b,c,d), the standard
RD estimand still identifies the RD LATE. (a,b) imply that the true
covariates that may affect $Y\left(d\right)$ still satisfy the balance
condition $\mu_{Z^{\star},+}=\mu_{Z^{\star},-}$. (c) assumes that
the RD LATE on $\zeta$ is $\delta\cdot l_{n}$, which generates the
local imbalance in the observed covariate: $\mu_{Z,+}-\mu_{Z,-}\asymp\left(nh\right)^{-1/2}$.
By using local asymptotic analysis, we analyze the performance of
our EL confidence set under such a local imbalance condition, which
is similar to using locally misspecified moment conditions in the
sense of \citet{armstrong2021sensitivity}. Our result differs from
\citet{armstrong2021sensitivity} and focuses on the coverage performance
of the confidence set when $\delta$ is close to 0.\footnote{The approach of \citet{armstrong2021sensitivity} specifies a set
in which $\delta$ possibly lies and then adjusts the critical value
to take into account the maximal misspecification bias. We take a
very different approach in this paper.}

Let $N\coloneqq Z-\left(\delta\cdot l_{n}\right)D=D\cdot N\left(1\right)+\left(1-D\right)N\left(0\right)$,
where $N\left(1\right)\coloneqq Z^{\star}\left(1\right)+\zeta\left(1\right)-\delta\cdot l_{n}$
and $N\left(0\right)\coloneqq Z^{\star}\left(0\right)+\zeta\left(0\right)$.
It now follows that $g_{N\left(1\right)\mid10}\left(0\right)=g_{N\left(0\right)\mid10}\left(0\right)$
and $\mu_{N,+}=\mu_{N,-}$. Let $\gamma_{N}\coloneqq\left(\mathrm{Var}_{\mid0^{\pm}}\left[N\right]\right)^{-1}\mathrm{Cov}_{\mid0^{\pm}}\left[N,M\right]$
and $\mathscr{V}_{N}\coloneqq\left(\omega_{p+1}^{0,2}\mathrm{Var}_{\mid0^{\pm}}\left[M-N^{\top}\gamma_{N}\right]\right)/\left(\varphi\mu_{D,\dagger}^{2}\right)$.
For simplicity, we assume that the distribution of $N$ does not vary
with $n$.\footnote{E.g., this holds if the measurement errors are the following form:
$\zeta\left(1\right)=\delta\cdot l_{n}+\zeta_{1}$ and $\zeta\left(0\right)=\zeta_{0}$,
for some zero-mean $\left(\zeta_{0},\zeta_{1}\right)$ that are independent
of other variables in the model. Relaxation of this assumption requires
more complicated arguments and suitable modification of the assumptions.} CCFT shows that the covariate-adjusted estimator is inconsistent
and the confidence interval fails to have asymptotically correct coverage
probability under ``global imbalance'' $\mu_{Z,+}\neq\mu_{Z,-}$.
Under local imbalance in Assumption \ref{assu:imbalance}, CCFT's
estimator and the generalized EB estimators are still consistent.\footnote{We can show that $\widehat{\gamma}_{Y}^{\mathsf{CCFT}}$ (see Section
6 of the online supplement of CCFT for its expression) in the representation
(\ref{eq:CCFT estimator 2}) converges in probability to $\left(\mathrm{Var}_{\mid0^{\pm}}\left[N\right]\right)^{-1}\mathrm{Cov}_{\mid0^{\pm}}\left[N,Y\right]$.
Then, since $\mu_{N,+}=\mu_{N,-}$, we have
\[
\widehat{\vartheta}_{Y,p}^{\mathsf{CCFT}}=\frac{1}{nh}\sum_{i}\widehat{W}_{p,i}\left(Y_{i}-N_{i}^{\top}\widehat{\gamma}_{Y}^{\mathsf{CCFT}}\right)-\left(\delta\cdot l_{n}\right)\left(\frac{1}{nh}\sum_{i}\widehat{W}_{p,i}D_{i}\right)=\mu_{Y,\dagger}+o_{p}\left(1\right).
\]
Under global imbalance, the EB estimator has a probabilistic limit
different from that of CCFT's estimator (see Lemma 1 of CCFT). Neither
of them is equal to $\vartheta$.} Inference suffers from the undercoverage problem, since the coverage
probabilities of the confidence sets (CCFT's or the EL) converge to
a limit in $\left(0,1-\tau\right)$.

We now consider $\mathrm{Pr}\left[\vartheta\in\mathit{CS}_{p+1,\tau}^{\mathsf{bc}}\left(h\right)\right]$
as a function of $\delta$ under local imbalance. A measure of sensitivity
of the coverage accuracy to local imbalance (i.e., how the coverage
probability drops relative to that under $\delta=0$) is given by
the slope of $\mathrm{Pr}\left[\vartheta\in\mathit{CS}_{p+1,\tau}^{\mathsf{bc}}\left(h\right)\right]$
as a function of $\delta$ at $\delta=0$. We extend Theorem \ref{thm:coverage}
and derive a two-term asymptotic expansion $\mathrm{Pr}\left[\vartheta\in\mathit{CS}_{p+1,\tau}^{\mathsf{bc}}\left(h\right)\right]=R\left(\delta\right)+o\left(l_{n}\right)$,
where $R\left(\delta\right)$ is the sum of the leading terms as an
approximation to $\mathrm{Pr}\left[\vartheta\in\mathit{CS}_{p+1,\tau}^{\mathsf{bc}}\left(h\right)\right]$
in finite samples. We show that $R\left(0\right)=1-\tau$ and the
gradient $\nabla R\left(\delta\right)\coloneqq\left(\partial/\partial\delta\right)R\left(\delta\right)$
at $\delta=0$ is equal to 0, so that $R\left(\delta\right)$ is locally
constant around $\delta=0$.

Let $F\left(\cdot\mid\iota\right)$ denote the CDF of a $\chi_{1}^{2}\left(\iota\right)$
(non-central $\chi^{2}$ with one degree of freedom and non-centrality
parameter $\iota\geq0$) random variable. Let $F^{\left(k\right)}\left(x\mid\iota\right)\coloneqq\left(\partial/\partial\iota\right)^{k}F\left(x\mid\iota\right)$
be the $k$-times partial derivative of $F\left(x\mid\iota\right)$
with respect to $\iota$. 
\begin{thm}
\label{thm:thm local imbalance}Suppose that Assumptions \ref{assu:identification}(a,b,c,d),
\ref{assu:data generating process}, \ref{assu:kernel}, \ref{assu:kernel 2}
and \ref{assu:imbalance} hold. Suppose that $h$ satisfies $h=H\cdot n^{-1/\left(p+2\right)}$
for some constant $H>0$. Then, 
\begin{eqnarray*}
\mathrm{Pr}\left[\vartheta\in\mathit{CS}_{p+1,\tau}^{\mathsf{bc}}\left(h\right)\right] & = & F\left(c_{\tau}\mid H\cdot\frac{\left(\bar{\gamma}_{N}^{\top}\delta\right)^{2}}{\bar{\mathscr{V}}_{N}}\right)+\left\{ \mathscr{P}_{1}\left(\delta\right)F^{\left(1\right)}\left(c_{\tau}\mid H\cdot\frac{\left(\bar{\gamma}_{N}^{\top}\delta\right)^{2}}{\bar{\mathscr{V}}_{N}}\right)\right.\\
 &  & \left.+\mathscr{P}_{2}\left(\delta\right)F^{\left(2\right)}\left(c_{\tau}\mid H\cdot\frac{\left(\bar{\gamma}_{N}^{\top}\delta\right)^{2}}{\bar{\mathscr{V}}_{N}}\right)\right\} l_{n}+o\left(l_{n}\right),
\end{eqnarray*}
where $\bar{\gamma}_{N}=\gamma_{N}+o\left(1\right)$ and $\bar{\mathscr{V}}_{N}=\mathscr{V}_{N}+o\left(1\right)$
and $\left(\mathscr{P}_{1},\mathscr{P}_{2}\right)$ are homogeneous
cubic polynomials with constant coefficients. The expressions of $\left(\bar{\gamma}_{N},\bar{\mathscr{V}}_{N},\mathscr{P}_{1},\mathscr{P}_{2}\right)$
are in the supplement.
\end{thm}
\begin{rembold}\label{Rmk: sensitivity perturbation}The first-order
term $F\left(c_{\tau}\mid H\left(\bar{\gamma}_{N}^{\top}\delta\right)^{2}/\bar{\mathscr{V}}_{N}\right)$
is an even function of $\delta$, and the second-order term is an
odd function of $\delta$. Clearly, we have $\nabla R\left(0\right)=0$
and therefore $R\left(\cdot\right)$ is locally constant around the
origin.\footnote{Let $\mathit{LR}_{p+1}^{\varrho}\left(\theta\mid h\right)$ denote
the likelihood ratio with KL divergence replaced by the Cressie-Read
divergence (\ref{eq:Cressie-Read}). Under the same assumptions as
in Theorem \ref{thm:thm local imbalance}, we can show that $\mathrm{Pr}\left[\mathit{LR}_{p+1}^{\varrho}\left(\vartheta\mid h\right)\leq c_{\tau}\right]$
admits a similar two-term asymptotic expansion with the same first-order
term $F\left(c_{\tau}\mid H\left(\bar{\gamma}_{N}^{\top}\delta\right)^{2}/\mathscr{\bar{V}}_{N}\right)$
and a second-order term with a non-zero gradient at 0 if $\varrho\neq0$.} We expect that the coverage accuracy of the $\mathit{CS}_{p+1,\tau}^{\mathsf{bc}}\left(h\right)$
is highly insensitive to local imbalance in finite samples. If $\left\Vert \nabla R\left(0\right)\right\Vert $
is large in magnitude, a slight perturbation will incur severe undercoverage.
To see that the slope is a measure of sensitivity to local imbalance,
we consider the approximate minimal coverage $\mathrm{min}_{\delta\in\mathbb{S}_{\iota}}R\left(\delta\right)$
on $\mathbb{S}_{\iota}$, where $\iota$ is a positive constant and
$\mathbb{S}_{\iota}\coloneqq\left\{ \delta\in\mathbb{R}^{d_{z}}:\left\Vert \delta\right\Vert =\iota\right\} $
represents perturbations with equal magnitude $\iota$ in all directions.
$\delta_{R}^{*}\coloneqq\mathrm{argmin}_{\delta\in\mathbb{S}_{\iota}}R\left(\delta\right)$
corresponds to the direction in which the perturbation results in
the most severe undercoverage. Clearly, $R\left(\delta_{R}^{*}\right)<1-\tau$,
and we have the approximation $R\left(\delta_{R}^{*}\right)=\left(1-\tau\right)-\left\Vert \nabla R\left(0\right)\right\Vert \iota+o\left(\iota\right)$
when $\iota$ is small.\footnote{By using the Lagrange multiplier method to solve the constrained minimization
problem $\mathrm{min}_{\delta\in\mathbb{S}_{\iota}}R\left(\delta\right)$
and mean value expansion, $\delta_{R}^{*}=-\left(\nabla R\left(\delta_{R}^{*}\right)/\left\Vert \delta_{R}^{*}\right\Vert \right)\iota$
and therefore, $R\left(\delta_{R}^{*}\right)=\left(1-\tau\right)-\left(\nabla R\left(\dot{\delta}_{R}\right)^{\top}\nabla R\left(\delta_{R}^{*}\right)/\left\Vert \nabla R\left(\delta_{R}^{*}\right)\right\Vert \right)\iota$,
where $\dot{\delta}_{R}$ is the mean value that lies between $\delta_{R}^{*}$
and 0. Clearly, $\nabla R\left(\dot{\delta}_{R}\right)^{\top}\nabla R\left(\delta_{R}^{*}\right)/\left\Vert \nabla R\left(\delta_{R}^{*}\right)\right\Vert \rightarrow\left\Vert \nabla R\left(0\right)\right\Vert =0$,
as $\iota\downarrow0$.}\end{rembold}

\begin{rembold}In some real applications (see, e.g., \citealp{Cattaneo2019}
and \citealp[Section 4.1]{Cattaneo2022} for discussion and examples),
the researcher may have access to observations on outcomes $\tilde{Y}$
determined after treatment but considered unaffected by the treatment
and to have no effect on the outcome of interest $Y$. \citet{Cattaneo2022}
note that \textquotedblleft the principle of covariate balance can
be extended beyond pre-determined covariates to variables that are
determined after the treatment is assigned but are known to be unaffected
by the treatment...\textquotedblright . The balance condition $\mu_{\tilde{Y},+}=\mu_{\tilde{Y},-}$
should also hold for \textquotedblleft unaffected\textquotedblright{}
outcomes. We can also augment the list of covariates in (\ref{eq:entropy balancing})
to include unaffected outcomes. While expanding the set of covariates
may improve the efficiency, it bears the risk that the prior belief
$\mu_{\tilde{Y},+}=\mu_{\tilde{Y},-}$ is wrong. Imbalance for $\tilde{Y}$
does not falsify the RD design (the continuity assumption for $Y$),
since $\tilde{Y}$ does not affect $Y$ by assumption. Theorem \ref{thm:thm local imbalance}
with $Z$ replaced by $\tilde{Y}$ still holds, under the assumption
that the potential unaffected outcomes satisfy the continuity assumption
and our prior belief is imperfect so that the balance condition is
just slightly violated (RD LATE on $\tilde{Y}$ is $\delta\cdot l_{n}$).\end{rembold}

\section{Covariate-adjusted estimation of the treatment effect derivative\label{sec:TED}}

The EB approach for covariate adjustment applies to parameters of
interest other than the standard RD LATE parameter. This section applies
EB to covariate-adjusted estimation of the treatment effect derivative
(TED). To focus on the main ideas, we consider the sharp design first.
\citet{Dong2015} propose using the TED defined as $\left.\left(\mathrm{d}/\mathrm{d}x\right)\mathrm{E}\left[Y\left(1\right)-Y\left(0\right)\mid X=x\right]\right|_{x=0}$
for evaluating the external validity of RD. A large TED suggests that
the LATE would be quite different if the score changes slightly, raising
more concern about external validity. The researcher can check whether
the RD LATE is likely to have external validity by testing for zero
TED. Under the assumption that $g_{Y\left(d\right)}$ is continuously
differentiable on a neighborhood of 0 (\citealp[Assumption A2]{Dong2015}),
the TED is identified: $\left.\left(\mathrm{d}/\mathrm{d}x\right)\mathrm{E}\left[Y\left(1\right)-Y\left(0\right)\mid X=x\right]\right|_{x=0}=\pi_{\mathsf{srd}}\coloneqq\mu_{Y,+}^{\left(1\right)}-\mu_{Y,-}^{\left(1\right)}$.
This section proposes an EB estimator for the TED. An inferential
procedure and standard errors can be found in Section S10 of the online
supplement.

Let $\dot{W}_{p;-,i}$ be defined by the right-hand side of (\ref{eq:regression weight})
with $\mathrm{e}_{p+1,1}^{\top}$ replaced by $\mathrm{e}_{p+1,2}^{\top}$.
Similarly, we define $\dot{W}_{p;+,i}$ and in addition, let $\dot{W}_{p,i}=\dot{W}_{p;+,i}-\dot{W}_{p;-,i}$.
In our notation, the standard LP estimator proposed in \citet{Dong2015}
for the TED is given by $\widehat{\pi}_{p}^{\mathsf{lp}}\coloneqq\left(nh^{2}\right)^{-1}\sum_{i}\dot{W}_{p,i}Y_{i}$
($p\geq2$). As the EB estimator $\widehat{\vartheta}_{p}^{\mathsf{eb}}$
for the RD LATE proposed in Section \ref{sec:Empirical-likelihood-method},
the EB-based TED estimator $\widehat{\pi}_{p}^{\mathsf{eb}}$ with
covariate adjustment can also be obtained by replacing the uniform
weights with the EB weights $\left(w_{1}^{\mathsf{eb}},...,w_{n}^{\mathsf{eb}}\right)$
defined by (\ref{eq:entropy balancing weight definition}):
\begin{equation}
\widehat{\pi}_{p}^{\mathsf{eb}}\coloneqq\frac{1}{h^{2}}\sum_{i}w_{i}^{\mathsf{eb}}\dot{W}_{p,i}Y_{i}.\label{eq:entropy balancing estimator ted definition}
\end{equation}
The above construction illustrates the convenience of EB-based covariate
adjustment: One can start with the standard estimator (without covariates)
for a parameter of interest in an RD-related context and then replace
its standard uniform weights with the EB weights. The EB weights are
computed using the covariates only, and are independent of the standard
estimator. Such an adjustment strategy also works straightforwardly
in other RD-related settings, such as the (nonlinear) estimators of
\citet{xu2017regression,Xu2018} in the scenarios with limited outcome
variables.

Let $\gamma_{\mathsf{ted}}\coloneqq\left(\mathrm{Var}_{\mid0^{\pm}}\left[Z\right]\right)^{-1}\left(\mathrm{Cov}_{\mid0^{+}}\left[Z,Y\right]-\mathrm{Cov}_{\mid0^{-}}\left[Z,Y\right]\right)$
and $\mathcal{\dot{K}}_{p;\mathrm{s}}\left(t\right)\coloneqq\mathrm{e}_{p+1,2}^{\top}\mathrm{V}_{p;\mathrm{s}}^{-1}r_{p}\left(t\right)K\left(t\right)$,
for $\mathrm{s}\in\left\{ -,+\right\} $. One can easily verify that
$\mathcal{\dot{K}}_{p;+}\left(t\right)=-\mathcal{\dot{K}}_{p;-}\left(-t\right)$
and $\int_{-1}^{0}\mathcal{K}_{p;-}\left(t\right)\dot{\mathcal{K}}_{p;-}\left(t\right)\mathrm{d}t=-\varpi_{p}$,
where $\varpi_{p}\coloneqq\int_{0}^{1}\mathcal{K}_{p;+}\left(t\right)\dot{\mathcal{K}}_{p;+}\left(t\right)\mathrm{d}t$.
Also denote $\dot{\omega}_{p;+}^{j,k}\coloneqq\int_{0}^{1}t^{j}\dot{\mathcal{K}}_{p;+}^{k}\left(t\right)\mathrm{d}t$
and $\dot{\omega}_{p;-}^{j,k}\coloneqq\int_{-1}^{0}t^{j}\dot{\mathcal{K}}_{p;-}^{k}\left(t\right)\mathrm{d}t$.
It can be checked that $\dot{\omega}_{p;+}^{0,2}=\dot{\omega}_{p;-}^{0,2}\eqqcolon\dot{\omega}_{p}^{0,2}$.
The following theorem shows the asymptotic normality of the EB estimator.
\begin{thm}
\label{thm:normality ted}Suppose that Assumptions \ref{assu:identification},
\ref{assu:data generating process} and \ref{assu:kernel} hold. Assume
that $g_{\left\Vert B\right\Vert ^{4}}$ is bounded on $\mathbb{B}\setminus\left\{ 0\right\} $.
Assume that the bandwidth satisfies $nh^{2p+3}=O\left(1\right)$ and
$nh^{3}\rightarrow\infty$. Then,
\[
\sqrt{nh^{3}}\left(\widehat{\pi}_{p}^{\mathsf{eb}}-\pi_{\mathsf{srd}}-\mathscr{B}_{p}^{\mathsf{\mathbf{\mathsf{ted}}}}h^{p}\right)\rightarrow_{d}\mathrm{N}\left(0,\mathscr{V}_{p}^{\mathsf{ted}}\right),
\]
where
\begin{eqnarray*}
\mathscr{B}_{p}^{\mathsf{\mathbf{\mathsf{ted}}}} & \coloneqq & \left(\dot{\omega}_{p;+}^{p+1,1}\frac{\mu_{Y,+}^{\left(p+1\right)}}{\left(p+1\right)!}-\dot{\omega}_{p;-}^{p+1,1}\frac{\mu_{Y,-}^{\left(p+1\right)}}{\left(p+1\right)!}\right)-\left(\frac{\varpi_{p}}{\omega_{p}^{0,2}}\right)\gamma_{\mathsf{ted}}^{\top}\left(\omega_{p;+}^{p+1,1}\frac{\mu_{Z,+}^{\left(p+1\right)}}{\left(p+1\right)!}-\omega_{p;-}^{p+1,1}\frac{\mu_{Z,-}^{\left(p+1\right)}}{\left(p+1\right)!}\right),\textrm{ }\\
\mathscr{V}_{p}^{\mathsf{ted}} & \coloneqq & \frac{\dot{\omega}_{p}^{0,2}\mathrm{Var}_{\mid0^{\pm}}\left[Y\right]-\left(\frac{\varpi_{p}^{2}}{\omega_{p}^{0,2}}\right)\gamma_{\mathsf{ted}}^{\top}\left(\mathrm{Var}_{\mid0^{\pm}}\left[Z\right]\right)\gamma_{\mathsf{ted}}}{\varphi}.
\end{eqnarray*}
\end{thm}
\begin{rembold}\label{rem:TED efficiency}The standard LP regression
theory shows $\sqrt{nh^{3}}\left(\widehat{\pi}_{p}^{\mathsf{lp}}-\pi_{\mathsf{srd}}-\dot{\mathscr{B}}_{p}^{\mathsf{\mathbf{\mathsf{lp}}}}h^{p}\right)\rightarrow_{d}\mathrm{N}\left(0,\dot{\mathscr{V}}_{p}^{\mathsf{lp}}\right)$,
where $\dot{\mathscr{B}}_{p}^{\mathsf{\mathbf{\mathsf{lp}}}}\coloneqq\left(\dot{\omega}_{p;+}^{p+1,1}\mu_{Y,+}^{\left(p+1\right)}-\dot{\omega}_{p;-}^{p+1,1}\mu_{Y,-}^{\left(p+1\right)}\right)/\left(p+1\right)!$
and $\dot{\mathscr{V}}_{p}^{\mathsf{lp}}\coloneqq\dot{\omega}_{p}^{0,2}\mathrm{Var}_{\mid0^{\pm}}\left[Y\right]/\varphi$.
The asymptotic variance of $\widehat{\pi}_{p}^{\mathsf{lp}}$ is larger
than $\mathscr{V}_{p}^{\mathsf{ted}}$ provided that $\gamma_{\mathsf{ted}}\neq0$.
Therefore, the EB method leads to efficiency gain in the case of estimating
TED. Consider the simulation design (the case with one covariate)
in Section \ref{sec:Monte-Carlo-Simulations}. We get $\gamma_{\mathsf{ted}}=1.5$
and$\sqrt{\mathscr{V}_{2}^{\mathsf{ted}}}=63.5$ by straightforward
calculation, while the asymptotic standard deviation $\sqrt{\dot{\mathscr{V}}_{2}^{\mathsf{lp}}}$
without covariate adjustment is $74.4$.\end{rembold}

\begin{rembold}\label{rem:TED CCFT}CCFT's regression-based method
can also be applied to obtain a covariate-adjusted estimator of the
TED, i.e., the regression coefficient of $I_{i}\cdot X_{i}$ in (\ref{eq:CCFT estimator}).
Let $\widehat{\pi}_{p}^{\mathsf{CCFT}}$ be defined by the right-hand
side of (\ref{eq:CCFT estimator}) with $\mathrm{e}_{2\left(p+1\right)+d_{z},p+2}^{\top}$
replaced by $\mathrm{e}_{2\left(p+1\right)+d_{z},p+3}^{\top}$. Consistency
of $\widehat{\pi}_{p}^{\mathsf{CCFT}}$ requires covariate balance
in the first derivative $\mu_{Z,+}^{\left(1\right)}=\mu_{Z,-}^{\left(1\right)}$.\footnote{See the discussion on Page 6 in the supplemental appendix of CCFT.
They also commented, ``This requirement is not related to typical
falsification conducted in empirical work, that is, $\mu_{Z,+}=\mu_{Z,-}$,
but a different feature of the conditional distribution at the cutoff.''} Indeed, an extension of Theorem \ref{thm:normality} shows that $\widehat{\pi}_{p}^{\mathsf{CCFT}}$
is first-order equivalent to an EB estimator using weights defined
by the right-hand side of (\ref{eq:entropy balancing weight definition})
with $\widehat{W}_{p,i}$ replaced by $\dot{W}_{p,i}$. Under continuous
differentiability of $g_{Z\left(1\right)}$ and $g_{Z\left(0\right)}$,
$\mu_{Z,+}^{\left(1\right)}=\mu_{Z,-}^{\left(1\right)}$ is equivalent
to the predeterminedness-type assumption $\left.\left(\mathrm{d}/\mathrm{d}x\right)\mathrm{E}\left[Z\left(1\right)\mid X=x\right]\right|_{x=0}=\left.\left(\mathrm{d}/\mathrm{d}x\right)\mathrm{E}\left[Z\left(0\right)\mid X=x\right]\right|_{x=0}$
(i.e., zero TED on covariates). In comparison, consistency and efficiency
gain of $\widehat{\pi}_{p}^{\mathsf{eb}}$ require the same predeterminedness
condition $\mathrm{E}\left[Z\left(1\right)\mid X=0\right]=\mathrm{E}\left[Z\left(0\right)\mid X=0\right]$
as the covariate-adjusted estimators for the RD LATE do. As the TED
estimator is often used to evaluate the external validity of RD LATE,
it is more natural to impose the same assumptions as those underlying
estimation of the RD LATE in an RD design with covariates.\footnote{Note that in CCFT, $\widehat{\pi}_{p}^{\mathsf{CCFT}}$is proposed
as an estimator of $\mu_{Y,+}^{\left(1\right)}-\mu_{Y,-}^{\left(1\right)}$
in the regression kink design whose identification assumptions include
$\mu_{Z,+}^{\left(1\right)}=\mu_{Z,-}^{\left(1\right)}$ (\citealp{Card:2015gc}).
Here, we emphasize that the RD design with covariates considered in
CCFT, where the same object $\mu_{Y,+}^{\left(1\right)}-\mu_{Y,-}^{\left(1\right)}$
is interpreted as the TED, does not require $\mu_{Z,+}^{\left(1\right)}=\mu_{Z,-}^{\left(1\right)}$,
although such an assumption holds if $Z\left(1\right)-Z\left(0\right)=0$
(zero individual treatment effect on covariates). In this case, researchers
need to be careful when using the CCFT covariate-adjusted estimator
for the derivative difference.}\end{rembold}

\begin{rembold}Suppose that a researcher believes that both of the
predeterminedness assumption $\mathrm{E}\left[Z\left(1\right)\mid X=0\right]=\mathrm{E}\left[Z\left(0\right)\mid X=0\right]$
(the usual covariate balance condition $\mu_{Z,+}=\mu_{Z,-}$) and
zero TED $\left.\left(\mathrm{d}/\mathrm{d}x\right)\mathrm{E}\left[Z\left(1\right)-Z\left(0\right)\mid X=x\right]\right|_{x=0}=0$
(the derivative version $\mu_{Z,+}^{\left(1\right)}=\mu_{Z,-}^{\left(1\right)}$
of covariate balance) are likely to hold. In this case, CCFT's estimator
$\widehat{\pi}_{p}^{\mathsf{CCFT}}$ does not fully exploit the information
in the covariates. An estimator linearly combining $\widehat{\pi}_{p}^{\mathsf{eb}}$
and $\widehat{\pi}_{p}^{\mathsf{CCFT}}$ in the form of $\varsigma\cdot\widehat{\pi}_{p}^{\mathsf{eb}}+\left(1-\varsigma\right)\cdot\widehat{\pi}_{p}^{\mathsf{CCFT}}$
achieves further efficiency improvement. We can show the following
joint asymptotic normality result:
\[
\sqrt{nh^{3}}\left(\begin{array}{c}
\widehat{\pi}_{p}^{\mathsf{eb}}-\pi_{\mathsf{srd}}-h^{p}\mathscr{B}_{p}^{\mathbf{\mathsf{ted}}}\\
\widehat{\pi}_{p}^{\mathsf{CCFT}}-\pi_{\mathsf{srd}}-h^{p}\dot{\mathscr{B}}_{p}^{\mathsf{CCFT}}
\end{array}\right)\rightarrow_{d}\mathrm{N}\left(\begin{array}{cc}
\mathscr{V}_{p}^{\mathsf{ted}} & \mathscr{C}_{p}\\
\mathscr{C}_{p} & \mathscr{\dot{V}}_{p}^{\mathsf{CCFT}}
\end{array}\right),
\]
where $\mathscr{\dot{V}}_{p}^{\mathsf{CCFT}}=\dot{\omega}_{p}^{0,2}\sigma^{2}/\varphi$
and
\begin{eqnarray*}
\dot{\mathscr{B}}_{p}^{\mathsf{CCFT}} & \coloneqq & \frac{\mu_{Y-Z^{\top}\gamma_{Y},+}^{\left(p+1\right)}\dot{\omega}_{p;+}^{p+1,1}-\mu_{Y-Z^{\top}\gamma_{Y},-}^{\left(p+1\right)}\dot{\omega}_{p;-}^{p+1,1}}{\left(p+1\right)!}\\
\mathscr{C}_{p} & \coloneqq & \mathscr{\dot{V}}_{p}^{\mathsf{CCFT}}-\frac{\varpi_{p}^{2}}{\omega_{p}^{0,2}}\left(\frac{\mathrm{Cov}_{\mid0^{+}}\left[Z,Y\right]-\mathrm{Cov}_{\mid0^{-}}\left[Z,Y\right]}{\varphi}\right)^{\top}\gamma_{\mathsf{ted}}+\frac{\varpi_{p}^{2}}{\omega_{p}^{0,2}}\cdot\frac{\gamma_{\mathsf{ted}}^{\top}\left(\mu_{ZZ^{\top},\dagger}\right)\gamma_{Y}}{\varphi}.
\end{eqnarray*}
Therefore, the optimal linear combination that has the smallest asymptotic
variance will assign to $\widehat{\pi}_{p}^{\mathsf{eb}}$ the following
optimal weight $\varsigma^{*}\coloneqq\left(\mathscr{\dot{V}}_{p}^{\mathsf{CCFT}}-\mathscr{C}_{p}\right)/\left(\mathscr{V}_{p}^{\mathsf{ted}}+\mathscr{\dot{V}}_{p}^{\mathsf{CCFT}}-2\mathscr{C}_{p}\right)$.
Once again, consider the simulation design in Section \ref{sec:Monte-Carlo-Simulations}
with slight modification to ensure $\mu_{Z,+}^{\left(1\right)}=\mu_{Z,-}^{\left(1\right)}=1.06$.\footnote{In this numerical example, letting $\mu_{Z,+}^{\left(1\right)}=\mu_{Z,-}^{\left(1\right)}=1.06$
does not change the asymptotic variance and covariance $\mathscr{V}_{2}^{\mathsf{ted}}$,
$\mathscr{C}_{2}$ and $\mathscr{\dot{V}}_{2}^{\mathsf{CCFT}}$.} The optimal weight $\varsigma^{*}=0.35$ and the resulting asymptotic
standard deviation is $47.5$, which is smaller than $\sqrt{\mathscr{V}_{2}^{\mathsf{ted}}}=63.5$
calculated in Remark \ref{rem:TED efficiency}. Another approach to
exploiting the information in both balance conditions is based on
the EB weights with a new set of constraints $\sum_{i}w_{i}\dot{W}_{p,i}\bar{Z}_{i}=0$
being added to (\ref{eq:entropy balancing}). We can show that this
estimator is first-order equivalent to the optimal combination. However,
such a method is more computationally costly.\end{rembold}

\begin{rembold}In the fuzzy RD, \citet{Dong2015} show that the
TED is identified:
\begin{equation}
\left.\frac{\mathrm{d}}{\mathrm{d}x}\mathrm{E}\left[Y\left(1\right)-Y\left(0\right)\mid X=x,\mathsf{co}\right]\right|_{x=0}=\frac{\mu_{Y,+}^{\left(1\right)}-\mu_{Y,-}^{\left(1\right)}}{\mu_{D,\dagger}}-\left(\mu_{D,+}^{\left(1\right)}-\mu_{D,-}^{\left(1\right)}\right)\cdot\frac{\mu_{Y,\dagger}}{\mu_{D,\dagger}^{2}}.\label{eq:TED fuzzy}
\end{equation}
The same equality with $Y$ replaced by $Z$ also holds. Covariate-adjusted
estimation of TED based on (\ref{eq:TED fuzzy}) and our EB approach
is straightforward. Under covariate balance, $\mu_{Z,+}^{\left(1\right)}-\mu_{Z,-}^{\left(1\right)}=0$
is implied by $\left.\left(\mathrm{d}/\mathrm{d}x\right)\mathrm{E}\left[Z\left(1\right)-Z\left(0\right)\mid X=x,\mathsf{co}\right]\right|_{x=0}=0$.
EB-based estimation exploiting both predeterminedness and zero TED
conditions is also straightforward.\end{rembold}

\section{Monte Carlo simulations\label{sec:Monte-Carlo-Simulations}}

We conduct simulations to evaluate the finite sample performance of
the proposed EL-based inference methods for sharp RD designs with
covariates. The data-generating process (DGP) of the outcome variable
$Y_{i}$, the score $X_{i}$ and the first covariate $Z_{i}^{\left(1\right)}$
is based on the simulation design of CCFT. The incorporation of additional
covariates $Z_{i}^{\left(2\right)},...,Z_{i}^{\left(l\right)}$ follows
that of \citet{arai2021regression}. Let 
\begin{eqnarray*}
\mu_{y0}\left(x\right) & \coloneqq & 0.36+0.96x+5.47x^{2}+15.28x^{3}+15.87x^{4}+5.14x^{5}\\
\mu_{y1}\left(x\right) & \coloneqq & 0.38+0.62x-2.84x^{2}+8.42x^{3}-10.24x^{4}+4.31x^{5}\\
\mu_{z0}\left(x\right) & \coloneqq & 0.49+\theta_{l}x+5.74x^{2}+17.14x^{3}+19.75x^{4}+7.47x^{5}\\
\mu_{z1}\left(x\right) & \coloneqq & 0.49+\theta_{r}x-0.23x^{2}-3.46x^{3}+6.43x^{4}-3.48x^{5}
\end{eqnarray*}
and
\begin{eqnarray}
\mu_{y}\left(x,z_{1}\right) & \coloneqq & \begin{cases}
\mu_{y0}\left(x\right)+\gamma_{l}z_{1} & \text{if }x<0\\
\mu_{y1}\left(x\right)+\gamma_{r}z_{1} & \text{if }x\geq0
\end{cases}\label{simu:DGP_y}\\
\mu_{z}\left(x\right) & \coloneqq & \begin{cases}
\mu_{z0}\left(x\right) & \text{if }x<0\\
\mu_{z1}\left(x\right) & \text{if }x\geq0,
\end{cases}\label{simu:DGP_x}
\end{eqnarray}
where with the coefficients $\gamma_{l}=0.22$, $\gamma_{r}=0.28$,
$\theta_{l}=1.06$ and $\theta_{r}=0.61$, all following CCFT. Then,
$Y_{i}=\mu_{y}\left(X_{i},Z_{i}^{\left(1\right)}\right)+\sum_{j=2}^{l}\pi^{j-1}Z_{i}^{\left(j\right)}+\varepsilon_{y,i}$
and $Z_{i}^{\left(1\right)}=\mu_{z}\left(X_{i}\right)+\varepsilon_{z,i}$.
Error terms $\left(\varepsilon_{y,i},\varepsilon_{z,i}\right)$ are
bivariate normal with mean $0$, standard deviation $1$ and correlation
coefficient $\rho=0.269.$ Additional covariates $\left(Z_{i}^{\left(2\right)},...,Z_{i}^{\left(l\right)}\right)$
have a multivariate normal distribution with mean zero and covariance
matrix given by $\mathrm{Cov}\left[Z_{i}^{\left(j\right)},Z_{i}^{\left(k\right)}\right]=0.5^{\left|j-k\right|},$
for all $j,k\geq2$. We take $\pi=0.2$. We consider three scenarios
with $l=0,2,4$, corresponding to the total number of covariates $d_{z}=l+1$
being $1,3,5$. CCFT uses local linear regression with bias correction,
equivalent to local quadratic regression. Our EL approach parallels
CCFT in that the degree of the LP is set to be $p=2$. The sample
sizes are $n=1000,2000$. The number of Monte Carlo replications is
$5000$.

Table \ref{tab: simu_size} presents the bias, root mean square error
(RMSE) of the MC-EL estimator $\widehat{\vartheta}_{p}^{\mathsf{mc}}$
defined in Section \ref{subsec:Connection-to-empirical}, as well
as the empirical coverage probability and the average length of the
EL confidence sets $\mathit{CS}_{p,\tau}^{\mathsf{bc}}\left(h\right)$
and $\mathit{CS}_{p+1,\tau}^{\mathsf{bc}}\left(h\right)$ defined
in Remark \ref{Rmk: Bartlett correction}. Following Remark \ref{Rmk: coverage},
we select a bandwidth of the form $h=H\cdot n^{-1/\left(p+2\right)}$,
replace $H$ with a consistent estimator $\widehat{H}$, and use the
bandwidth $\widehat{h}\coloneqq\widehat{H}\cdot n^{-1/\left(p+2\right)}$.
\citet[Section 5.3]{calonico2018optimal} propose an approach that
takes the estimated AMSE optimal bandwidth and rescales it to make
it obey the coverage optimal rate (see Section IV(C) of CCFT). One
choice of bandwidth $\widehat{h}$ is to follow this approach and
use CCFT's bandwidth, denoted as CCFT in Table \ref{tab: simu_size}.
CCFT's bandwidth is computed from $\mathtt{R}$ function \textit{$\mathtt{rdrobust}$
}with the options $\mathtt{p=1}$, $\mathtt{rho=1}$, and $\mathtt{bwselect}=$``$\mathtt{cerrd}$''.\footnote{The rate of CCFT's bandwidth is $n^{-1/4}$, which matches the rate
of $\widehat{h}=\widehat{H}\cdot n^{-1/\left(p+2\right)}$ with $p=2$.} Another simpler choice is a rescaled rule of thumb (ROT) bandwidth
that uses the constant part $\widehat{H}$ computed according to \citet[Chapter 21.6]{Hansen2021}'s
ROT bandwidth. For comparison, Table \ref{tab: simu_size} also includes
results from CCFT's method that uses the CCFT bandwidth and restricts
$\rho=h/b=1$, where $b$ stands for the pilot bandwidth used for
bias estimation. Table \ref{tab: simu_size} shows that both EL and
CCFT approaches perform well for estimation and inference. A closer
look reveals that EL with $p=2$ and using the rescaled ROT bandwidth
yields similar bias and RMSE compared with CCFT, but slightly better
coverage (especially for $d_{z}=3$ and $5$), and shorter confidence
intervals. On the other hand, EL that uses CCFT's bandwidth, which
amounts to half of the ROT bandwidth, yields smaller bias but larger
RMSE and longer confidence intervals. In particular, the length of
$\mathit{CS}_{p+1,\tau}^{\mathsf{bc}}\left(\widehat{h}\right)$ is
longer than those of other confidence sets. In sum, all the methods
we consider deliver satisfactory finite-sample performances.\footnote{The EL confidence intervals are also well-centered. E.g., the average
center (across all $5,000$ simulation replications) of the EL intervals
with ROT bandwidth ($\mathrm{EL}_{p}$ in Table 1), $d_{z}=5$ and
$n=1,000$ is only $0.0015$ away from the true treatment parameter
($0.0494$).} Computing the EB weights (for the point estimator) and the EL likelihood
ratio statistic (for the confidence set) only requires solving convex
optimization problems (corresponding to the ``inner loop'' in the
standard EL computation) and thus is very fast. E.g., computing the
row $\mathrm{EL}_{p}$ in Table 1 with $d_{z}=5$, $n=1,000$ and
CCFT bandwidth costs $0.06$ to $1.08$ seconds for one replication,
with the average computation time per replication about $0.23$ second
on an Intel Core i7 processor with 32 GB of RAM.

\begin{table}[H]
\centering \caption{Performance of EL and Wald-type confidence sets in sharp RD with covariates:
$\mathrm{EL}_{p}$ corresponding to the point estimator $\widehat{\vartheta}_{p}^{\mathsf{mc}}$
and the confidence set $\mathit{CS}_{p,\tau}^{\mathsf{bc}}\left(\widehat{h}\right)$,
$\mathrm{EL}_{p+1}$ corresponding to the point estimator $\widehat{\vartheta}_{p+1}^{\mathsf{mc}}$
and the confidence set $\mathit{CS}_{p+1,\tau}^{\mathsf{bc}}\left(\widehat{h}\right)$;
$\tau=0.05$ $p=2$, the bandwidth $\widehat{h}=$ rescaled rule of
thumb (ROT) or CCFT's (CCFT) bandwidth with the average bandwidth
length for $n=1,000$ reported in the parenthesis. CCFT's Wald-type
inference uses the CCFT bandwidth, CP $=$ the coverage probability,
CIL $=$ the average length of the confidence intervals, $n=$ sample
size, $d_{z}=$ the number of covariates.}
\label{tab: simu_size} {\small{}}%
\begin{tabular}{cllcccccccc}
\toprule 
\multicolumn{1}{c}{} & \multicolumn{1}{c}{} & \multicolumn{1}{c}{} & \multicolumn{2}{c}{{\small{}Bias}} & \multicolumn{2}{c}{{\small{}RMSE}} & \multicolumn{2}{c}{{\small{}$0.95$ CP}} & \multicolumn{2}{c}{{\small{}$0.95$ CIL}}\tabularnewline
\cmidrule{4-11} \cmidrule{5-11} \cmidrule{6-11} \cmidrule{7-11} \cmidrule{8-11} \cmidrule{9-11} \cmidrule{10-11} \cmidrule{11-11} 
\multicolumn{1}{c}{{\small{}$d_{z}$}} & \multicolumn{1}{c}{{\small{}Methods}} & \multicolumn{1}{c}{{\small{}$\widehat{h}$}} & \multicolumn{1}{c}{{\small{}$n=1,000$}} & \multicolumn{1}{c}{{\small{}$2,000$}} & \multicolumn{1}{c}{{\small{}$n=1,000$}} & \multicolumn{1}{c}{{\small{}$2,000$}} & \multicolumn{1}{c}{{\small{}$n=1,000$}} & \multicolumn{1}{c}{{\small{}$2,000$}} & \multicolumn{1}{c}{{\small{}$n=1,000$}} & \multicolumn{1}{c}{{\small{}$2,000$}}\tabularnewline
\midrule 
{\small{}1} & {\small{}$\mathrm{EL}_{p}$} & {\small{}ROT (0.301)} & {\small{}0.011} & {\small{}0.014} & {\small{}0.336} & {\small{}0.246} & {\small{}0.960} & {\small{}0.964} & {\small{}1.472} & {\small{}1.044}\tabularnewline
 &  & {\small{}CCFT (0.147)} & {\small{}0.007} & {\small{}0.008} & {\small{}0.420} & {\small{}0.331} & {\small{}0.946} & {\small{}0.949} & {\small{}1.931} & {\small{}1.326}\tabularnewline
 & {\small{}$\mathrm{EL}_{p+1}$} & {\small{}ROT} & {\small{}0.011} & {\small{}0.012} & {\small{}0.461} & {\small{}0.326} & {\small{}0.948} & {\small{}0.949} & {\small{}1.790} & {\small{}1.282}\tabularnewline
 &  & {\small{}CCFT} & {\small{}0.003} & {\small{}0.010} & {\small{}0.481} & {\small{}0.400} & {\small{}0.940} & {\small{}0.938} & {\small{}2.503} & {\small{}1.809}\tabularnewline
\midrule 
 & {\small{}CCFT} & {\small{}CCFT} & {\small{}0.013} & {\small{}0.014} & {\small{}0.334} & {\small{}0.238} & {\small{}0.945} & {\small{}0.951} & {\small{}1.822} & {\small{}1.285}\tabularnewline
\midrule 
{\small{}3} & {\small{}$\mathrm{EL}_{p}$} & {\small{}ROT (0.303)} & {\small{}0.014} & {\small{}0.011} & {\small{}0.349} & {\small{}0.248} & {\small{}0.957} & {\small{}0.954} & {\small{}1.460} & {\small{}1.040}\tabularnewline
 &  & {\small{}CCFT (0.145)} & {\small{}0.007} & {\small{}0.004} & {\small{}0.421} & {\small{}0.334} & {\small{}0.933} & {\small{}0.946} & {\small{}1.917} & {\small{}1.345}\tabularnewline
 & {\small{}$\mathrm{EL}_{p+1}$} & {\small{}ROT} & {\small{}0.007} & {\small{}0.002} & {\small{}0.499} & {\small{}0.335} & {\small{}0.934} & {\small{}0.940} & {\small{}1.784} & {\small{}1.279}\tabularnewline
 &  & {\small{}CCFT} & {\small{}0.001} & {\small{}-0.000} & {\small{}0.471} & {\small{}0.414} & {\small{}0.932} & {\small{}0.937} & {\small{}2.484} & {\small{}1.822}\tabularnewline
\midrule 
 & {\small{}CCFT} & {\small{}CCFT} & {\small{}0.012} & {\small{}0.010} & {\small{}0.347} & {\small{}0.241} & {\small{}0.936} & {\small{}0.949} & {\small{}1.806} & {\small{}1.280}\tabularnewline
\midrule 
{\small{}5} & {\small{}$\mathrm{EL}_{p}$} & {\small{}ROT (0.302)} & {\small{}-0.001} & {\small{}0.001} & {\small{}0.358} & {\small{}0.253} & {\small{}0.946} & {\small{}0.957} & {\small{}1.452} & {\small{}1.035}\tabularnewline
 &  & {\small{}CCFT (0.143)} & {\small{}0.001} & {\small{}-0.006} & {\small{}0.438} & {\small{}0.346} & {\small{}0.917} & {\small{}0.929} & {\small{}1.956} & {\small{}1.336}\tabularnewline
 & {\small{}$\mathrm{EL}_{p+1}$} & {\small{}ROT} & {\small{}-0.004} & {\small{}-0.007} & {\small{}0.507} & {\small{}0.346} & {\small{}0.916} & {\small{}0.933} & {\small{}1.793} & {\small{}1.274}\tabularnewline
 &  & {\small{}CCFT} & {\small{}0.000} & {\small{}-0.004} & {\small{}0.486} & {\small{}0.426} & {\small{}0.915} & {\small{}0.915} & {\small{}2.509} & {\small{}1.826}\tabularnewline
\midrule 
 & {\small{}CCFT} & {\small{}CCFT} & {\small{}0.000} & {\small{}0.004} & {\small{}0.354} & {\small{}0.242} & {\small{}0.924} & {\small{}0.938} & {\small{}1.787} & {\small{}1.270}\tabularnewline
\bottomrule
\end{tabular}
\end{table}

\begin{figure}[H]
\caption{Sensitivity of coverage probabilities of EL-based confidence sets
$\mathit{CS}_{p,\tau}^{\mathsf{bc}}\left(\widehat{h}\right)$, $\mathit{CS}_{p+1,\tau}^{\mathsf{bc}}\left(\widehat{h}\right)$
and the CCFT confidence set with respect to a local imbalance of magnitude
$\delta$, $n=2000$, $p=2$, bandwidth $\widehat{h}=$ CCFT's bandwidth.}
\label{fig:misspe}%
\begin{tabular}{cc}
\includegraphics[scale=0.45]{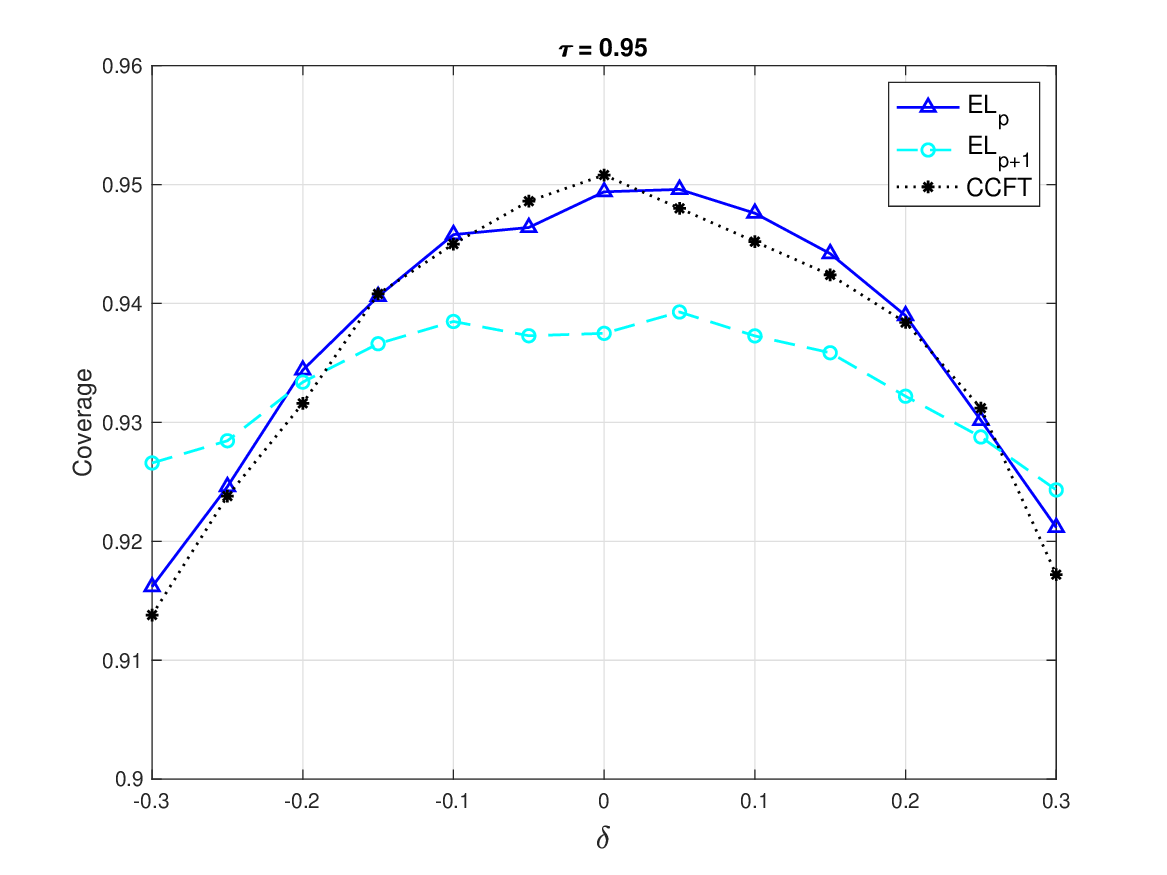} & \includegraphics[scale=0.45]{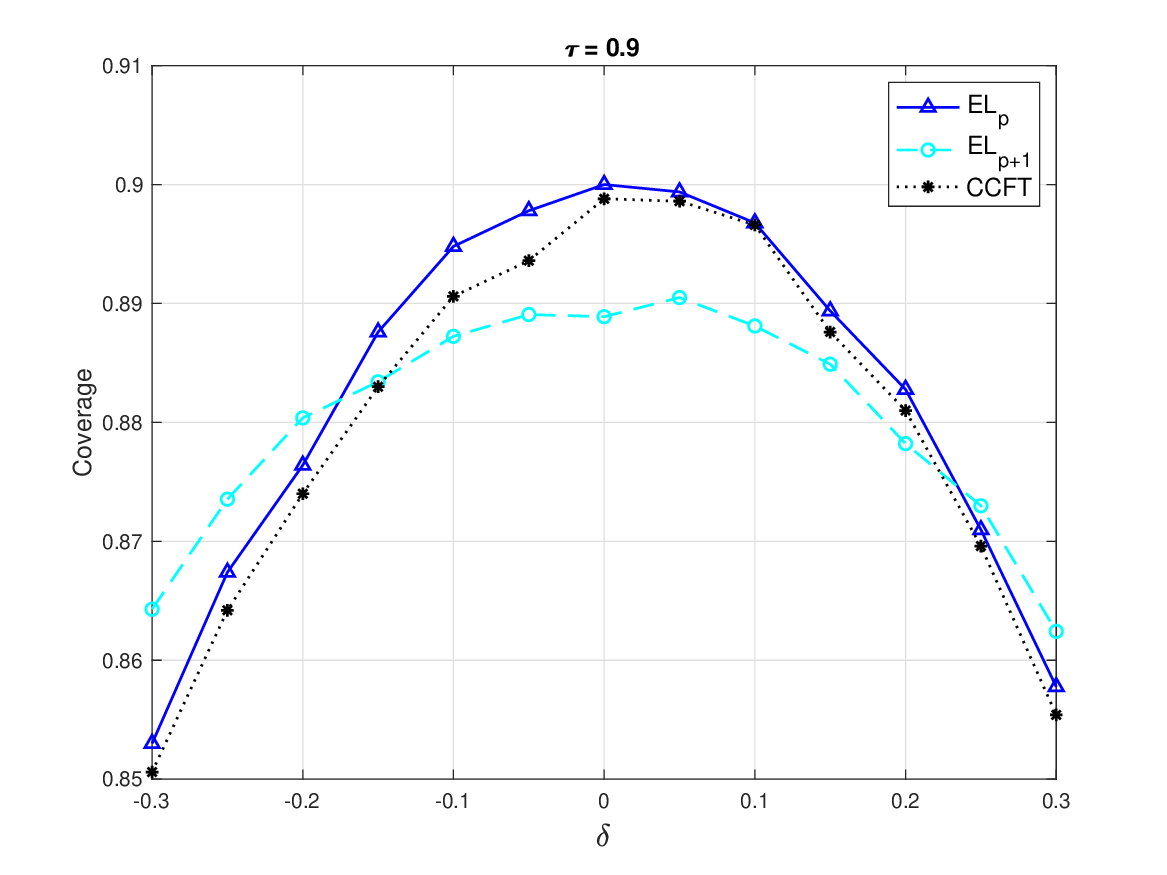}\tabularnewline
\end{tabular}
\end{figure}

We also examine how the coverage performance of EL and CCFT confidence
sets changes when the covariate balance condition is slightly violated.
We consider the case with one covariate ($d_{z}=1$). The data-generating
process for $\left(Y_{i},X_{i},Z_{i}^{\left(1\right)}\right)$ remains
the same but the incorporated covariate is given by $\tilde{Z}_{i}^{\left(1\right)}\coloneqq Z_{i}^{\left(1\right)}+\mathbbm{1}\left(X_{i}<0\right)\delta$,
so that the local covariate imbalance is measured by the perturbation
$\delta$. Figure \ref{fig:misspe} plots the simulated coverage probabilities
of the EL and CCFT confidence sets as a function of $\delta\in\left[-0.3,0.3\right]$.
We observe that the coverage probability of $\mathit{CS}_{p+1,\tau}^{\mathsf{bc}}\left(\widehat{h}\right)$
is less sensitive to the change of $\delta$, which parallels the
discussion in Remark \ref{Rmk: sensitivity perturbation}.

We then investigate the performance of the EB approach in the covariate-adjusted
estimation of TED. We consider the case with one covariate ($d_{z}=1$)
and modify the coefficients $\gamma_{l}$, $\gamma_{r}$, $\theta_{l}$
and $\theta_{r}$ in the design in order to highlight two features
of covariate-adjusted estimation of TED. First, the magnitude of efficiency
gain from incorporating the single covariate is determined by $|\gamma_{l}-\gamma_{r}|$.
We choose $\gamma_{l}=3$ and $\gamma_{r}=0$ to highlight the efficiency
contribution of the covariate adjustment. Second, note that $\theta_{r}-\theta_{l}$
corresponds to $\mu_{Z,+}^{(1)}-\mu_{Z,-}^{(1)}$, which is required
to be zero for CCFT's augmented regression estimator of TED (Remark
\ref{rem:TED CCFT}). More specifically, the asymptotic bias of CCFT's
estimator is proportional to $|\theta_{l}-\theta_{r}|$. We set $\theta_{l}=3$
and $\theta_{r}=0$ to highlight such a bias. Table \ref{tab: simu_ted}
reports the finite-sample performances of three TED estimators: the
EB estimator with $p=2$ given by (\ref{eq:entropy balancing estimator ted definition})
and its confidence interval constructed following the procedure in
Section S10 in our online supplement, CCFT's TED estimator with $p=2$
in Remark \ref{rem:TED CCFT}, and the standard local quadratic (LQ)
TED estimator (\citealp{Dong2015}) without using covariate information,
all three methods using \citet[CCT, hereafter]{calonico2014robust}'s
bandwidth for the first derivative computed from the $\mathtt{R}$
function \textit{$\mathtt{rdrobust}$}.\footnote{The rate of CCT's bandwidth here is still $n^{-1/4}$, which is also
the CO rate for Wald-type inference on the TED. See \citet[Theorem 3.1]{calonico2018optimal}.} As expected, Table \ref{tab: simu_ted} shows that the CCFT's TED
estimator leads to a substantial bias and undercoverage for TED inference,
given that $\mu_{Z,+}^{(1)}-\mu_{Z,-}^{(1)}$ is away from zero. This
problem can be solved by the EB estimator, which incorporates the
correctly specified covariate balance condition $\mu_{Z,+}=\mu_{Z,-}$
rather than the misspecified condition $\mu_{Z,+}^{(1)}=\mu_{Z,-}^{(1)}$
. The standard LQ TED estimator without covariates remains valid but
has a larger RMSE and yields a longer confidence interval than those
from the EB estimator, which reflects the efficiency gain of EB from
covariate adjustment.

\begin{table}[H]
\centering \caption{Treatment effect derivative (TED) inference with covariates: the EB
method compared with the standard LQ or CCFT, COV indicates whether
a method adjusts for covariates; CP $=$ the coverage probability,
CIL $=$ the average length of the confidence intervals, $n=$ sample
size, $d_{z}=$ the number of covariates. The true TED $=-9.34$.
All rows use CCT's bandwidth. The average bandwidth length equals
$0.151$ for $n=1,000$ and $0.123$ for $n=2,000$.}
\label{tab: simu_ted} %
\begin{tabular}{lcccccccccccc}
\toprule 
\multicolumn{1}{c}{} & \multicolumn{1}{c}{} & \multicolumn{2}{c}{Bias} & \multicolumn{1}{c}{} & \multicolumn{2}{c}{RMSE} & \multicolumn{1}{c}{} & \multicolumn{2}{c}{$0.95$ CP} & \multicolumn{1}{c}{} & \multicolumn{2}{c}{$0.95$ CIL}\tabularnewline
\cmidrule{3-4} \cmidrule{4-4} \cmidrule{6-7} \cmidrule{7-7} \cmidrule{9-10} \cmidrule{10-10} \cmidrule{12-13} \cmidrule{13-13} 
\multicolumn{1}{c}{Methods} & \multicolumn{1}{c}{COV} & \multicolumn{1}{c}{$n=1,000$} & \multicolumn{1}{c}{$2,000$} & \multicolumn{1}{c}{} & \multicolumn{1}{c}{$n=1,000$} & \multicolumn{1}{c}{$2,000$} & \multicolumn{1}{c}{} & \multicolumn{1}{c}{$n=1,000$} & \multicolumn{1}{c}{$2,000$} & \multicolumn{1}{c}{} & \multicolumn{1}{c}{$n=1,000$} & \multicolumn{1}{c}{$2,000$}\tabularnewline
\midrule 
EB & YES & 0.961 & 0.710 &  & 11.542 & 8.266 &  & 0.934 & 0.942 &  & 40.034 & 30.357\tabularnewline
CCFT & YES & 5.894 & 5.830 &  & 11.046 & 9.094 &  & 0.839 & 0.791 &  & 32.444 & 25.042\tabularnewline
Standard & NO & 0.221 & 0.362 &  & 12.359 & 9.254 &  & 0.936 & 0.946 &  & 43.325 & 33.762\tabularnewline
\bottomrule
\end{tabular}
\end{table}

\section{Empirical illustration: Finnish municipal election data\label{sec:Empirical}}

We apply our estimation/inference method to analyze the individual
incumbent advantage in Finnish municipal elections, which was first
studied by \citet{hyytinen2018does}. The outcome variable $Y$ indicates
whether the candidate is elected in an election, and the score $X$
is the vote share margin in the previous election. Table \ref{empi: RDD_Finnish}
presents the RD LATE point estimate $\widehat{\vartheta}$, the $p$-value
for testing the null hypothesis $\vartheta=0$, the $95\%$ confidence
intervals (CI), and the CI length. The first row of Table \ref{empi: RDD_Finnish}
presents the standard LQ regression estimator that ignores the covariates.
Then, we incorporate four covariates $Z$: candidates' age, gender,
age squared, and $\mathrm{age}\times\mathrm{gender}$. EL estimation
and inference use CCT's bandwidth ($h_{\mathsf{CCT}}=0.396$) and
the rescaled ROT bandwidth (equal to $2.917$). The last row of Table
\ref{empi: RDD_Finnish} reproduces the ``experiment benchmark''
reported originally by \citet{hyytinen2018does}(see their Table 2,
Column 4, the $p$-value is imputed by us).\footnote{The dataset includes 1351 candidates ``for whom the (previous) electoral
outcome was determined via random seat assignment due to ties in vote
counts'' (\citealp[Page 1020]{hyytinen2018does}), which constitutes
an experiment benchmark to evaluate the credibility of the RD treatment
effect estimated from the non-experimental data (candidates with previous
electoral ties are excluded from the RD sample).} Apparently, all RD estimates, with or without covariates, are small
in magnitude and statistically insignificant, which agrees with the
finding in the experiment benchmark. By comparing the covariate-adjusted
estimates (EL and CCFT) with the standard LQ regression without covariates,
we see that incorporating covariates helps to reduce the CI length
for four out of five confidence intervals, except for $\mathit{CS}_{p+1,\tau}^{\mathsf{bc}}\left(h_{\mathsf{CCT}}\right)$.
Among them, the EL confidence set $\mathit{CS}_{p,\tau}^{\mathsf{bc}}\left(h_{\mathsf{CCT}}\right)$
that uses the same bandwidth as the standard LQ regression is $7.2\%$
shorter than the standard method and is $5.5\%$ shorter than CCFT.
Here, the efficiency improvement is moderate, probably because the
election outcome is only weakly correlated with age and gender.

\begin{figure}[H]
\caption{A sensitivity analysis of the EL-based covariate-adjusted inference
using the Finnish municipal election data: uniform (solid) and pointwise
(dotted) confidence bands as functions of the bandwidth $h$ with
the $\underline{h}=h_{\mathsf{CCT}}/3\approx0.13$ and $\overline{h}=h_{\mathsf{CCT}}\times2\approx0.78$,
$\tau=0.05$. Bandwidth snooping corrected critical value = $\mathrm{2.413}^{2}$.
Vertical line indicates the CCT's bandwidth $h_{\mathsf{CCT}}=0.396$.}
\center \label{fig:Finnish_CB} \includegraphics[scale=0.8]{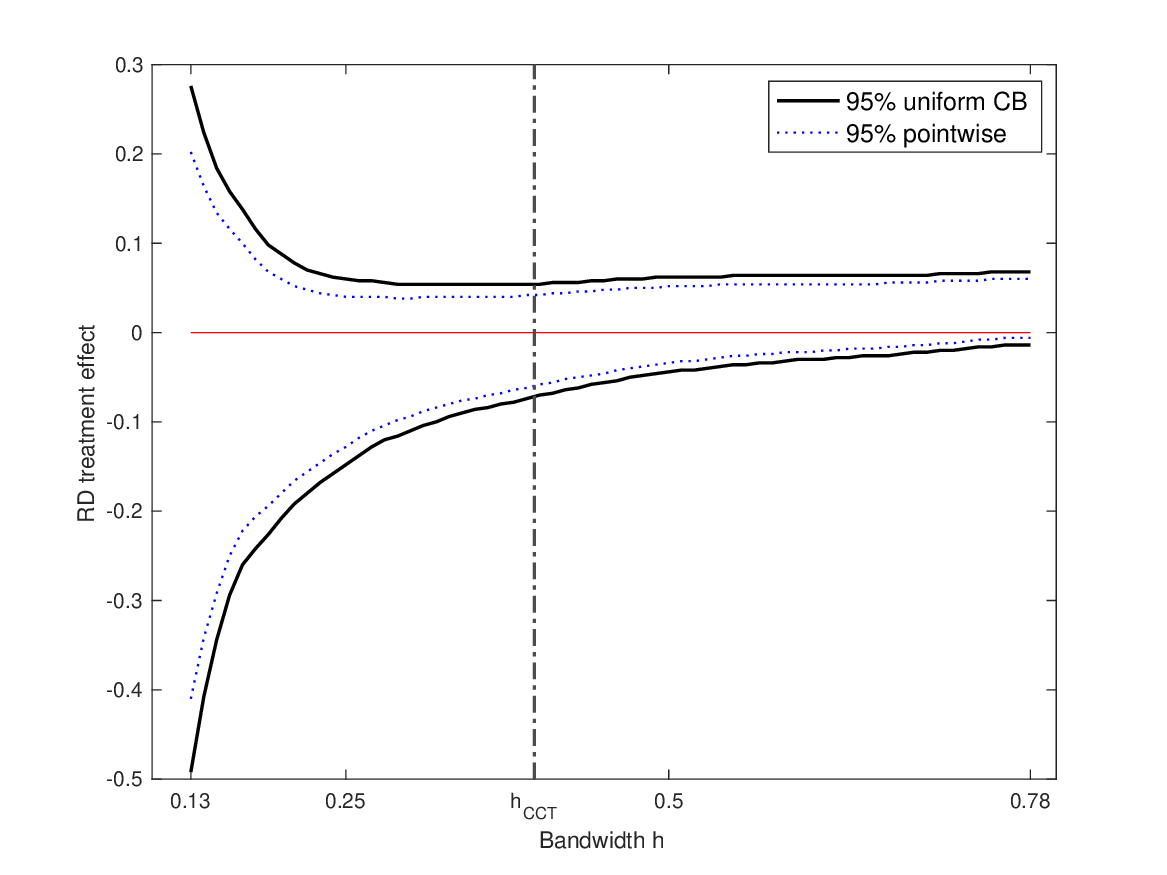}
\end{figure}

We then conduct a sensitivity analysis of the EL-based covariate-adjusted
inference with respect to the bandwidth choice by plotting the confidence
band (Remarks \ref{Rmk: sensitivity} and \ref{Rmk: doubly corrected}).
We consider the continuous range of bandwidths $h\in\left[\underline{h},\overline{h}\right]$
with the lower bound $\underline{h}=h_{\mathsf{CCT}}/3\approx0.13$
and the upper bound $\overline{h}=h_{\mathsf{CCT}}\times2\approx0.78$.
The rate of $h_{\mathsf{CCT}}$ is $n^{-1/4}$, which satisfies the
conditions for $\underline{h}$ and $\overline{h}$ in Theorem \ref{thm:Wilks}.
Using the $\mathtt{R}$ package $\mathtt{BWSnooping}$, we calculate
the snooping corrected critical value $\mathrm{2.413}^{2}$ for the
triangular kernel and bandwidth ratio $\overline{h}/\underline{h}=6$.
In Figure \ref{fig:Finnish_CB}, the solid (or dotted) lines correspond
to a $95\%$ uniform (or pointwise) confidence band. For small bandwidth
(say, less than $0.2$), the uniform confidence band is wide. However,
as long as the bandwidth is not so small, the confidence band appears
stable. Moreover, the confidence band includes zero over the entire
bandwidth range, demonstrating the robustness of the finding of no
incumbency advantage with respect to the bandwidth choice.

Lastly, we evaluate the external validity by testing the null hypothesis
that the TED is zero. It will tell us whether the RD estimate, which
by design only applies to the ``local'' incumbents whose previous
vote share margin resides at the $0$ cutoff, can be applied to incumbents
whose previous vote share margins are slightly higher than $0$. When
estimating the TED, we maintain the usual covariate balance condition
$\mu_{Z,+}=\mu_{Z,-}$ but do not impose the balance condition $\mu_{Z,+}^{(1)}=\mu_{Z,-}^{(1)}$
for the derivatives, so the CCFT's augmented regression estimator
for TED is not a proper choice, as discussed in Remark \ref{rem:TED CCFT}.
Our EB method gives a point estimate of TED equal to $-0.631$, and
a $p$-value for testing a zero TED equal to $0.064$. In comparison,
the standard estimate (without covariates) of TED is $-0.634$ with
the $p$-value equal to $0.023$.\footnote{Both EB and the standard estimates use the CCT bandwidth for the first
derivative, which equals $0.462$.} Therefore, both methods raise the concern of external validity of
applying the RD estimate to incumbents with share margins above $0$,
as the treatment effect is likely to significantly decrease in response
to a marginal increase in the score.

\begin{table}[H]
\centering \caption{Incumbency Advantage in Finnish Municipal Election: $\widehat{\vartheta}$
= RD LATE estimate, COV: NO = without covariate; YES = with covariate,
bandwidth selector being CCT's bandwidth ($h_{\mathsf{CCT}}=0.396$)
or the ROT bandwidth $=2.917$, the $p$-value for testing $\vartheta=0$.
The sample size $n=154,543$ for all RD methods, and $n=1,351$ for
the experimental data in the last row.}
\label{empi: RDD_Finnish} %
\begin{tabular}{lclccccc}
\toprule 
Methods & COV & $\widehat{h}$ & $\widehat{\vartheta}$ & $p$-value & $95\%$ CI & CI length & \tabularnewline
\midrule 
Standard & NO & CCT & 0.012 & 0.675 & $\left[-0.067,0.044\right]$ & 0.111 & \tabularnewline
\midrule 
$\mathrm{EL}_{p}$ & YES & ROT & 0.031 & 0.187 & $\left[-0.014,0.075\right]$ & 0.089 & \tabularnewline
 &  & CCT & -0.009 & 0.741 & $\left[-0.060,0.043\right]$ & 0.103 & \tabularnewline
$\mathrm{EL}_{p+1}$ & YES & ROT & 0.009 & 0.447 & $\left[-0.014,0.033\right]$ & 0.047 & \tabularnewline
 &  & CCT & -0.049 & 0.294 & $\left[-0.141,0.042\right]$ & 0.183 & \tabularnewline
CCFT & YES & CCT & 0.014 & 0.621 & $\left[-0.068,0.041\right]$ & 0.109 & \tabularnewline
\midrule 
Experimental data & NO &  & -0.010 & 0.516 & $\left[-0.060,0.040\right]$ & 0.100 & \tabularnewline
\citet{hyytinen2018does} &  &  &  &  &  &  & \tabularnewline
\bottomrule
\end{tabular}
\end{table}

\section{Conclusion and further discussion\label{sec:Conclusion}}

This paper proposes a balancing approach to covariate adjustment for
RD. The covariate balance condition can be viewed as over-identifying
restrictions, which the EB estimator incorporates when formulated
as an EL estimator. By establishing the first-order equivalence between
the EB estimator and CCFT's regression estimator, we show that the
efficiency gain can be attributed to incorporating covariate balance
as side information.

The EB problem (\ref{eq:entropy balancing}) can be cast in a more
general framework under which several extensions can be considered.
The construction follows \citet{BenMichael2021}. Consider the following
imbalance measure $\mathrm{imbalance}_{\mathcal{M}}\left(w_{1},...,w_{n}\right)\coloneqq\mathrm{sup}_{f\in\mathcal{M}}\left|\sum_{i}w_{i}\widehat{W}_{p,i}f\left(Z_{i}\right)\right|$
with respect to a function space $\mathcal{M}$. Let $m:\mathbb{R}_{+}\rightarrow\mathbb{R}_{+}\cup\left\{ \infty\right\} $
be an increasing and convex function. Let $\mathrm{complexity}\left(w_{1},...,w_{n}\right)$
denote some complexity (or dispersion) measure of the weights. Consider
the following risk minimization problem similar to \citet[Equation (12)]{BenMichael2021}:
\begin{equation}
\underset{w_{1}+\cdots w_{n}=1}{\mathrm{min}}\,m\left(\mathrm{imbalance}_{\mathcal{M}}\left(w_{1},...,w_{n}\right)\right)+\varsigma\cdot\mathrm{complexity}\left(w_{1},...,w_{n}\right),\label{eq:risk minimization weights}
\end{equation}
for some tuning parameter $\varsigma>0$. Denote $\mathcal{M}_{0}\coloneqq\left\{ \mathbb{R}^{d_{z}}\ni z\mapsto a+z^{\top}b:\left|a\right|+\sum_{j=1}^{d_{z}}\left|b^{\left(j\right)}\right|\leq1\right\} $.
We may also take the Cressie-Read divergence $D_{\varrho}\left(w_{1},...,w_{n}\parallel1/n,...,1/n\right)$
defined by (\ref{eq:Cressie-Read}) as a complexity measure. It is
clear that under $\mathrm{complexity}\left(w_{1},...,w_{n}\right)=D_{\varrho}\left(w_{1},...,w_{n}\parallel1/n,...,1/n\right)$
the generalized balancing problem (\ref{eq:generalized balancing})
can be written in the form (\ref{eq:risk minimization weights}) with
$\mathcal{M}=\mathcal{M}_{0}$ (see \citealp[Equation (13)]{BenMichael2021})
and $m\left(\cdot\right)$ taken to be $\mathbb{R}_{+}\ni x\mapsto\infty\cdot\mathbbm{1}\left(x>0\right)$,
so that exact balance is required.\footnote{The risk minimization problem can now be written as
\begin{eqnarray*}
 & \underset{w_{1},...,w_{n}}{\mathrm{min}} & \varsigma\cdot\mathrm{complexity}\left(w_{1},...,w_{n}\right)\\
 & \textrm{subject to} & \mathrm{imbalance}_{\mathcal{M}}\left(w_{1},...,w_{n}\right)=0.
\end{eqnarray*}
Clearly, the optimal weights do not depend on the choice of $\varsigma$
in this case. Relaxation of the ``exact balance'' constraint by
using a strictly positive threshold (see \citealp[Section 9.1.2]{BenMichael2021})
is also straightforward.} The sieve balancing problem (\ref{eq:sieve balancing}) is also of
the form (\ref{eq:risk minimization weights}) with $\mathrm{complexity}\left(w_{1},...,w_{n}\right)=KL\left(w_{1},...,w_{n}\parallel1/n,...,1/n\right)$
and $\mathcal{M}$ taken to be the broader sieve space.

An alternative balancing scheme similar to \citet{Hirshberg2021}
is based on solving (\ref{eq:risk minimization weights}) with $\mathcal{M}$
taken to be the sieve space, $m\left(\cdot\right)$ taken to be $x\mapsto x^{2}$
and $\mathrm{complexity}\left(w_{1},...,w_{n}\right)$ taken to be
the ``square Euclidean'' divergence given by $D_{-2}\left(w_{1},...,w_{n}\parallel1/n,...,1/n\right)$.
Then, we expect to find a dual characterization of the optimal weights
by using results from \citet{Hirshberg2021}. An asymptotic normality
result similar to Theorem \ref{thm:sieve} is expected to hold under
a suitable choice of tuning parameters $\left(h,k,\varsigma\right)$.
With $\mathcal{M}$ taken to be a ball in a Reproducing Kernel Hilbert
Space (RKHS), we get a balancing scheme similar to \citet{Kallus2020,Wong2017}
(see \citealp[Equation (5)]{Wong2017}). We also expect an asymptotic
normality result similar to Theorem (\ref{thm:sieve}) holds under
a suitable choice of the three tuning parameters ($\left(h,\varsigma\right)$
and the radius of the ball) and the assumption that the ``optimal
adjustment function'' $\eta^{*}$ (see Section \ref{subsec:Balancing-over-functions})
lies in the RKHS.

Our EB approach avoids the selection of the additional tuning parameter.
Another advantage is the favorable second-order properties developed
in the EL literature carry over to our proposed method. These include
a small nonlinearity bias of the point estimator and a simple analytical
correction to improve coverage accuracy for the confidence set. We
also show a uniform-in-bandwidth Wilks theorem, which can be used
for sensitivity analysis and robust inference along the lines of AK.
We also derive the distributional expansion for the EL ratio statistics
under the local imbalance condition and analyze the sensitivity of
the coverage performance to the balance assumption. Lastly, we demonstrate
that our approach can address previously unsolved covariate adjustment
problems in RD by deriving an EB-based covariate-adjusted estimator
for the TED. We also expect the large-deviation optimality results
for EL (e.g., \citealp{otsu_2010_bahadur}) to carry over. In the
presence of high-dimensional covariates (\citealp{arai2021regression,Kreiss2022}),
resorting to the dual characterization (\ref{eq:entropy balancing weight definition}),
we apply appropriate penalization in (\ref{eq:lambda_hat}) (\citealp{Chang2018})
to reduce the effective number of covariates. Properties of the penalized
EB are left for future investigation.

\bibliographystyle{chicago}
\bibliography{btw_selection,Jun_Ma_EL,RD,balancing}

\end{document}